\documentclass[12pt]{article}
\usepackage{amsmath}
\usepackage{graphicx}
\usepackage{indentfirst}
\usepackage{setspace}
\usepackage{hyperref}
\hypersetup{
colorlinks=true,
    linkcolor=black,
    filecolor=black,      
    urlcolor=blue,
    citecolor=black,
}
\usepackage[letterpaper, portrait, margin=1in]{geometry}
\usepackage{bbm}
\usepackage{dcolumn}
\usepackage{appendix}
\usepackage{subcaption}

\usepackage[para,online,flushleft]{threeparttable}

\usepackage{natbib}

\title{Who heeds the call to conserve in an energy emergency?  Evidence from smart thermostat data}
\author{Dylan Brewer \& Jim Crozier\thanks{School of Economics, Georgia Institute of Technology, 221 Bobby Dodd Way, Atlanta, Georgia 30332, \href{mailto:brewer@gatech.edu}{brewer@gatech.edu} and \href{mailto:crozierrj@gmail.com}{crozierrj@gmail.com}.  We gratefully acknowledge the support and contribution of Ecobee and Ecobee customers to this research.  This material is based upon work supported by the Google Cloud Research Credits program with the award GCP19980904.  We thank Soren Anderson, Laura Taylor, A. Justin Kirkpatrick, Prabhat Barnwal, Alecia Cassidy, Casey Wichman, Nathan Chan, Elise Breshears, Matt Oliver, Cody Orr, and seminar participants at Virginia Tech, Ohio University, the Southern Economics Association Meetings, the Midwest Economics Association Meetings, the UCLA Climate Adaptation Research Symposium, and the AERE Summer Conference for useful discussion and feedback.  Thank you to Graham Lewis for research assistance.  Employees at Consumers Energy provided valuable perspective and total gas consumption data, for which we are grateful.}}
\date{January 31, 2025}

\begin{document}

\maketitle
\begin{abstract}
\singlespacing \noindent In 2019, a fire at a natural gas plant and historically low temperatures caused an emergency shortage of natural gas in Michigan. A statewide emergency text alert asked households to turn thermostats down to 65\(^\circ\)F. We analyze the effectiveness of this request using high-frequency smart-thermostat data from Michigan and four neighboring states. Using a difference-in-differences research design, we find that Michigan households reduced thermostat settings by 1.1 degrees on average. Our results suggest that the use of the wireless emergency alert system was critical in creating an effective emergency response.  We examine heterogeneity in responsiveness by whether a household's baseline thermostat setting was above or below the compliance target of 65\(^\circ\)F and by Democratic Party gubernatorial vote share.
\end{abstract}


\section{Introduction}

During emergencies, government officials often make appeals to citizens to contribute effort to a common goal.  These appeals have taken the form of requests to contribute effort to a public good (such as buying war bonds), to voluntarily ration consumption of scarce goods (such as reducing consumption of water during a drought), or to comply with safety protocols (such as taking certain precautions during a pandemic).  Requests have become sophisticated over time as the communication medium has evolved from print materials, to radio and television announcements, and now to digital alerts with the ability to target specific individuals in real time.  The strategy of requests has also evolved with advances in psychological research in how to influence behavior \citep{cialdini2006} and the use of the ``nudge'' framework to reduce the costs of compliance \citep{thalersunstein2008}. Modern requests for pro-social behavior often take the form of nudge reminders, informational treatments \citep{allcotttaubinsky2015}, appeals to morality or ``moral suasion'' \citep{itoidatanaka2018}, appeals to expert authority \citep{brezaetal2021}, or social comparison to peer behavior to induce compliance \citep{ferraromirandaprice2011,allcottrogers2014}. 

This paper analyzes the efficacy of an emergency request for households to reduce thermostat settings during a natural gas shortage caused by a fire at a natural gas facility.  During the cold wave of the 2019 polar vortex, outdoor temperatures were extremely low, causing high demand for natural gas for space heating.  At 10:30 am on January 30, 2019, a fire broke out at Consumers Energy's largest natural gas storage facility.  Consumers Energy is a gas and electric utility that serves roughly half of Michigan's households, and 75\% of Michigan households rely on natural gas for heating.  By 1:00 pm, officials at the utility had recognized that demand for natural gas might exceed supply, with the potential to cause the system to fail.  At 2:30 pm, the utility requested via emails, social media, and news media that all households reduce natural gas consumption.  At 10:00 pm as pipeline pressures continued to drop, the Michigan Governor tweeted a request to conserve natural gas, and at 10:30 pm the state government sent an emergency alert on behalf of the utility directly to cell phones in Michigan requesting that households reduce thermostat settings to 65$^\circ$F or below.  The utility communications ensured that at least some households were aware of the request, but the cell phone alert went out to all households within the lower peninsula of Michigan.  The next day at 4:30 pm, the utility issued an ``all clear'' time of midnight to its customers---thanks to voluntary reductions in demand by households and industrial consumers, the system did not fail and natural gas outages were avoided.

To measure household responses to the requests, we use smart thermostat data provided by Ecobee's Donate-Your-Data program.  The data include thermostat setting and furnace fan run time at 5-minute intervals.  This high-frequency, household-level data allows us to observe and measure each household's response to the requests as the emergency unfolded.  Our empirical strategy uses households in the surrounding states of Ohio, Indiana, Illinois, and Wisconsin as control units for a difference-in-differences approach.  We find that mean thermostat settings, the proportion of homes with a thermostat setting at or below 65\(^\circ\)F, and furnace fan run time exhibit parallel pre-trends across the treatment and control units, which support our interpretation of our estimates as a causal effect of the emergency request on changes in thermostat settings.

Using a difference-in-differences strategy with four control states, we estimate the average treatment effect of the emergency request.  We find that on average, households lowered their thermostats by 1.1\(^\circ\)F, roughly 25\% of the size of the typical variation in the average thermostat setting.  The request increased the proportion of household thermostat settings at or below 65\(^\circ\)F by 10 percentage points, a 45\% increase relative to the proportion of households whose thermostat settings are normally 65\(^\circ\)F or less.  Finally, we examine the effect of the request on furnace fan run time, which is our best available proxy for household natural gas consumption.  We find evidence that the emergency request reduced the furnace run time by 1.5 minutes per hour, a 6\% reduction relative to the predicted run time for Michigan households during the emergency if there had been no reduction.  These results are robust across a number of specifications and checks for spillover treatment to border counties, and a placebo test with an earlier cold wave shows that this behavior is not driven by outside temperature alone. An event-study analysis reveals that the amplification of the utility's earlier request using the emergency alert system was critical for achieving compliance.  Prior to the phone alert, only 0.4\% of additional households reduced thermostat settings to 65\(^\circ\)F or less; after the phone alert, the additional compliance rate was as high as 20 percentage points.  The five-minute thermostat setting data shows that meaningful compliance with the emergency requests only began within five minutes of the cell phone alert.

The utility and phone alert framed the emergency request with a clear reference point of 65$^\circ$F that affected household responses.  We develop a theoretical framework that models a nudge with a compliance target as a conservation nudge with an additional moral tax on behavior that exceeds the reference point and as a moral subsidy on behavior that is below the reference point, extending the work of \cite{levittlist2007}.  A nudge with a compliance goal creates two types of reference point heterogeneity:  households that normally set the thermostat below this point were given license to contribute less effort to the public good, and households that normally set the thermostat significantly higher were asked to deviate more from their typical consumption patterns. Our model predicts that households with baseline thermostat settings below the 65\(^\circ\)F target will be less responsive than otherwise (and can increase their thermostat setting if the reference point effect dominates the conservation nudge), and households with thermostat settings near the target will exactly comply with the request and therefore have a smaller response to the emergency request relative to households whose baseline thermostat settings are far from the compliance target.  Finally, the model implies that a nudge with a compliance target imposes unequal marginal incentives on households, which violates the equimarginal principle and suggests that a nudge with a compliance target does not achieve the least-cost behavior change.

Empirically, we observe perverse framing effects in the data consistent with our model \citep{kahnemantversky1981}.  Using an estimate of each household's expected baseline temperature, we find that households that typically set the thermostat below 65$^\circ$F were less responsive to the emergency request on average.  Households with baseline thermostat settings between 67-71\(^\circ\)F reduced their thermostat settings by approximately 30\% more than households with thermostat settings between 63-65\(^\circ\)F. For baseline thermostat settings above the reference point, the higher the baseline thermostat setting, the less likely a household was to meet the compliance target.  At first, the average thermostat reduction increases with distance from the compliance target, but for baseline settings 75$^\circ$F and above the average thermostat reduction decreases as the compliance rate effect begins to dominate.  The results suggest that setting a more aggressive reference point trades off an increased treatment effect for individuals near the reference point with decreased compliance from discouraged individuals far from the reference point.

We scrutinize the role of political ideology and polarization as a factor in determining compliance with the request.  We hypothesize that responsiveness to the emergency request may have differed based on political affiliation \citep{costakahn2013}.  Our analysis studies the differential compliance of households in counties that supported the Democratic Party gubernatorial candidate in the 2018 election, using data on county-level election returns.  We show that compliance rates and the average reduction in thermostat setting are increasing in Democratic Party vote share, but some of the differences are not statistically significant and results are mixed for furnace fan run time.  Households in counties where the Democratic Party candidate's vote share was above 70\% reduced thermostat settings by about twice as much relative to households in counties where the Democratic Party candidate's vote share was below 40\%.  We do not claim that this represents a causal relationship given the correlation of political affiliation with unobserved characteristics of households that likely affect responsiveness to an emergency conservation request.

We make three main contributions to the literature.  First, the results of this study are important for policymakers studying compliance with emergency requests in a broad range of fields.  For instance, during the COVID-19 pandemic, local, national, and international governmental bodies sought to coordinate behavior to reduce the spread of the virus through a combination of compulsory policies and requests for voluntary compliance.  Pandemic-related policies and requests were met with mixed compliance and even open defiance, and a growing literature seeks to understand the effects of political affiliation on cooperation with requests for social distancing and stay-at-home orders \citep[e.g.,][]{allcottetal2020,barrioshochberg2020}.  Related to compliance with energy and environmental policy, economists have studied firm strategic avoidance of air quality monitoring \citep{zou2021}, imperfect enforcement of emissions caps \citep{sigmanchang2011}, voluntary reductions of emissions \citep{fosteretal2009,fostergutierrez2013}, compliance with the US acid rain program \citep{montero1999}, and the use of regulatory loopholes to avoid compliance with fuel efficiency regulations \citep{andersonsallee2011}.  \cite{beattyshimshackvolpe2019} study household emergency preparedness for hurricanes, finding that household behavior is highly influenced by recent hurricane events and that households generally do not follow government preparedness recommendations.  Other work shows that an increased perception of risk and confidence in government institutions increases compliance with hurricane evacuation orders \citep{whiteheadetal2000,KimOh2015}.  In another context, \cite{wichmantaylorvonhaefen2016} find that households in North Carolina reduced consumption of water during a drought when both voluntary and mandatory non-price mechanisms were implemented to restrict water use.  \cite{costakahn2013} find in a field experiment that households with progressive political stances are more responsive to energy conservation nudges and less likely to opt out of home electricity reports. During or after energy emergencies, utilities and governments often resort to emergency appeals for conservation.  For example, \cite{luyben1982} studies a 1977 request by President Carter for US households to reduce thermostat settings to 65\(^\circ\)F or below, finding that compliance was low overall (27\%) and that self-reported compliance was higher than recorded compliance.  After the 2000 and 2001 California energy crisis, energy conservation campaigns were successful in reducing electricity consumption when electricity prices were capped \citep{reisswhite2008}.  

Our paper contributes to the emergency response literature by providing what we believe is the most granular data on household compliance with emergency requests.  These data allow us to distinguish which emergency communications trigger household actions and to test our hypotheses of heterogeneous responses to the requests based on heterogeneity in the baseline thermostat setting.  In addition, the unexpected nature of the emergency and its isolation to one state creates a credible natural experiment that allows us to pursue an identification strategy that takes advantage of plausibly exogenous time and cross-sectional variation, which is uncommon for this literature. The ability to test for differential effects by political affiliation in our setting is also noteworthy, given the proximity of the emergency to a recent Governor's election in a politically divided state which gives us a recent measure of political affiliation by county.

Second, our work contributes to the empirical literature studying reference points and economic behavior.  Research in this area examines labor supply behavior relative to earnings expectations \citep[see][for a review]{ferrarotracy2022}, retirement decisions relative to age reference points \citep{seibold2021}, and loss aversion in tax filing \citep{engstrom2015}.  Other work focuses on the use of social comparison as a reference point to influence behavior \citep[e.g.,][]{hallsworthetal2017} and is often applied to energy conservation (e.g., \cite{allcott2011}, \cite{ferraromirandaprice2011}, and \cite{ferraroprice2013}, \cite{brentetal2015}).  In the charitable giving literature, suggested donation amounts increase voluntary contributions and anchor donations to the suggested amount \citep{EDWARDS2014}. \cite{harding2014} study how non-binding goal setting for energy conservation leads to behavior consistent with reference-dependent preferences.  \cite{brownetal2013} find that factory-default thermostat settings substantially impacted subsequent thermostat levels chosen in the workplace. 

Our paper contributes to the reference point literature by studying a novel reference point created within the phrasing of a governmental emergency request.  Our results suggest that for the policymaker, setting a reference point more aggressively trades off an increase in the effect of meeting the reference point with the cost of meeting the reference point.  In our context, further lowering the requested thermostat setting would have reduced compliance from those with high baseline settings but would have increased the effort from those with medium and low baseline settings.  These findings imply that policies may be designed so as to have effect-maximizing reference point levels.  Finally, our theoretical framework extends the moral payoff model of \cite{levittlist2007} to nudges with a compliance target, providing new predictions for heterogeneous behavior based on a person's pre-treatment distance from the reference point. Our compliance target model also has parallels to the social pressure model \citep{akerlofkranton2000} and more broadly the class of reference-dependent utility functions from prospect theory reviewed by \cite{dellavigna2009}.  These models have been applied, for example, in the charitable giving literature to model donation levels relative to an expected donation level that is similar to our compliance target set by the policymaker \citep{dellavignaetal2012}.  Our model differs from this in that the reference levels here are exogenously determined by a nudge rather than as an endogenous feature of the agent’s preferences.  Unlike those models, however, the moral payoff of meeting the compliance target is more similar to the reputational payoff from the self-signaling interpretation of the model of prosocial behavior in \cite{benaboutirole2006}. This is particularly important because we show that compliance targets can cause households to respond perversely when they are already in compliance, undermining the goal of the policy.

Third, this paper is relevant to the literature in environmental and energy economics analyzing the use of non-price mechanisms to conserve household consumption of water, natural gas, and electricity.\footnote{See \cite{carlssonetal2021} for a recent overview of papers analyzing nudges and non-price mechanisms.}  Given political constraints on raising prices of these goods, regulators and suppliers have sought to curb consumption via mandatory restrictions and voluntary requests, which have seen varying levels of success.  In \cite{itoidatanaka2018}, the authors conduct a field experiment that provided Japanese households with voluntary appeals or price incentives to reduce electricity consumption.  Relative to a control group, voluntary appeals resulted in a short term reduction in electricity consumption of 8\% while price incentives resulted in a reduction in electricity consumption of 17\% that was sustained over a longer period.  In the United States, \cite{holladayetal2015} find that utility and media appeals for peak-hour conservation can perversely lead to increases in energy consumption in anticipation of an outage, \cite{burkhardtetal2019} find very little responsiveness to voluntary appeals to conserve during peak hours relative to price mechanisms, while \cite{brandonetal2018} find that utility-led requests resulted in a 4\% average reduction in consumption during peak hours.  

In our setting, we document that households were unresponsive to utility and media appeals, but the intervention of the state government via the wireless alert system resulted in compliance of a similar magnitude to the field experimental results in \cite{itoidatanaka2018} and \cite{brandonetal2018}.  Our paper contributes novel evidence that reach and authority of the messenger can substantially affect the salience of emergency appeals for conservation, and that political context substantially impacts the effectiveness of non-price mechanisms. \cite{allcottrogers2014} find evidence that households reduce electricity consumption in response to home energy reports and that repeated treatments induce additional reductions and enforce habits, while \cite{costagerard2021} find that nine-month quotas on energy consumption provided reductions in energy use up to nine years after the quotas ended.  In contrast, we find that without repeated treatments that household compliance wanes by the end of the request, and while some conservation persists after the all-clear, the effect is modest.

The paper proceeds by describing the polar vortex and natural gas fire events in greater detail.  We develop a theoretical model of thermostat setting choice in the presence of a nudge with a compliance target, which generates hypotheses for household behavior.  We then introduce the smart thermostat data used in the paper.  Next, we present our empirical strategy and analysis, which we subdivide into a section estimating the average treatment effect of the request, a section presenting an event-study analysis, a section examining the effects of the reference point on behavior, and a section examining the role of political ideology on compliance.  Finally, section \ref{sec:conclusion} summarizes the findings and concludes.

\section{Polar vortex and natural gas fire events}

Extreme cold weather events caused by disturbances to the polar vortex have recently received significant attention in the United States and Europe.  Perhaps most notable was the 2021 polar vortex event that overwhelmed the electricity grid in Texas, killing 172 people and resulting in damages valued at levels ranging from \$20 billion to \$295 billion \citep{noaadisasters,perryman2021}.  Since 1980, winter disasters have resulted in 19 ``billion-dollar climate disasters'' in the United States, causing 1,223 deaths \citep{noaadisasters}.  There is only weak evidence that climate change is contributing to the perceived increase in polar vortex events \citep{blackportscreen}; however, aging energy infrastructure in the United States and Europe may increase the costs of such events in the future.

Beginning on Tuesday, January 29, 2019, temperatures in the Midwest declined to nearly record-low levels as cold air in the stratosphere over the Arctic blew southward over North America \citep{noaa2019}.  Temperatures reached -23\(^\circ\)F in Chicago, -13\(^\circ\)F in Detroit, -11\(^\circ\)F in Indianapolis, and as low as -45\(^\circ\)F elsewhere in the United States \citep{eia2019}.  On Wednesday, January 30, 2019, single-day estimated natural gas consumption in the United States hit an all-time high with 37.9 billion cubic feet consumed in a single day, and electricity demand in the Midwest approached all-time peaks \citep{eia2019b}.  In 2017, over 75 percent of Michigan homes used natural gas as the primary heating fuel \citep{mpsc2019a}.

Coinciding with this extreme demand-side stress, a supply-side emergency caused a near system-wide natural gas delivery failure in Michigan.  On January 30, 2019 at 10:30 am, a fire broke out at the Ray Compressor Station in Macomb County, Consumers Energy's largest natural gas storage facility \citep{mpsc2019a}.  Immediately after the fire broke out, the utility drew upon standby natural gas reserves to stabilize pipeline pressures \citep{mpsc2019b}. By 1:00 pm, Consumers Energy recognized the possibility that demand could exceed supply, which could cause total system failure, and contacted their highest demand industrial and commercial customers with requests to reduce consumption of natural gas.  At 2:26 pm, Consumers issued a tweet (appendix figure \ref{fig:tweet1}) requesting households to reduce thermostat settings and sent emails to residential and business customers requesting reductions in natural gas use. Shortly thereafter, the CEO of Consumers Energy took to Facebook Live to urge households to reduce thermostat settings (appendix figure \ref{fig:facebooklive}). The utility ultimately sent over 500,000 external emails, made 21 social media posts, and responded to 130 media inquiries on January 30-31 \citep{mpsc2019b}.  State-operated buildings reduced thermostat settings by 5\(^\circ\)F and manufacturers reduced consumption of natural gas \citep{desormeau2019}.  In addition, the utility issued mandatory curtailment orders for industrial and commercial natural gas customers and requested that natural gas electricity generators reduce generation to preserve residential heating \citep{mpsc2019b}.  On the supply side, Consumers Energy purchased 925 MMcf/day worth of same-day supply of natural gas for January 30th, of which only 61\% was ultimately delivered due to supply constraints \citep{mpsc2019b}.\footnote{Same-day natural gas delivery is relatively rare compared to same-day electricity generation, for example.  This event was the first time that Consumers Energy had attempted to secure same-day delivery \citep{mpsc2019b}.  For extreme-weather events, utilities can increase pressure in natural gas pipelines ahead of time, storing gas within the system.  Given that the flow of gas is not instantaneous, same-day supply is not typically used to balance supply and demand.}

\begin{figure}
    \centering
    \includegraphics[scale = 0.7]{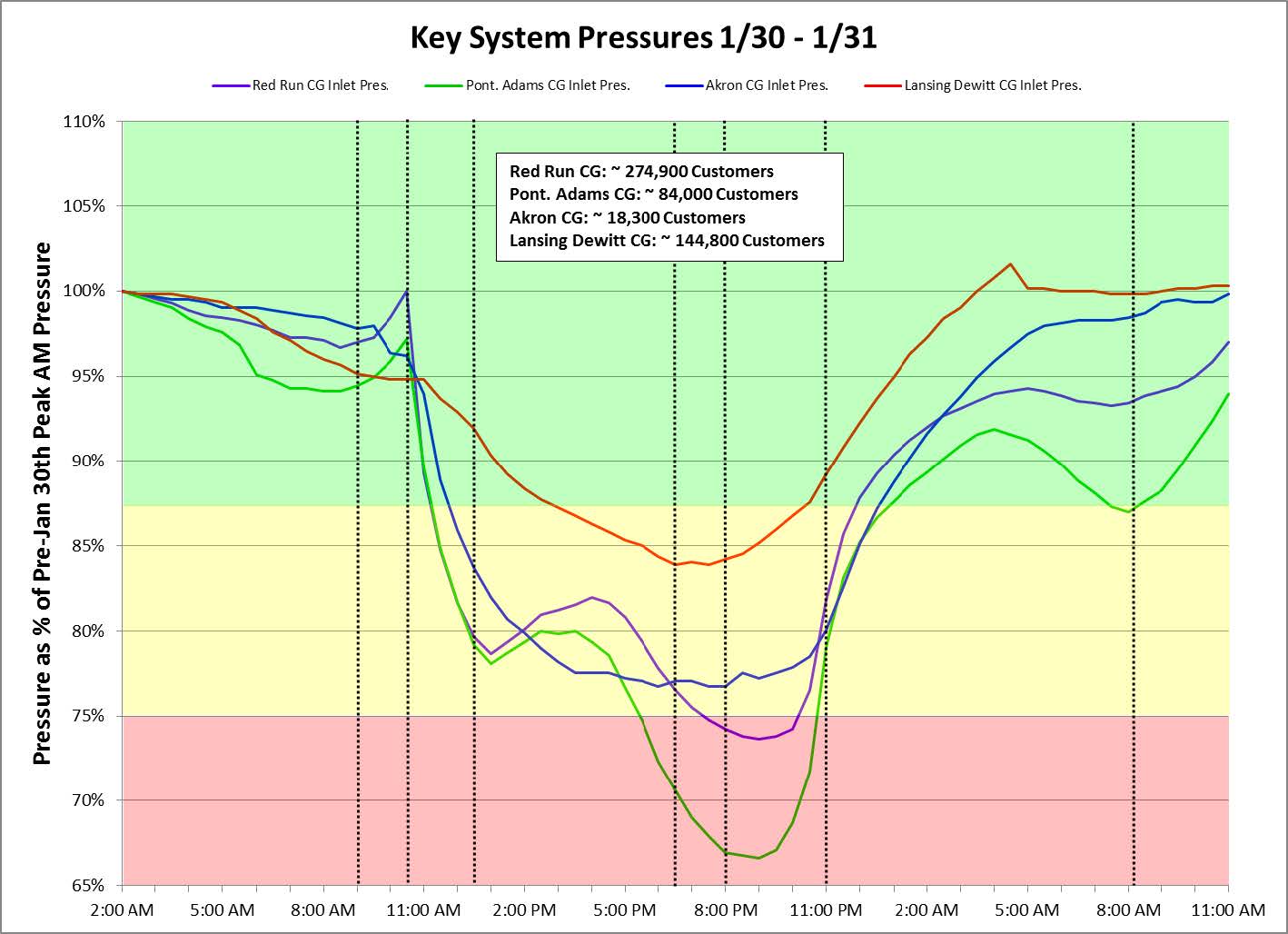}
    \caption{Each series corresponds to a Michigan natural gas pipeline instantaneous pressure, January 30-31, 2019.  Image source: Michigan Public Service Commission Case No. U-20463 \citep{mpsc2019b}.}
    \label{fig:pressures}
\end{figure}

Despite efforts to reduce non-residential consumption and to procure natural gas on the supply side, the system was still unpacking (losing pressure) going into the evening. Figure \ref{fig:pressures} displays Michigan natural gas pipeline pressures on January 30th and 31st.  Despite efforts to curb demand and increase supply, equilibrium pressures were dropping as the evening approached and temperatures continued to get colder.  At 8:00 pm, Consumers Energy reached out to the State of Michigan to make a final public appeal to households to reduce thermostat settings \citep{mpsc2019b}.

At 10:01 pm, the Governor of Michigan tweeted a request for households to reduce thermostats to 65\(^\circ\)F (appendix figure \ref{fig:tweet2}), and at 10:30 pm activated FEMA's Wireless Emergency Alert system to send a text alert (appendix figure \ref{fig:alert}) to all cell phones in Michigan asking households to reduce thermostat settings to 65\(^\circ\)F \citep{freep2019}.  The text of the cell phone alert message read ``Due to extreme temps Consumers asks everyone to lower their heat to 65 or less through Fri.''  Conversations with the Michigan State Police Emergency Management and Homeland Security department and Consumers Energy indicated that officials believed 65\(^\circ\)F was achievable, comfortable, and likely to be lower than the usual thermostat setting, but the number was chosen arbitrarily.

Shortly after the phone alert at 10:40 pm, 30\% of the Ray Compressor Station capacity came back online, which combined with demand reductions to begin to increase pressures \citep{mpsc2019b}.  Using data provided by Consumers Energy, forecasted natural gas demand using realized weather conditions was 3.3 billion cubic feet on January 30th and 2.9 billion cubic feet on January 31st.  After all reductions in consumption were accounted for, actual consumption was 3.0 billion cubic feet on January 30th and 2.6 billion cubic feet on January 31st, implying a 10.7\% and 10.5\% reduction in daily consumption from all sources (residential and non-residential).  On January 31st at 4:30 pm, Consumers Energy tweeted an ``all clear'' time of midnight that night, after which households could resume heating normally (appendix figure \ref{fig:tweet3}).

Did households listen and comply with the emergency requests issued by the utility and public officials?  Furthermore, how did the phrasing of the request around a thermostat setting of 65\(^{\circ}\)F affect household compliance?  Did political ideology and polarization affect which households were likely to comply?  We introduce a theoretical framework to generate hypotheses and test them using high-frequency smart thermostat data.

\section{Theoretical framework} \label{sec:theory}

Here we develop a theoretical model in which utility-maximizing households choose thermostat settings in the presence of an emergency request to reduce thermostat settings to a compliance target. Typically, households choose the thermostat setting that equates their marginal benefit with the marginal cost of increasing the thermostat by an additional degree Fahrenheit.\footnote{A choice modeled in \cite{brewer2022}.}  We model the emergency request to reduce thermostat settings to 65\(^\circ\)F as an additional nudge-motivated payoff term in the utility function similar to the moral payoff term introduced by \cite{levittlist2007} and considered in \cite{ferraroprice2013}.  In our case, we model the emergency request as a nudge-induced tax on thermostat settings that can arbitrarily differ above and below the requested reference level based on whether the household is in or out of compliance with the nudge. The nudge induces an additional moral tax on thermostat settings above the requested reference level and a moral subsidy on thermostat settings below the reference level.

Our model generates predictions that relative to a request without a compliance target, the emergency request will cause households with baseline thermostat settings above 65\(^\circ\)F to reduce thermostat settings more and households with thermostat settings below 65\(^\circ\)F to decrease thermostat settings less or even to increase thermostat settings.  Because of the discontinuity in the marginal incentive at the 65\(^\circ\)F target, households with baseline thermostat settings near the target may have an incentive to exactly comply with the request in a corner solution, while households far from the target will partially comply but have a larger treatment effect, all else equal.  The model thus suggests two perverse framing effects of the 65\(^\circ\)F target.  First, households heating at temperatures below the target will not conserve as much energy or may increase energy consumption.  Second, households with thermostat settings near 65\(^\circ\)F may only have an incentive to reduce the thermostat setting by a small amount if they exactly comply with the request.

\noindent \textit{Model setup:}

We consider a household \(i\) with characteristics \(\theta_i\) choosing the thermostat setting \(T_i\) in any given time period.  The regulator announces an energy emergency with requested thermostat setting \(R\) (which is equal to 65\(^\circ\)F in our context).  The utility function for household \(i\) is linearly separable in consumption benefits \(B_i(\cdot)\), energy heating costs \(C_i(\cdot)\), and the nudge-motivated payoff \(M_i(\cdot)\):\footnote{Linear separability in heating costs is not necessary to derive the comparative statics in this section but simplifies the notation and exposition.}
\begin{align}
    U_i(T_i,R,s;\theta_i) = B_i(T_i;\theta_i) - C_i(T_i;\theta_i) - M_i(T_i,R,s;\theta_i).
\end{align}
We assume that locally, households weakly prefer higher temperatures \(\partial B_i/\partial T_i \geq 0\) at a diminishing rate \(\partial ^2 B_i/\partial T_{i}^2 < 0\).  In addition, heating costs weakly increase with higher thermostat settings \(\partial C_i/\partial T_i \geq 0\) and are weakly convex \(\partial ^2 C_i/\partial T_{i}^2 \geq 0\).  The nudge-motivated payoff term exerts a marginal incentive on consumption that differs based on the requested reference point and the salience of the request.  Similar to \cite{ferraroprice2013}, \(s\) represents the salience or strength of the request, which can be impacted for instance by who makes the request, the household's perception of the stakes and consequentiality of its actions, or the wording of the request.  For nudges with compliance targets, the salience of the request for conservation may differ from the salience of the compliance target, so we differentiate the salience of each component and denote \(s=(s_C,s_R)\), where \(s_C\) is the salience of the overall request for conservation and \(s_R\) is the salience of the compliance target. Thus, the nudge-motivated payoff term has the following marginal effects on the household's utility:
\begin{align} \label{eq:marginalcost}
    \frac{\partial M_i(T_i,R,s;\theta_i)}{\partial T_i} = \begin{cases}
        \mu(T_i,s_C;\theta_i) + \tau(T_i,s_R;\theta_i) &\text{if } T_i > R, \\
        \mu(T_i,s_C;\theta_i) &\text{if } T_i = R, \\
        \mu(T_i,s_C;\theta_i) - \sigma(T_i,s_R;\theta_i) &\text{if } T_i < R.
    \end{cases}
\end{align}
The nudge-motivated marginal conservation incentive \(\mu(T_i,s_C;\theta_i)\geq 0\) is weakly increasing in salience \(\partial \mu/\partial s_C \geq 0\).  The marginal conservation incentive \(\mu\) reflects an incentive to comply with the request that is independent of the reference point.  This term might reflect a warm glow from contributing to a public good, a desire to avoid a negative reputation, or the household's belief about the additional private returns to reducing consumption during the emergency (in this case the benefits of avoiding a natural gas outage).\footnote{There are approximately 3 million households heating with natural gas in Michigan, meaning it is unlikely that any individual household would have a significant impact on the natural gas system, but it is possible that a household may \textit{believe} that its contribution is necessary for the system to remain stable. Indeed, a nudge that convinced the household of the consequentiality of its actions could be understood in this framework as having a stronger salience \(s_C\).}  The moral tax term \(\tau(T_i,s_R;\theta_i) \geq 0\) is weakly increasing in thermostat setting \(\partial \tau/\partial T_i \geq 0\) and salience \(\partial \tau/\partial s_R \geq 0\), and reflects a penalty for being out of compliance or nonconforming that might arise intrinsically or from social pressure.  The moral subsidy term \(\sigma(T_i,s_R;\theta_i) \geq 0\) is weakly decreasing in thermostat setting \(\partial \sigma/\partial T_i \leq 0\) and weakly increasing in salience \(\partial \sigma/\partial s_R \geq 0\), and reflects a reduction in the incentive to provide additional effort when the household is already in compliance with the request.  This reduction might be intrinsically motivated (for example, a household might feel that it has contributed its fair share) or might be due to a reduction in the social pressure to conform. 

Our model nests three types of nudges: a pure conservation nudge, a pure reference point nudge, and a nudge with a compliance target. The model in \cite{ferraroprice2013} is a pure conservation nudge with \(\mu > 0\) and \(\tau = \sigma = 0\).\footnote{Although the model in \cite{ferraroprice2013} matches the pure conservation nudge, the social comparison in their field experiment could be considered a compliance target.} Conservation nudges increase the marginal cost of consumption without referring to a target level of consumpion. A pure reference point nudge would have \(\mu = 0\) and \(\tau, \sigma > 0\), and would correspond to a nudge that recommends a particular level of consumption without any additional incitement to reduce consumption.  Finally, a nudge with a compliance target would have \(\mu > 0\) and \(\tau, \sigma > 0\).  The nudge with a compliance target combines both a recommended level of consumption with an appeal for conservation and fits the emergency request in our context as households were asked to conserve energy and to set the thermostat to a specific level.  

A more salient request will increase the strength of the marginal conservation incentive \(\mu\) if the conservation component of the request is made more salient (\(s_C\)) and may result in a larger moral tax \(\tau\) on thermostat settings above the reference point and a larger moral subsidy \(\sigma\) on thermostat settings below the reference point if the compliance target is made more salient (\(s_R\)).  Changing the requested thermostat setting \(R\) will change the impact of the nudge for all households depending on their baseline thermostat setting.  For example, setting a more aggressive requested thermostat setting will increase the moral tax on households with high baseline thermostat settings and decrease the moral subsidy on households with low baseline thermostat settings.  For households with baseline thermostat settings just below the initial requested level, a more aggressive requested thermostat setting might change the moral subsidy to a moral tax.

When there is no emergency request, the household maximizes utility by choosing the thermostat setting that equates the household's marginal benefit of indoor temperature with the marginal energy cost in the first-order condition: 
\begin{align}
\frac{\partial B_i(T_i^0;\theta_i)}{\partial T_i} = \frac{\partial C_i(T_i^0;\theta_i)}{\partial T_i}.
\end{align}  
The thermostat setting \(T_i^0\) that solves this equation is the household's utility-maximizing baseline thermostat setting.

When there is an emergency request, the household's marginal cost of heating includes both the marginal nudge-motivated cost and the marginal energy cost of heating.  The household chooses thermostat setting \(T_i'\) that maximizes utility by equating the marginal benefit of indoor temperature with the new marginal cost of heating in the first order condition:
\begin{align}
\frac{\partial B_i(T_i';\theta_i)}{\partial T_i} = \frac{\partial C_i(T_i';\theta_i)}{\partial T_i} + \frac{\partial M_i(T_i',R,s;\theta_i)}{\partial T_i},
\end{align}  
or by choosing a corner solution of exact compliance with the request: \(T_i' = R\).  If the request specifies a compliance target \(R\) and induces an additional nudge-motivated payoff so that \(\mu(T_i,s_C;\theta_i) > 0\), \(\tau(T_i,s_R;\theta_i)>0\), and \(\sigma(T_i,s_R;\theta_i)>0\), the reference-independent nudge-motivated payoff term \(\mu\) will provide a marginal incentive for households to reduce thermostat settings while the moral tax \(\tau\) and subsidy \(\sigma\) will on the margin push all thermostat settings toward the compliance target. For a conservation nudge with a compliance target, there are two cases that generate distinct predictions for household behavior depending on the relative magnitude of the conservation incentive and the reference point.  

\noindent \textit{Case 1} (\textit{conservation-dominant nudge}, \(\mu > \sigma\)):

If the request-induced conservation incentive is stronger than the compliance target's moral subsidy \(\mu(T_i',s_C;\theta_i) > \sigma(T_i',s_R;\theta_i)\), the household will reduce its thermostat setting regardless of whether its baseline thermostat setting \(T_i^0\) is above or below the compliance target \(R\).  The reason for this is that the total nudge-induced marginal incentive \(\partial M_i/\partial T_i\) will be positive, increasing the marginal cost of heating and leading to a reduction in thermostat settings.  Because of the potential discontinuity in the marginal incentive at the compliance target, households with baseline thermostat settings above the compliance target may choose the compliance target as a corner solution. If a compliance target is below the household's baseline thermostat setting, the household will reduce the thermostat setting less than if the compliance target is above the household's baseline thermostat setting:  \((T_i^0-T_i' | T_i^0 < R ) < (T_i^0-T_i' | T_i^0 > R)\).

Relative to a pure conservation nudge (with \(\tau=\sigma=0\)), the addition of a relatively weak compliance target can induce greater reductions in thermostat settings for households with baseline thermostat settings above the reference level \(T_i^0>R\). A household with baseline thermostat setting above the reference point that chooses the reference level as a corner solution would have conserved more if facing only the marginal conservation incentive because the discontinuity in marginal incentives at the reference level made it optimal to stop at exact compliance with the request. At the same time, the compliance target can cause perverse behavior for households with baseline thermostat settings below the reference level relative to a pure conservation nudge because the moral subsidy term \(\sigma\) will incentivize households on the margin to increase their thermostat settings. 

\begin{figure}
    \begin{subfigure}[t]{.49\textwidth}
        \centering
        \includegraphics[width=\linewidth]{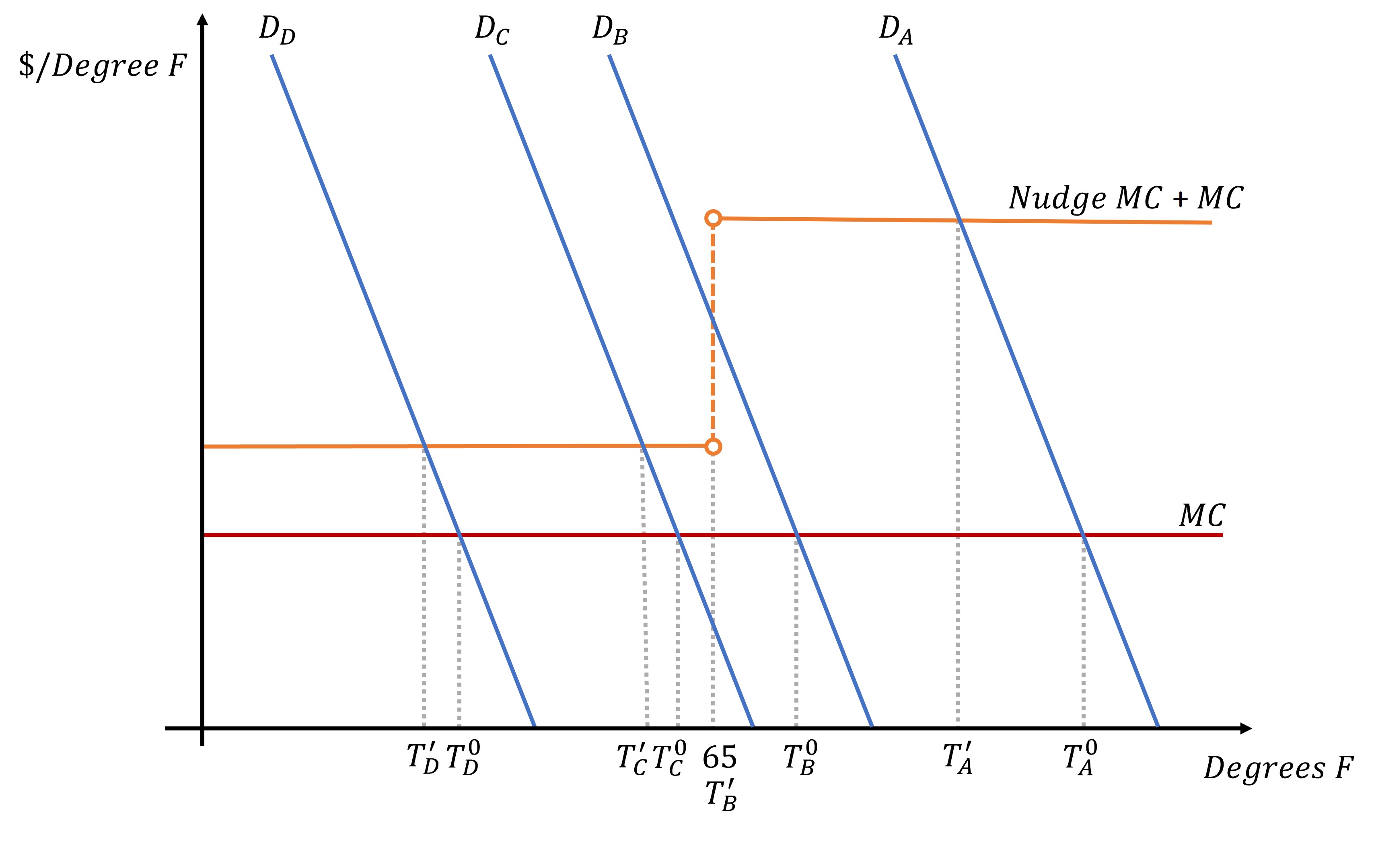}
        \caption{}\label{fig:theory_a}
    \end{subfigure}
    \begin{subfigure}[t]{.49\textwidth}
        \centering
        \includegraphics[width=\linewidth]{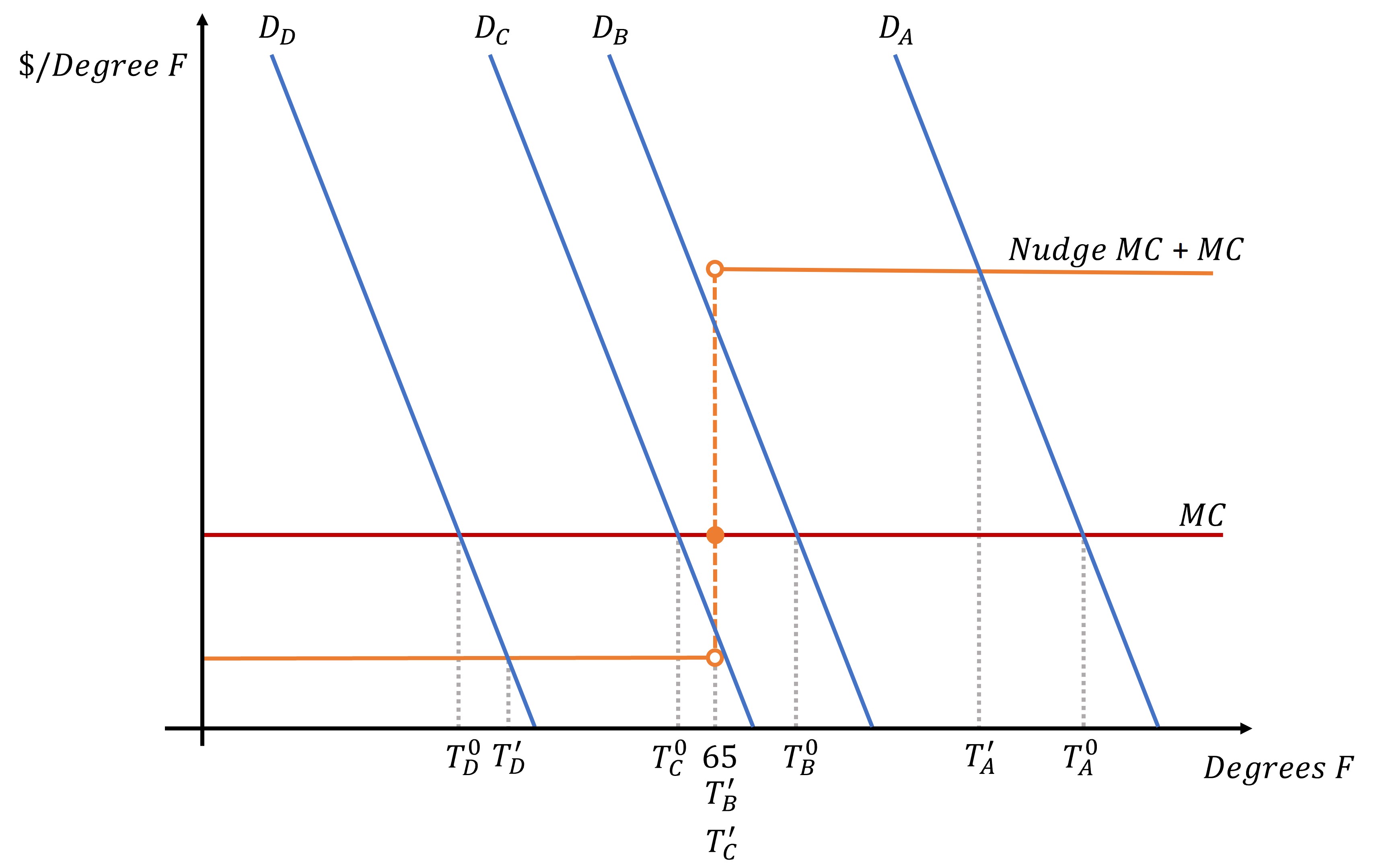}
        \caption{}\label{fig:theory_b}
    \end{subfigure}
    \caption{Theoretical effect of a request to reduce thermostats to 65\(^\circ\)F for four household types with marginal willingness to pay for heating \(D_i\) for \(i \in \lbrace A, B, C, D \rbrace\).  Before the emergency request, households equate marginal willingness to pay with the marginal cost of heating and choose thermostat setting \(T_i^0\) and after the request households choose \(T_i'\). Panel (a) illustrates the conservation-dominant nudge case, where the conservation incentive is stronger than the compliance target's moral subsidy.  Panel (b) illustrates the reference-point-dominant nudge case, where the compliance target's moral subsidy is stronger than the conservation incentive.}
    \label{fig:theory_ab}
\end{figure}

Figure \ref{fig:theory_a} illustrates the model with constant marginal cost and linear demand curves for four hypothetical households.  Households \(i \in \lbrace A, B, C, D \rbrace\) have marginal willingness to pay for heating \(D_i\).  Prior to the emergency request, each household equates marginal willingness to pay with marginal cost of heating and chooses thermostat setting \(T_i^0\).  After the request, the marginal cost of heating now includes the moral marginal cost.  Household \(A\) partially complies, reducing the thermostat setting to \(T_A' > 65\), while household \(B\) fully complies, choosing the corner solution \(T_B' = 65\).  The treatment effect for household \(B\) is thus limited by the reference point.  Households \(C\) and \(D\) are already under the compliance target but choose to reduce thermostat settings further because the nudge still increases the marginal cost of heating relative to the baseline.  For \(C\) and \(D\), their response to the nudge is dampened by the moral subsidy term \(\sigma\), but the overall impact on behavior is still for these households to reduce the thermostat setting.

\noindent \textit{Case 2} (\textit{reference-point-dominant nudge}, \(\mu \leq \sigma\)):

If the request-induced conservation incentive is weaker than the compliance target's moral subsidy \(\mu(T_i',s_C;\theta_i) \leq \sigma(T_i',s_R;\theta_i)\), a salient emergency request causes \textit{all} households to move the thermostat closer to the reference level, whether they initially were heating above or below the reference point. If the requested thermostat setting is less than the household's baseline level \(R < T_i^0\), the nudge-induced marginal cost of heating is positive, causing the household to reduce its thermostat setting: \((T_i^0-T_i'|R < T_i^0) > 0\).  If the requested thermostat setting is greater than the household's baseline level \(R > T_i^0\), the nudge-induced marginal cost of heating is a net subsidy, causing the nudge to backfire as the household increases its thermostat setting: \((T_i^0-T_i'|R > T_i^0) < 0\).  If the requested thermostat setting is equal to the household's baseline level \(R = T_i^0\), the nudge has no marginal effect on the household's thermostat setting: \((T_i^0-T_i'|R = T_i^0) =0\).  Households with baseline thermostat settings either above or below the compliance target may choose the compliance target as a corner solution due to the discontinuity in marginal incentives at the compliance target.  

As in the conservation-dominant nudge case, adding a relatively strong compliance target to a conservation nudge can induce greater reductions in thermostat settings for households with baseline thermostat settings above the reference level \(T_i^0>R\).  However, in the reference-point-dominant nudge case, the compliance target is completely counterproductive for households with baseline thermostat settings below the reference level \(T_i^0<R\) as these households will not comply or will increase thermostat settings relative to their baseline.  

Figure \ref{fig:theory_b} illustrates the model in the reference-point-dominant nudge case.  Similar to the previous case, household \(A\) partially complies, reducing the thermostat setting to \(T_A' > 65\), while household \(B\) fully complies, choosing the corner solution \(T_B' = 65\).  Households \(C\) and \(D\) respond perversely to the emergency request and increase thermostat settings in response to the moral subsidy.  Household \(C\) mirrors the behavior of household \(B\) and chooses the corner solution \(T_C'=65\), while household \(D\) mirrors the behavior of household \(A\) and increases the thermostat setting to \(T_D' > T_D^0\) toward the reference level.  

\noindent \textit{Empirical hypotheses}:

The behavior of hypothetical households \(A\) and \(B\) generates our first two empirical hypotheses.  The first hypothesis is that households with baseline thermostat settings \(T_i^0\) near the reference point will be more likely to comply fully with the request relative to households far above the reference point.  The second hypothesis is that households with baseline thermostat settings far from the reference point will have a larger average treatment effect as they are less likely to achieve the corner solution.  We believe this effect is likely to attenuate for households with the highest baseline thermostat settings, as the compliance target may appear out of reach or unrealistic.  This attenuation would be consistent with a moral tax function decreasing in thermostat setting, or \(\partial \tau(T_i) / \partial T_i < 0\).

For households with baseline thermostat settings below 65\(^\circ\)F such as hypothetical households \(C\) and \(D\), case 1 (conservation-dominant nudge) predicts that the emergency request will lead to reduced thermostat settings while case 2 (reference-point-dominant nudge) predicts that the emergency request will lead to increased thermostat settings.  Thus, we hypothesize that the emergency request will lead to weaker treatment effects for households with baseline thermostat settings below 65\(^\circ\)F relative to those with baseline thermostat settings above 65\(^\circ\)F, while the theoretical prediction for the overall sign of the treatment effect is ambiguous.

Increasing the salience \(s=(s_C,s_R)\) of the request will increase the magnitude of each of the nudge payoff terms, causing households to reduce the thermostat setting more and to move the thermostat setting closer to the reference level on the margin.  For households facing a moral tax, a more salient request will increase both the conservation incentive and the moral tax, causing the household to reduce the thermostat setting further unless the household is already at a corner solution.  For households facing a moral subsidy, a more salient request will increase both the conservation incentive and the moral subsidy, which could result in either a reduction in the thermostat setting or an increase in the thermostat setting depending on the relative magnitude of \(\partial \mu / \partial s_C\) and \(\partial \sigma / \partial s_R\).  

In the context of the polar vortex and natural gas shortage, the strength or salience varies based on the platform of the request and identity of the messenger.  The utility company broadcast the initial emergency request, which was also taken up by local traditional news media from 2:30 pm - 10:00 pm and included the 65\(^\circ\)F compliance target.  Beginning at 10:01 pm, the Governor took up the emergency appeal via Twitter, and the cell phone alert occurred at 10:30 pm.  We hypothesize that the initial request was a very low-salience request, but that the cell phone alert substantially increased the salience of both the conservation component of the request \(s_C\) and the compliance target \(s_R\). Prior work studying emergency appeals for energy conservation have found that appeals via channels such as local news fail to induce reductions in energy consumption or can perversely cause households to increase energy consumption in expectation of a potential outage \citep{holladayetal2015}.  Thus, we expect that most reductions in thermostat setting will occur after 10:30 pm, reflecting the change in salience. Furthermore, prior work has shown that energy conservation requests have a stronger impact for households that are more politically progressive \citep{costakahn2013}, suggesting that the request may have been more salient for these households.  We hypothesize that the treatment effect will be larger for households in areas that voted for the Democratic Party Governor in the 2018 election.

Finally, this model suggests that an emergency request tied to a reference point is not the least-cost nudge required to achieve a given reduction in energy consumption.  Equation \ref{eq:marginalcost} shows that households will face heterogeneous marginal costs of energy consumption, violating the equimarginal principle (also called the equal marginal principle or Gossen's second law), immediately implying this nudge is of higher cost relative to a nudge that exerts a moral marginal cost of consumption that applies no matter the household's baseline thermostat setting.  Perhaps more intuitively, a nudge that perversely subsidizes additional consumption of energy for some households when energy is scarce increases the mismatch between the retail price of energy and the wholesale price of energy for those households. If a policy goal is to maximize conservation rather than economic surplus, the addition of a compliance target can create perverse incentives for some households but may still be optimal if the moral tax incentive is strong and the moral subsidy incentive is weak. While it is possible that the compliance target increases the strength or salience of the request by providing households with a concrete action to perform, a request for a uniform reduction in the thermostat setting (e.g., a request to reduce the thermostat setting by 5\(^\circ\)F) would still be concrete.  Such a uniform request would rule out perverse behavior, but would still not satisfy the equimarginal principle.  We revisit these implications for the design of an emergency request in the conclusion.

\section{Smart thermostat data and research design}

We use data on smart thermostat temperature settings provided by Ecobee as part of the 2022 release of the ``Donate Your Data'' program.\footnote{This paper is among a few others studying the effects of smart thermostats or using smart thermostat data.  \cite{geho2018} study how households change the thermostat in response to warm and cold weather and assess the degree of habit formation in thermostat settings.  A working paper by \cite{brandonetal2021} find that smart thermostats alone do not result in energy savings, partially due to users overriding smart thermostat algorithms.  Another working paper by \cite{blonzetal2021} studies an energy-efficiency program implemented by Ecobee that automatically reduces thermostat settings during peak pricing periods.}  The raw data include 5-minute interval observations of thermostat settings and the amount of time the furnace fan was running.  In addition, a small amount of self-reported information about the household is available, including the location up to city and state, the number of occupants, the size, age, and number of floors of the home, and when the smart thermostat was first connected.  We augment these data with hourly outdoor temperature, humidity, wind speed, precipitation, snow depth, and cloud cover at the city level purchased from Visual Crossing.  Consumers Energy only serves households in Michigan.  We limit the sample of households to those in Michigan and the surrounding four states for controls: Ohio, Indiana, Illinois, and Wisconsin.\footnote{We exclude households in the Upper Peninsula of Michigan because these households are on a separate natural gas network and it is unclear whether they were treated or were controls.  The Upper Peninsula accounts for about 3\% of the population of Michigan; dropped households represent 1\% of the Michigan sample in the data.}  We include all observations between January 2nd and February 3rd, 2019. There are 3,036 households from Michigan and 9,221 control households in the final sample.  

\begin{table}
    \centering
    \begin{threeparttable}
    \caption{Summary statistics \label{tab:summary} }
    {\def\sym#1{\ifmmode^{#1}\else\(^{#1}\)\fi} \begin{tabular}{l*{3}{cc}} \hline
                    &\multicolumn{1}{c}{(1)}&\multicolumn{1}{c}{(2)}&\multicolumn{1}{c}{(3)}\\
                    &\multicolumn{1}{c}{Michigan}&\multicolumn{1}{c}{Controls}&\multicolumn{1}{c}{Difference}\\
                    &     Mean/SD&     Mean/SD&Diff./t-stat  \\
\hline
Sq ft               &    2,387.18&    2,545.96&      158.78**\\
                    &  (1,033.94)&  (1,138.21)&      (6.83)  \\
Age of home (years) &       32.73&       33.96&        1.23* \\
                    &     (29.00)&     (31.25)&      (1.98)  \\
Number of occupants &        1.19&        1.34&        0.15**\\
                    &      (1.70)&      (1.75)&      (4.27)  \\
January 2 - 29 thermostat setting&       66.87&       67.45&        0.58**\\
                    &      (3.60)&      (3.32)&      (7.83)  \\
January 30 thermostat setting before event&       67.38&       68.20&        0.82**\\
                    &      (3.93)&      (3.79)&      (9.88)  \\
January 30 - 31 thermostat setting during event&       67.09&       68.67&        1.58**\\
                    &      (3.48)&      (3.56)&     (21.35)  \\
January 2 - 29 outside temperature&       23.86&       25.08&        1.22**\\
                    &      (2.33)&      (4.18)&     (20.02)  \\
January 30 - 31 outside temperature&       -0.29&       -0.50&       -0.22**\\
                    &      (0.24)&      (0.55)&    (-29.69)  \\
 & & & \\
Households          &       3,036&       9,221&      12,257  \\
Observations        &     581,163&   1,780,876&   2,362,039  \\
\hline \end{tabular} }

    \begin{tablenotes}
        \footnotesize Each observation is a household in a four-hour period in Michigan or in the control states (Ohio, Indiana, Illinois, and Wisconsin) from January 2nd-February 3rd. Difference in means t-test performed assuming unequal variances collapsed to the household level: ** p$<$0.01, * p$<$0.05.
    \end{tablenotes}
    
    \end{threeparttable}
\end{table}

It is possible that the households in our sample responded to the emergency request differently than the general population due to selection into smart thermostat ownership and the Donate-Your-Data program.  On observable characteristics, the Ecobee Donate-Your-Data households are comparable to the average household in the nationally representative Residential Energy Consumption Survey sample, though the Ecobee households have slightly more members \citep{meieretal2019}.  In our sample, there are two notable differences in observable characteristics between the Ecobee sample and a random sample of households.  First, 99.89 percent of sample households heat with natural gas, compared to 75 percent of population households in Michigan.  Second, the average number of occupants per home in the United States was 2.52 in 2019 versus 1.19 of Michigan Ecobee households and 2.34 control households \citep{households}.  Our primary selection concern is that Ecobee households who join the Donate-Your-Data program may be more likely to contribute to other public goods and therefore more likely to comply with the emergency request.  Another concern we have is that the Ecobee smart thermostat may make compliance with the request easier than compliance using a conventional thermostat because Ecobee thermostats can be controlled remotely via an app.  While these issues are not a problem for our research design because treatment and control households are the same (i.e., our research design is internally valid), it may be that our estimated treatment effects overstate the response of the average household.  In appendix section \ref{sec:externalvalidity} we discuss the external validity of the estimates in more detail.  We find that our estimated treatment effect is smaller than that estimated using aggregate natural gas consumption data provided by the utility, the opposite of what we would expect if smart thermostat users were more likely to comply with the request.\footnote{Aggregate consumption data includes residential, commercial and industrial consumption.}  We also show in that section that early adopters of smart thermostats responded similarly to late adopters.  These findings mitigate our concerns that our estimates may not generalize to those with conventional thermostats. 

For computational tractability and to reduce noise, we aggregate the data into four-hourly time intervals, resulting in 581,163 household-time observations in Michigan and 1,780,876 household-time observations in the controls.  We compute the four-hourly average thermostat setting and minutes per hour the furnace was running.  Analysis at the hourly level or lower does not substantially change the estimates, which we display in appendix \ref{app:hourly}.  Table \ref{tab:summary} displays summary statistics for the treatment and control groups.  Due to the large sample size, most differences in means for treatment and control are statistically significant, but are not practically meaningful and do not pose a threat to our empirical strategy.  The primary differences we see are that sample homes in Michigan are slightly smaller and have fewer occupants on average.  During January 2 - 29, Michigan and the control states experienced average temperatures around 24 and 25\(^\circ\)F.  The outside average temperature in both treatments and controls dropped to just under 0\(^\circ\)F during the event.  From January 2nd through 29th, Michigan household thermostat settings were 0.6\(^\circ\)F lower than the control household thermostat settings.  In the hours before the first appeal to lower thermostats, the gap in thermostat settings had increased to 0.78\(^\circ\)F.  After households were asked to reduce the thermostat, the gap increased to 1.58\(^\circ\)F.  

\begin{figure}
    \begin{subfigure}[t]{.49\textwidth}
        \centering
        \includegraphics[width=\linewidth]{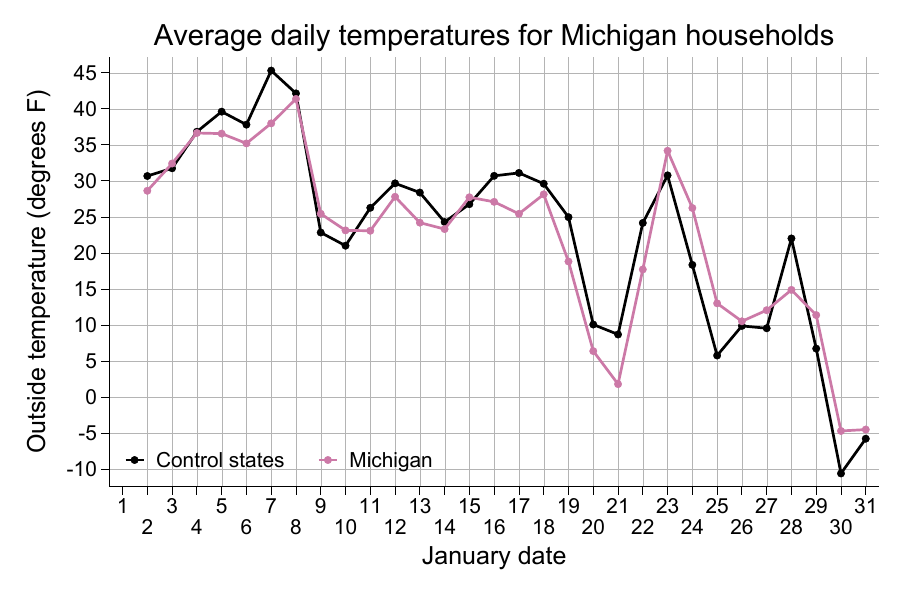}
        \caption{}\label{fig:temps_a}
    \end{subfigure}
    \begin{subfigure}[t]{.49\textwidth}
        \centering
        \includegraphics[width=\linewidth]{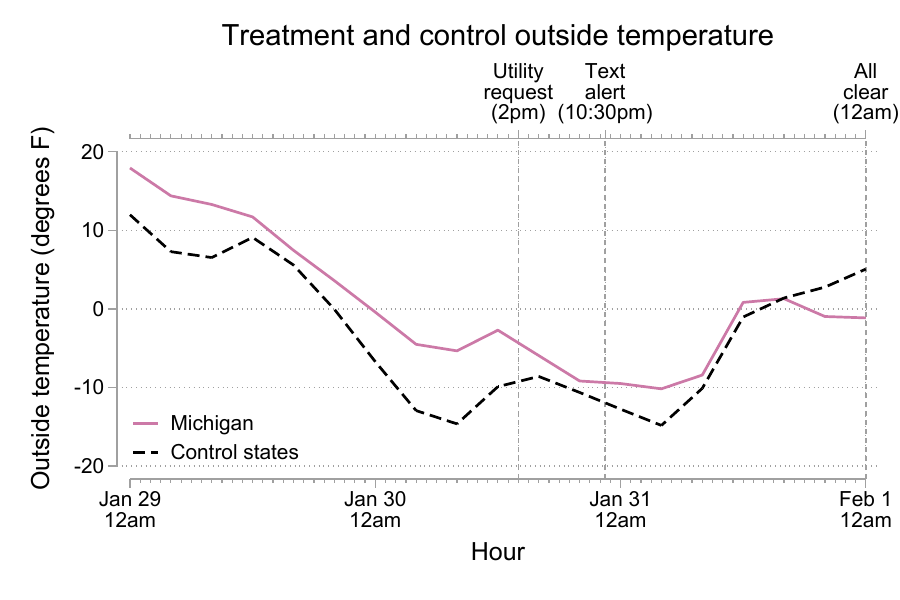}
        \caption{}\label{fig:temps_b}
    \end{subfigure}
    \caption{(a): Sample mean daily temperatures for Michigan and control households in January 2019. January 20-21 are used as a ``placebo'' event for the January 30-31 polar vortex and emergency request. (b): Sample mean four-hour average outdoor temperatures for Michigan and control households in the hours before and during the emergency.}
    \label{fig:temps}
\end{figure}

Our research design compares outcomes in Michigan to those in control states where there were no appeals to reduce natural gas consumption.  We consider three outcomes:  the thermostat setting, a binary variable equal to one if the thermostat setting is at or below 65\(^\circ\)F, and the amount of time the furnace fan ran during the hour. Standard furnaces run at essentially one speed.\footnote{Two-stage furnaces can run at full-speed and half-speed depending on the scenario to reduce energy use and ramping costs.}  When the thermostat setting is reduced, the home cools to the new setting and the furnace does not run, saving energy.  When the indoor temperature goes below the new thermostat setting, the furnace runs again at full speed for a short period to maintain the indoor temperature.  Thus, furnace fan running time is our best proxy for natural gas consumption \citep{meieretal2019}.  Consumers Energy shared daily aggregate natural gas consumption and forecasts of expected consumption from their internal forecasting model, which we use to construct an estimate of total demand response from all sources.  

Households in the control states (the ``Great Lakes states'': Ohio, Indiana, Illinois, and Wisconsin) have weather patterns and housing stocks similar to  Michigan.  Furthermore, these states also experienced extreme cold during the polar vortex event.  Figure \ref{fig:temps_a} plots daily average temperatures during January for Michigan and control states, showing the polar vortex event at the end of the month in addition to a similar cold wave on January 20th and 21st that we study in a placebo exercise in the appendix.  Figure \ref{fig:temps_b} plots the four-hour average outdoor temperature before and during the emergency, demonstrating that both treatment and control groups experienced similar conditions during the polar vortex.

\begin{figure}
\centering
\begin{subfigure}[t]{.49\textwidth}
    \centering
    \includegraphics[width=\linewidth]{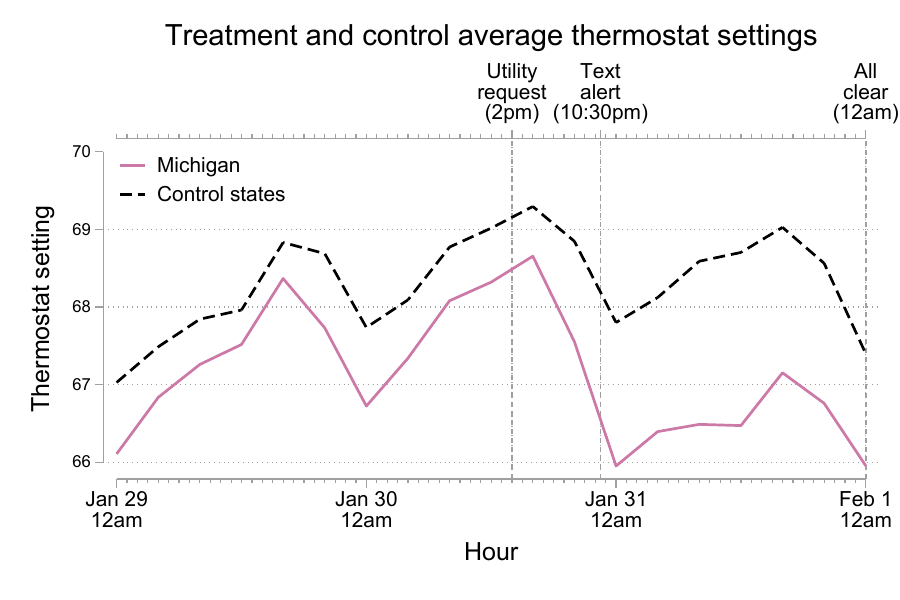}
            \caption{}\label{fig:thermostatparallel}
    \end{subfigure}
\begin{subfigure}[t]{.49\textwidth}
    \centering
    \includegraphics[width=\linewidth]{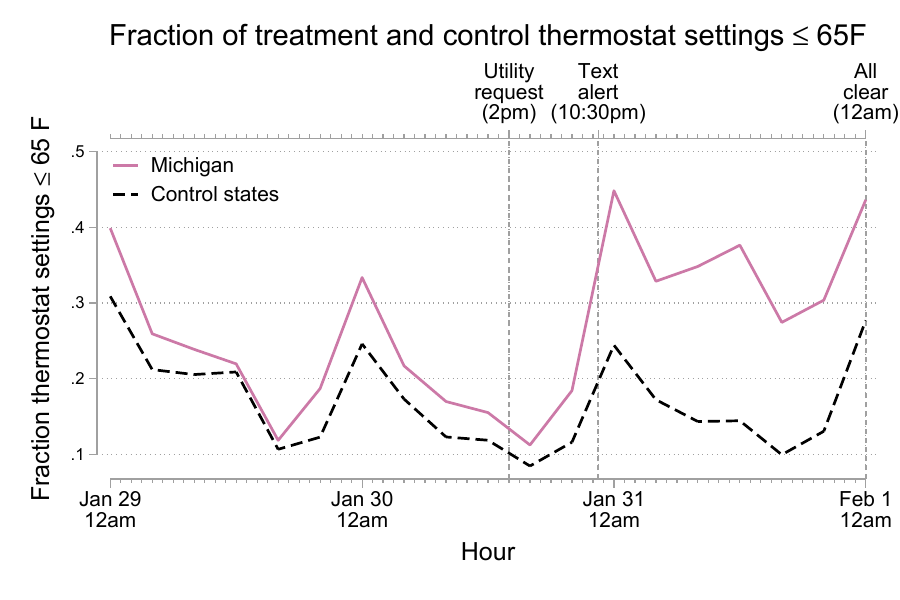}
        \caption{}\label{fig:complianceparallel}
\end{subfigure}
\medskip
\begin{subfigure}[t]{.49\textwidth}
    \centering
    \vspace{0pt}
    \includegraphics[width=\linewidth]{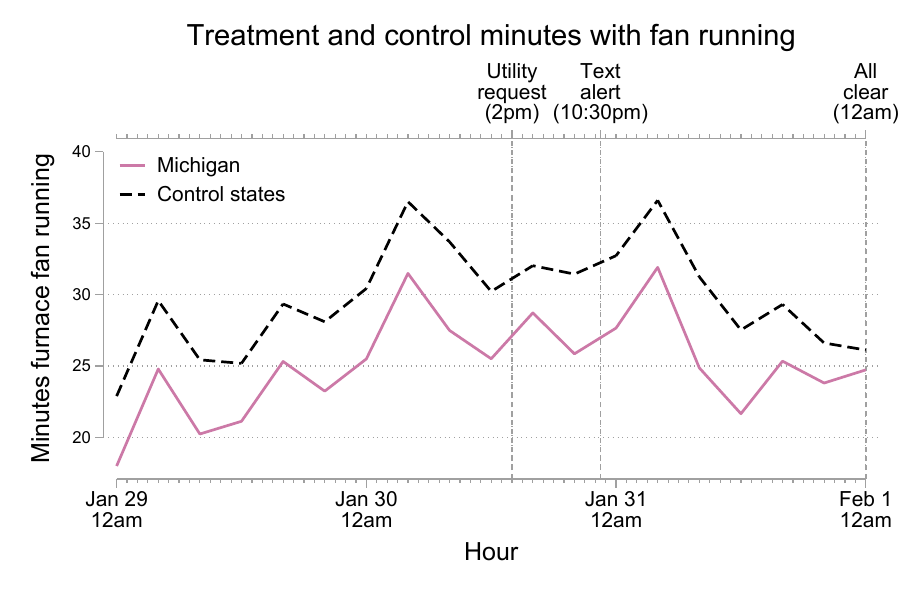}
        \caption{}\label{fig:fanparallel}
\end{subfigure}
\begin{minipage}[t]{.49\textwidth}
    \medskip \medskip \medskip
    \caption{Sample average values of the outcome variables for treatment and control households, January 29 - February 1.  Panel (\subref{fig:thermostatparallel}) plots average thermostat settings, panel (\subref{fig:complianceparallel}) plots the fraction of households with thermostat settings at or below 65\(^\circ\)F, and panel (\subref{fig:fanparallel}) plots the average furnace fan run time in each four-hour period. } \label{fig:paralleltrends}
\end{minipage}
\end{figure}

Because we observe treatment and control households before and during the event, the difference-in-differences framework is a natural candidate to estimate the effect of the emergency request on thermostat settings.  The key assumption needed in a difference-in-differences design is a parallel trends assumption in the evolution of the potential untreated outcome.  To demonstrate the validity of the difference-in-differences assumption, we demonstrate visually that our outcome variables exhibit parallel trends prior to the event (we also provide event-study estimates with 10 days of pre-trend estimates in section \ref{sec:eventstudy}).  Figure \ref{fig:thermostatparallel} plots four-hour sample average thermostat settings, figure \ref{fig:complianceparallel} plots the fraction of households with thermostat settings at or below 65\(^\circ\)F, and figure \ref{fig:fanparallel} plots the number of minutes per hour the fan was running in Michigan and the control states from January 29th through February 1st.  The first vertical dashed line indicates when the utility company first broadcast a request to residential customers to reduce natural gas consumption by reducing thermostats, the second vertical dashed line indicates the emergency phone alert, and the final vertical dashed line indicates the all-clear time.

Prior to the event, the treatment and control thermostat settings, fraction of households with thermostat settings at or below 65\(^\circ\)F, and the number of minutes the fan was running in Michigan exhibit roughly parallel trends even without conditioning on covariates.  When the event begins, a take-up lag can be observed where households have either not received the message or are not home and able to respond.  A few hours after the event begins, the average thermostat setting in Michigan breaks trend and significantly decreases.  Despite the request that households reduce thermostats to 65\(^\circ\)F or lower, the average observed smart thermostat setting in Michigan is above 65\(^\circ\)F during the entire event.  It appears that compliance with the request does not begin until after the phone alert.  The differences in furnace fan running time after the request are more difficult to detect visually than thermostat setting and fraction of compliant households.

\section{Empirical analysis and results}

We compare differences in outcomes between households in Michigan and surrounding states before and after the emergency requests.  We consider three outcomes.  The household's thermostat setting is a continuous measure of the household's compliance with the emergency request and encompasses the thermal discomfort that the household incurred to contribute to the public good.  The second outcome is a binary variable equal to one when the thermostat setting is at or below 65\(^\circ\)F.  This binary variable captures whether households complied with the request to the letter.  The final outcome variable, average number of minutes the fan ran during the hour, is the best proxy available for the amount of energy conserved.

The analysis is divided into four subsections.  We begin with a standard difference-in-differences framework to estimate the average treatment effect of the program.  We then move to an event-study framework that allows for dynamic effects by sample time period as word of the emergency reached more households.  Next, we test whether Democratic party vote share affected compliance rates.  Finally, we study how the phrasing of the emergency request around 65\(^{\circ}\)F influenced household behavior.

\subsection{Average treatment effect estimates}

\begin{table}
    \centering
    \begin{threeparttable}
    \caption{Two-way fixed effects regressions \label{tab:twfe}}
    \begin{tabular}{lccc} \hline
 & (1) & (2) & (3) \\
VARIABLES & Thermostat setting & Thermostat $\leq$ 65F & Fan run time \\ \hline
 &  &  &  \\
Michigan x Post & -1.072** & 0.101** & -1.470* \\
 & (0.133) & (0.007) & (0.514) \\
 &  &  &  \\
Observations & 2,126,336 & 2,126,336 & 2,135,114 \\
R-squared & 0.710 & 0.504 & 0.752 \\
Weather controls & YES & YES & YES \\
Household FE & YES & YES & YES \\
Time FE & YES & YES & YES \\
Day of week $\times$ hour of day $\times$ state & YES & YES & YES \\
 Pre-treatment mean & 66.87 & 0.270 & 18.09 \\ \hline
\end{tabular}

    \begin{tablenotes}
    \footnotesize Estimates of the regressions from equation \ref{eq:twfe}. The sample includes 4-hour-average household observations from January 2nd-January 31st. Pre-treatment means reported for Michigan households.  Standard errors clustered at the state level.  ** p$<$0.01, * p$<$0.05
    \end{tablenotes}
    \end{threeparttable}
\end{table}

The first set of regressions we consider are two-way fixed effects specifications on all January 2019 observations.  We consider outcomes \(Y_{i,t}\) and code a binary variable \(D_{i,t} = 1\) for all Michigan observations beginning January 30th at 2:00 pm and zero beforehand.  Our preferred specification takes the following form:
\begin{align} \label{eq:twfe}
    Y_{i,t} = \alpha_i + \lambda_t + \beta D_{i,t} + \gamma X_{i,t} + \delta_{s,h,d} + \varepsilon_{i,t},
\end{align}
where \(\alpha_i\) are household fixed effects, \(\lambda_t\) are time-of-sample indicator variables, \(X_{i,t}\) are controls for weather variables (including outside temperature, humidity, wind speed, precipitation, snow depth, and cloud cover), \(\delta_{s,h,d}\) are state by day-of-week by time-of-day indicator variables, and \(\varepsilon_{i,t}\) is mean-zero heterogeneity.  The state by day-of-week by time-of-day indicator variables \(\delta_{s,h,d}\) allow us to control flexibly for differences in time-varying heterogeneity across states. Equation \ref{eq:twfe} is a two-way fixed-effects specification.  The ordinary-least-squares estimate \(\hat{\beta}\) is a difference-in-differences estimate that identifies the causal average treatment effect on the treated under the standard parallel trends, no spillovers, and strict exogeneity assumptions.  We provide empirical evidence that there are parallel trends and test for spillovers into neighboring states, finding no evidence that these assumptions are violated.  We believe the strict exogeneity assumption is likely to hold given the surprise nature of the emergency and subsequent emergency appeal.

Table \ref{tab:twfe} presents the coefficient estimates of the difference-in-differences regressions for each of the three outcome variables: thermostat setting, a binary variable for setting the thermostat at or below 65\(^\circ\)F, and the average number of minutes per hour the furnace fan ran.  We cluster the standard errors at the state level.\footnote{We cluster at the state level because that is the level of treatment variation \citep{abadieatheyimbenswooldridge20}, but given that there are only five state clusters, it is possible that the cluster-robust standard error estimators will not asymptotically converge \citep{camerongelbachmiller2008} and may either overstate or understate the precision of the estimates.  Furthermore, the wild cluster bootstrap is not appropriate for only one treated cluster \citep{mackinnonwebb}.  As a robustness check, we implement the \cite{donaldlang2007} estimator of the average treatment effect in appendix \ref{app:donaldlang}, which provides valid inference for five clusters and find that the average treatment effects are still statistically significant and more precise than the estimates in the main paper, so we are not concerned about the precision of our estimates and we are likely using too conservative of standard errors in the main paper.} When the thermostat setting is the outcome variable (column 1), the coefficient on \(D_{i,t}\) is an estimate of the average treatment effect and is the mean difference in thermostat settings for Michigan and control states before and after the treatment.  We estimate a reduction of 1.1\(^{\circ}\)F after the emergency request for Michigan households relative to neighboring state households.  This reduction is about 0.31 standard deviations in the thermostat setting from January 2 - 29.

Column 2 presents estimates using an indicator variable for having the thermostat at or below 65\(^{\circ}\)F as the outcome variable, which we interpret as full compliance with the request.  Given that at any time, some fraction of Michigan households would already have thermostat settings at 65\(^{\circ}\)F or below, the difference-in-differences estimate accounts for this by differencing out the within-household and within-time average incidental compliance.  The coefficient on \(D_{i,t}\) is an estimate of the additional fraction of households induced to set the thermostat at or below 65\(^{\circ}\)F.  We estimate a 10.1 percentage point increase in the fraction of households with thermostat settings at or below 65\(^{\circ}\)F for Michigan households relative to neighboring state households.  The pre-treatment average fraction of households in Michigan that already would have had thermostat settings less than or equal to 65\(^{\circ}\)F was 27 percent, which we consider the incidental compliance.

Finally, column 3 presents the coefficient estimates of the difference-in-differences regressions using furnace fan run time as the dependent variable, which is the closest proxy to energy consumption in the smart thermostat data.  We estimate a 1.5 minute per hour average reduction in furnace fan run time for Michigan households relative to neighbor state households.  Relative to the predicted mean furnace fan run time for Michigan households during the emergency period, this is a 6 percent decrease.  Given the lack of natural gas consumption data at the household level, this is the best estimate of the amount of natural gas savings caused by the emergency request.\footnote{Natural gas consumption at the household level is measured at the monthly level by the utility.}  Using aggregate daily consumption data provided by Consumers Energy, the total reduction in natural gas consumption from all sources was about 10 percent, which is in line with our estimates.

\subsubsection{Robustness checks, alternative specifications, and placebo analysis}

In this section and the appendices, we consider alternative behavioral responses (appendix \ref{app:alternates}) and discuss the sensitivity of the main estimates to estimation at the hourly level (appendix \ref{app:hourly}), a series of robustness checks and alternative specifications (appendix \ref{sec:robustness}), a placebo analysis of an earlier cold wave with no emergency response (appendix \ref{sec:placebo}), and alternative \cite{donaldlang2007} inference (appendix \ref{app:donaldlang}). We find little evidence that households responded to the request via alternative behavioral channels such as turning off the heat or using smart thermostat settings, and we find that the results are generally consistent across the sensitivity tests.  As such, the results of these checks are located in the appendices.

First, we consider the possibility that households responded to the emergency request in ways other than the thermostat setting.  For instance, households may have turned off the heat, changed the thermostat mode to ``hold'' to override previously programmed thermostat setting changes, changed the thermostat to an automated program designed by Ecobee, or may have spent more or less time at home because of the emergency request.  The smart thermostat data contains information on whether the furnace was turned off, whether the thermostat was on ``hold'' (which selects a constant temperature setting), or whether the thermostat was utilizing Ecobee's automation settings which adjust automatically to the household's schedule.\footnote{Ecobee's automation includes ``smart recovery mode,'' which will adjust the temperature in anticipation of a user's normal schedule and usual time it takes to heat or cool the home.  For instance, if a user arrives home from work at 6 pm and typically increases the thermostat setting upon arrival, the smart recovery program may begin heating the home at 5:45 pm.  This is the only automation mode described in the data.}  In addition, the thermostat contains a motion sensor that registers when there is motion in front of the thermostat, which allows us to test whether households were at home more or less due to the emergency request.  We use these variables as outcomes in the main specification as described in equation \ref{eq:twfe} and display the estimates in appendix section \ref{app:alternates}.  

We find that the emergency request induced a 2 percentage point increase in ``hold'' thermostat settings (relative to approximately 27\% baseline) and a 3 percentage point increase in the use of thermostat automation (relative to an approximately 20\% baseline).  The small size of the estimated effect on hold thermostat settings does not rule out the use of a thermostat setting override as a mechanism for compliance with the emergency request.  We find that in Michigan the fraction of hold thermostat settings increased from 27\% to 40\% during the emergency, while in the control states the fraction of hold thermostat settings increased from 28\% to 38\%.  These findings suggest that during extreme cold waves, households are likely to override their typical thermostat settings, regardless of whether there is an energy emergency.  Similarly, the fraction of households using Ecobee's automation settings declined in both Michigan and the control states during the emergency.  We find no evidence that households turned off the heat or that households were activating the motion sensor more or less during the event.  In addition, we include all mode variables and motion sensor variables in a robustness check for the thermostat setting, compliance, and fan regressions, finding that adding these controls does not substantially impact the average treatment effect estimates in table \ref{tab:twfe}.  Overall, these results suggest that thermostat setting was the main behavioral channel through which households responded to the request.

Next, we repeat the analysis using hourly data to test the sensitivity of the results to the choice of time interval in appendix \ref{app:hourly}.  We find that the point estimates of our main treatment effects are almost identical, but the estimates on furnace fan run time are less precise and are not statistically different from zero.  This is because aggregating to four-hour intervals reduces noise not captured by the fixed effects and controls, improving the precision of the estimates.  The robustness checks in section \ref{sec:robustness} test the sensitivity of the average treatment effects to alternative difference-in-differences specifications, omitting households who join the sample late or leave early (i.e., using a balanced panel), and allowing for spillovers to counties bordering Michigan. We find that the estimated effects do not change substantially.  In section \ref{sec:placebo},  the placebo test analyzes a cold wave in Michigan that occurred ten days earlier on January 20-21, 2019, where temperatures dropped by a similar magnitude.  We find that Michigan's heating behavior remains parallel to the control households during this placebo event, and that estimation using regression equation \ref{eq:twfe} with the placebo treatment yields estimates of zero, suggesting that our findings are not an artifact of differential responses by Michigan households to cold waves.  In our regular specifications, this cold wave is included in the data and thus serves as a control, lending credibility to the research design.  Finally in section \ref{app:donaldlang}, we calculate the \cite{donaldlang2007} estimate of the treatment effect (which has valid inference for five clusters), and find comparable and statistically significant estimates using this approach.

\subsection{Event-study estimates} \label{sec:eventstudy}

\begin{figure}
\centering
\begin{subfigure}[t]{.49\textwidth}
    \centering
    \includegraphics[width=\linewidth]{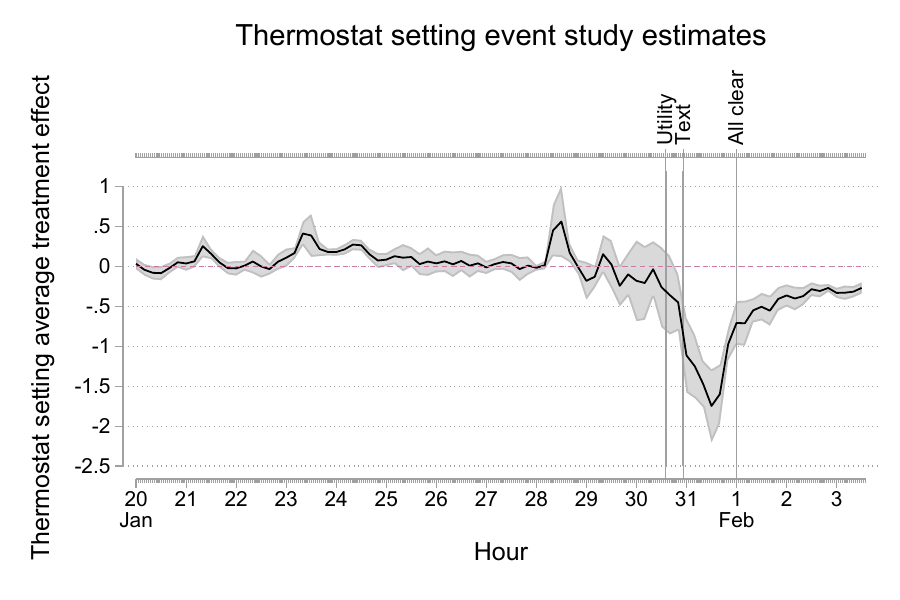}
            \caption{}\label{fig:eventtemperature}
    \end{subfigure}
\begin{subfigure}[t]{.49\textwidth}
    \centering
    \includegraphics[width=\linewidth]{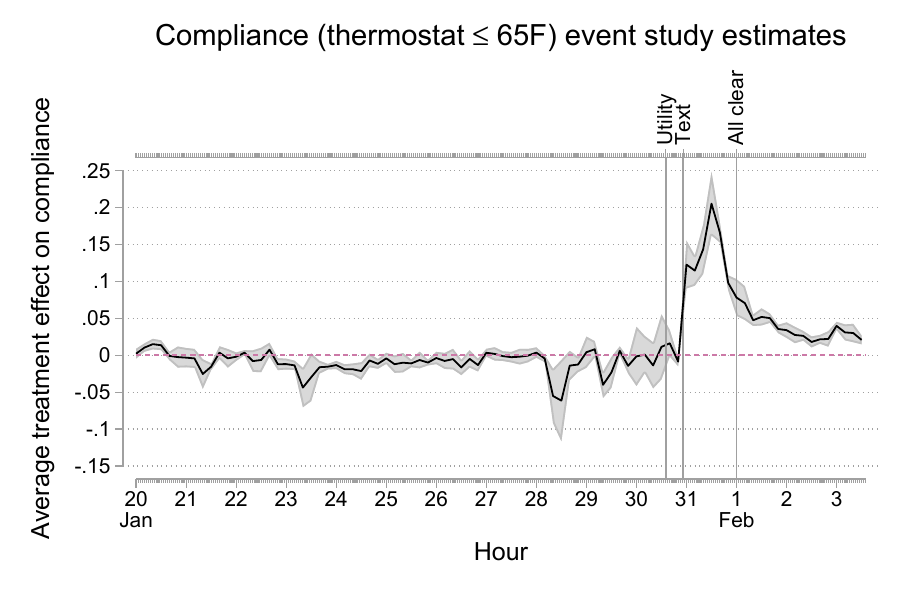}
        \caption{}\label{fig:eventcompliance}
\end{subfigure}
\medskip
\begin{subfigure}[t]{.49\textwidth}
    \centering
    \vspace{0pt}
    \includegraphics[width=\linewidth]{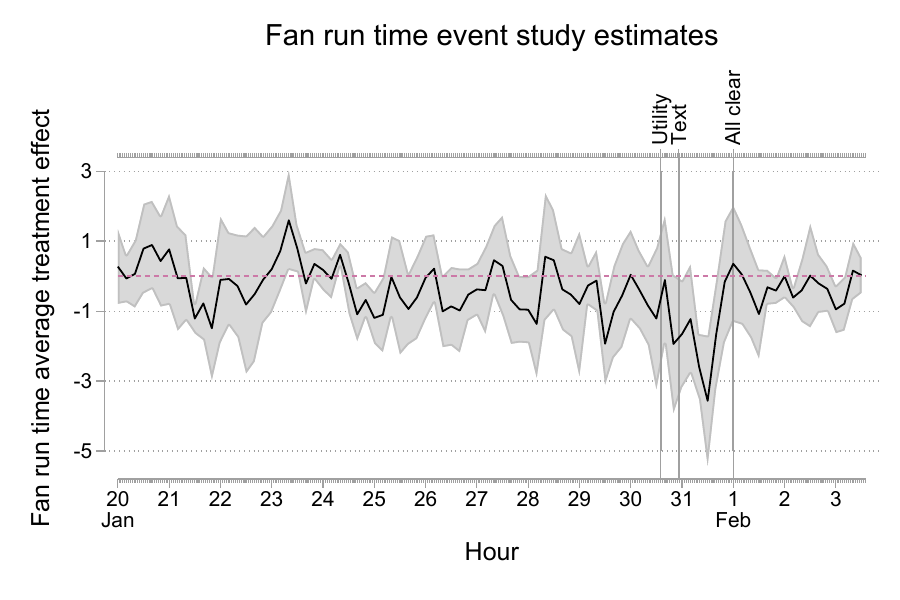}
        \caption{}\label{fig:eventfan}
\end{subfigure}
\begin{minipage}[t]{.49\textwidth}
    \medskip \medskip \medskip
    \caption{Event-study coefficients estimated using regression equation \ref{eq:eventstudy} with (\subref{fig:eventtemperature}) thermostat setting, (\subref{fig:eventcompliance}) compliance, and (\subref{fig:eventfan}) minutes of furnace fan run time as the dependent variables.  95 percent confidence intervals constructed from standard errors cluster-robust to heteroskedasticity.} \label{fig:eventstudy}
\end{minipage}
\end{figure}

Given the repeated requests for reductions in thermostat settings over time, we next account for a dynamic response in an event-study framework.  We estimate a two-way fixed effects regression using the following specification: 
\begin{align} \label{eq:eventstudy}
Y_{i,t} = \alpha_i + \lambda_{t} + \sum_{k=-57}^{-1} \beta_{k}^{lead} \mathbf{1}[k=t-g] + \sum_{k=0}^{29} \beta^{lag}_k \mathbf{1}[k=t-g] + \gamma X_{i,t} + \delta_{s,h,d} + \varepsilon_{i,t},
\end{align}
where \(g\) is the time period the utility made its first emergency request to households.  Thus, we estimate 57 lead coefficients and 30 lag coefficients to include 10 and a half days of pre-trends and four days of dynamic treatment effects (including two days after the treatment ends).\footnote{Choosing greater or fewer leads and lags does not substantially change the coefficient estimates, but increases computational cost.  We chose the window to allow us to test for the presence of pre-trends, observe when thermostat settings returned to a normal level after the event, and to keep computational times reasonable.} We hypothesize that prior to the end of working hours, household responses will be muted and that the largest responses will occur after the emergency text message alert at 10:30 pm.  Further, we suspect that the treatment effect persisted after the ``all clear" time due to barriers to receiving the all-clear message or adjusting the thermostat (e.g., if the home is vacant or all occupants are sleeping).

Figure \ref{fig:eventstudy} plots the dynamic treatment effects estimated using the event-study regressions specified in equation \ref{eq:eventstudy} using thermostat setting, compliance with the request, and fan run time as outcome variables.  Leading up to the emergency, the difference in thermostat settings between Michigan and the control states is typically small and positive when the confidence interval does not overlap zero.  To the extent that the pre-trends are not parallel, we expect that our event-study estimates may underestimate the treatment effect on thermostat setting during some time periods, though this effect is likely to be small given the size of the estimated treatment effect relative to the pre-treatment noise.  We see similar results for the compliance estimates and note that the compliance estimates may also be conservative.  The pre-treatment trends are more noisy for fan run time, but do not display systematic trends, and the confidence interval contains zero for most pre-period estimates.

The first finding of note is that the emergency text alert was essential to increasing compliance.  Averaging over the event-study coefficients for the eight hours prior to the emergency text alert, the utility's emergency request only resulted in an average additional compliance rate of 0.4 percent, resulting in an average thermostat reduction of just 0.4\(^\circ\)F.  Following the emergency text alert, the average additional compliance rate was 14.2 percent (peaking at over 20 percent), resulting in an average thermostat reduction of 1.4\(^\circ\)F (with a peak of 1.7\(^\circ\)F).  Given the utility's actions of sending emails to customers, posting on social media, and reaching out to traditional news media, we do not think the lack of responsiveness was solely due to a lack of reach, although we cannot rule this out entirely.  While it is possible that households were not at home, causing the initial lukewarm response, we see this as unlikely because the Ecobee smart thermostat setting can be changed remotely via app.  Instead, it is likely that households did not take the request seriously until it became clear that there was a true emergency.  The additional authority of the emergency text alert and the repeated request to reduce thermostat settings likely increased the salience of the request, inducing additional compliance.  Another potential explanation is that the discomfort of lower-than-normal temperatures may be lower overnight, but given the continued compliance during the daytime of the next day, we do not see this as playing a large role.




After the emergency request and before the ``all clear,'' thermostat settings begin to trend upward and compliance begins to fall.  Sustained compliance is likely increasingly costly, so participation rates decline toward the end of the event.  This trend may also be due to households choosing low thermostat settings when sleeping and leaving the home for work in the morning.  Upon returning from work, households may increase the thermostat setting to a slightly higher level.  This behavior is similar to the ``backsliding'' dynamic reported by \cite{allcottrogers2014} in which households conserve electricity after receiving a home energy report, but the effect lessens over time.  To sustain high levels of compliance in an emergency, repeated requests are likely necessary.

Another interesting finding is that the effect persists after the ``all clear.''  This persistence suggests that some households which had set their thermostats to reduce the temperature setting in keeping with the emergency request had not yet re-set them after the all clear due to the fixed cost of interacting with the thermostat.  This result is consistent with previous work that finds that changes in thermostat settings in response to a cold or hot period tend to persist after the cold or hot period ends \citep{geho2018}.

In addition, one can see that the reductions in furnace fan running time are only transitory.  Because furnaces essentially run at one speed, reducing the thermostat at night or when out of the home will reduce energy use while the home cools, but upon increasing the thermostat, the furnace will need to run again and incur a ramp-up cost to increase the temperature.  We can see in the fan run time event study estimates that on January 31st, there were savings during the early morning and day, but when households returned home in the evening that furnaces had to run at essentially full intensity to warm the home again.  After the all clear, the estimates return to mean zero more quickly than the thermostat setting and compliance rate estimates.

In appendix section \ref{app:hourly}, we replicate the event-study analysis at the hourly level.  We find similar results across all three variables, and one can see that the treatment effect begins the hour of the emergency text alert, providing additional evidence that the emergency text was key to increasing the strength and salience of the nudge.  The main difference is that the pre-treatment coefficients are more noisy, which leads us to favor the four-hour analysis.

\begin{figure}
    \begin{subfigure}[t]{.49\textwidth}
        \centering
        \includegraphics[width=\linewidth]{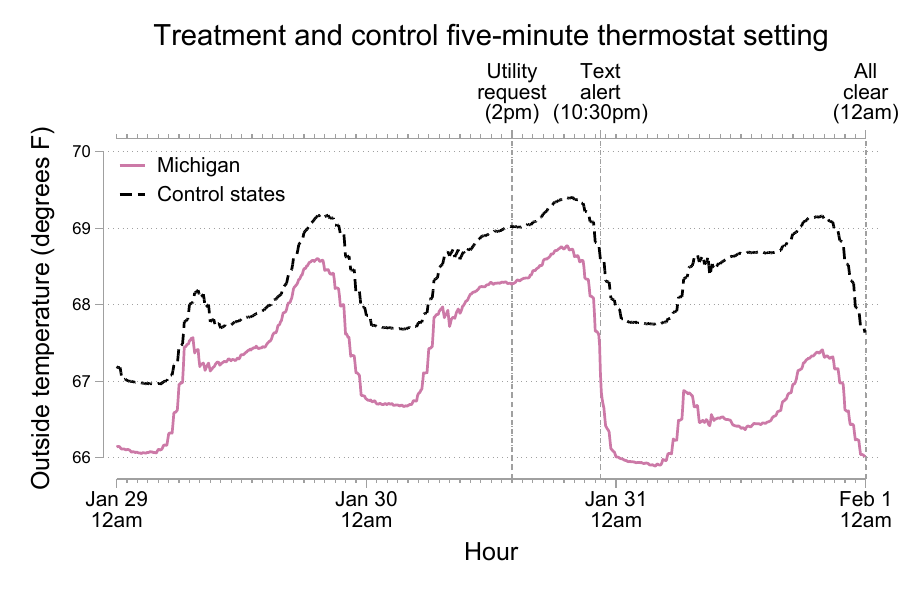}
        \caption{}\label{fig:fivemin_a}
    \end{subfigure}
    \begin{subfigure}[t]{.49\textwidth}
        \centering
        \includegraphics[width=\linewidth]{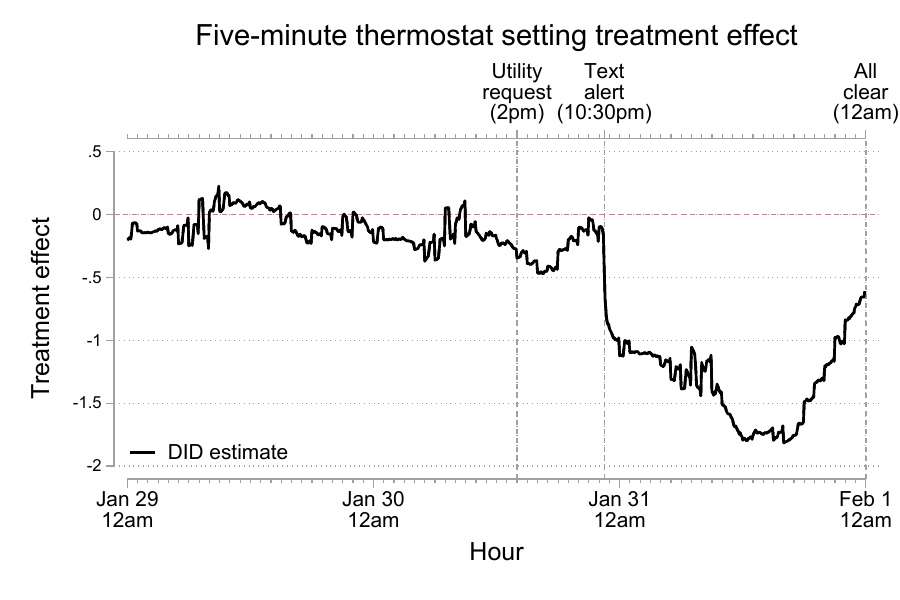}
        \caption{}\label{fig:fivemin_b}
    \end{subfigure}
    \medskip
    \begin{subfigure}[t]{.49\textwidth}
        \centering
        \vspace{0pt}
        \includegraphics[width=\linewidth]{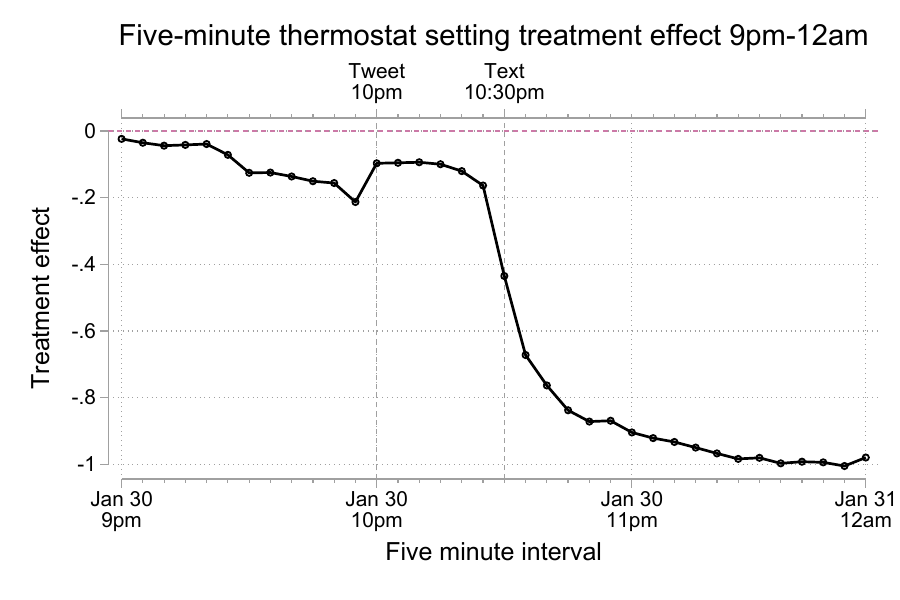}
            \caption{}\label{fig:fivemin_c}
    \end{subfigure}
    \begin{minipage}[t]{.49\textwidth}
        \medskip \medskip
        \caption{(a): Five-minute sample mean thermostat settings for Michigan and control households from January 30th 12 pm to January 31st 11:59 pm.  (b): Five-minute difference-in-differences estimates. (c): Five minute difference-in-differences estimates zooming in on January 30th 9 pm to 11:59 pm.} \label{fig:fivemin}
    \end{minipage}
\end{figure}

We supplement the event study with a graphical analysis of the five-minute thermostat-setting data.  In figure \ref{fig:fivemin_a}, we plot five-minute thermostat setting data for Michigan and the control states from January 29th through February 1st.  In figure \ref{fig:fivemin_b}, we plot a difference-in-differences estimate of the treatment effect, which we construct as the average difference between Michigan and the control states in five-minute thermostat settings less same day-of-week and hour-of-day thermostat settings from before the event.  In these figures, one can only see a clear decline in Michigan thermostat settings after the emergency text alert.  In appendix section \ref{sec:fiveminplacebo}, we replicate the five-minute analysis during the placebo cold wave.  The difference-in-differences estimates are zero throughout most of the placebo period other than a slight increase in thermostat settings for Michigan after the placebo all clear time, lending credibility to the difference-in-differences estimates in figure \ref{fig:fivemin_b}.  Figure \ref{fig:fivemin_c} zooms in on the difference-in-differences estimates from 9 pm to 11:59 pm on January 30th, which is the period when the Governor's tweet (10:00 pm) and emergency text alert (10:30 pm) occurred.  These results show very little change in the treatment effect after the Governor's tweet and before the phone alert, but large declines in thermostat settings that begin in the five minute period directly after the phone alert, suggesting that the phone alert was pivotal in increasing compliance.

\subsection{Heterogeneity analysis} \label{sec:heterogeneity}

Next, we analyze the effect of political ideology and the effect of the reference point on household behavior in a triple-differences framework.  Our approach analyzes both forms of heterogeneity in the same estimating equation with additional controls to account for the possibility that county-level vote share is correlated with baseline thermostat setting or other demographic factors.  We supplement the smart thermostat data with data on gubernatorial election county vote shares for each state's most recent election obtained from the Voting and Elections Collection maintained by the \cite{cqdatacounty}.  We include household-level controls available in the Ecobee data as well as demographic controls at the county level obtained from the American Community Survey \citep{acs2019}.  Below, we discuss how we define the baseline thermostat setting and political affiliation variables separately before presenting a single regression equation where we estimate the effect of baseline thermostat setting and Democratic Party Governor vote share on the treatment effect.  We estimate the effects in the same regression in case homes in Republican-voting counties have differing baseline thermostat settings than homes in Democrat-voting counties (or vice versa).

Households whose thermostats would have been at 65\(^{\circ}\)F or lower essentially received information that they were already keeping the thermostat low enough and may have felt that they did not need to reduce the thermostat further.\footnote{This analysis generally treats larger reductions as welfare-improving, but we note that thermostat settings that are too low increase the risk of frozen pipes.}  Furthermore, we hypothesize that the distance from the reference point may also affect household behavior as outlined in the theoretical framework in section \ref{sec:theory}.  To test our hypotheses, we estimate the effect of the emergency request allowing for different responses by expected baseline thermostat settings.   We construct a non-parametric estimate of baseline expected thermostat setting \(\hat{T}_{i,t}\) for each household by calculating the household's sample average thermostat setting for each day-of-week and time-of-day combination from the pre-treatment period.  Denote \(\mathcal{T} = \left\lbrace [0,59),[59,61),[61,63),...,[73,75),[75,100]\right\rbrace \) as the collection of 2-degree intervals from 59\(^{\circ}\)F to 75\(^{\circ}\)F with binned endpoints for higher and lower temperatures, and \(b \in \mathcal{T}\) the interval with upper bound \(b\).\footnote{Appendix figure \ref{fig:expected_distribution} displays a histogram of the expected thermostat setting bins.}  We interact indicator variables for belonging in each interval with the treatment variable to create a third difference and estimate heterogeneous effects by baseline temperature category.  

In the same regression, we analyze heterogeneity by Democratic Party Governor vote share.  The data on gubernatorial election returns is at the county level.  In the 2018 election, the distribution of the Michigan Governor's county vote share ranged from 31 percent to 73 percent.  We expect the effect of political support to be non-linear so we create 5-percentile indicator variables between 30 and 75 percent.  Denote \(\mathcal{P} = \left\lbrace [30,40),[40,45),[45,50),...,[70,75]\right\rbrace \) as the collection of a 10-percentile interval between 30 and 40 and 5-percentile intervals between 40 and 75, and \(a \in \mathcal{P}\) the interval with upper bound \(a\).\footnote{We combined the 30-35 and 35-40 percentile intervals because only 0.38 percent of households lived in counties with a Democratic Party vote share between 30 and 35 percent, which lead to extremely imprecise estimates.  Appendix figure \ref{fig:voteshare_distribution} displays a histogram of the vote share bins.} Similarly to the baseline thermostat setting, we interact indicator variables for belonging in each interval with the treatment variable.

Thus, our estimating equation is
\begin{align} 
    \begin{split} \label{eq:heterogeneity}
        Y_{i,t} =\text{ } & \alpha_i + \lambda_{b,t} + \sum_{b \in \mathcal{T}} \beta_b D_{i,t} \times \mathbf{1}[\hat{T}_{i,t} \in b] + \phi_b \mathbf{1}[\hat{T}_{i,t} \in b] + \sum_{a \in \mathcal{P}} \beta_a D_{i,t} \times \mathbf{1}[P_{county} \in a] \\
        &+ \gamma_1 D_{i,t} \times Z_{i} + \gamma_2 X_{i,t} + \delta_{s,h,d} + \varepsilon_{i,t},
    \end{split}
\end{align}
where \(Z_{i}\) is a vector of controls for household-level characteristics available in the smart thermostat data as well as county-level demographics to address correlation between county vote share and demographics, which we interact with the treatment indicator.\footnote{From the smart thermostat data, we include square feet of the home, number of occupants, whether the home is detached or an apartment, and the age of the home.   From the ACS, we include county level median income, median age, population, fraction male, fraction white, fraction with a high school education or more, and fraction of non-US citizen residents.} In addition to weather controls, household fixed effects, time-of-sample indicators, and state by day-of-week by time-of-day indicators included in our previous regressions, we also add expected thermostat setting indicators interacted with time-of-sample indicators (denoted \(\lambda_{b,t}\)) to control for common trends in thermostat setting by expected thermostat setting such as mean reversion.

The estimate of \(\beta_b\) is the effect of the treatment on the outcome conditional for baseline thermostat setting \(b\) relative to the omitted category of 63-65\(^\circ\)F. Our theory predicts that the thermostat setting treatment effect should be higher for households with baseline thermostat settings above the compliance target and lower for households with baseline thermostat settings below the compliance target. Furthermore, we expect that the effect will increase with a higher baseline expected thermostat setting.  Given that the compliance target is the reference category, we expect coefficients below the compliance target to be zero or positive and coefficients above the compliance target to be negative and growing in magnitude.  For our compliance outcome, we expect compliance to fall with an increased distance from the compliance target, which is the omitted category in the regression.  Thus, we expect \(\beta_b\) to be negative and growing as \(b\) deviates from the compliance target.  

Moreover, it is possible that very cold baseline households may increase thermostat setting when introduced to the reference level of 65 if the nudge's reference point incentive is stronger than the conservation incentive (theory case 2). Thus, it is possible for the overall effect of the treatment to be zero or positive for \(b \leq 65\).\footnote{This backfire is called the ``boomerang effect'' in the social-comparison literature where if a household receives information that they are consuming less than the average they may decide they were being too conservative and increase consumption \citep{allcott2011}.}  The estimated coefficients are all relative to the omitted category, so to evaluate whether the overall effect of the nudge was to induce a backfire for baseline thermostat settings below the compliance target, we also estimate the average partial effect of treatment on the outcomes by baseline thermostat setting.  Formally, the average partial effect of treatment by baseline thermostat setting \(\hat{T}_{i,t} \in b\) is 
\begin{align}
    E[Y_{i,t}|W_{i,t},\hat{T}_{i,t} \in b,D_{i,t}=1]-E[Y_{i,t}|W_{i,t},\hat{T}_{i,t} \in b,D_{i,t}=0], \label{eq:ape_temp}
\end{align}
where \(W_{i,t}\) is a vector of all other control variables in equation \ref{eq:heterogeneity}.

The estimate of \(\beta_a\) is the effect of the treatment on the outcome for vote share category \(a\) relative to the ommitted vote share category of 30-40 percent (the least favorable vote share category for the Democratic Party Governor). For the thermostat setting and compliance outcome variables, we hypothesize that the coefficients on the interaction with vote share \(\beta_a\) will be increasing as Democratic vote share increases and that the opposite will be true for the fan running outcome variable regression, indicating that the appeal was more effective for households in counties that supported the Governor's election.  Similarly to the baseline thermostat setting, we estimate the average partial effect of treatment by vote share category to evaluate the overall effect of the nudge on the outcomes by vote share category.  The average partial effect of treatment by baseline county-level vote share \(P_{county} \in a\) is 
\begin{align}
    E[Y_{i,t}|W_{i,t},P_{county} \in a,D_{i,t}=1]-E[Y_{i,t}|W_{i,t},P_{county} \in a,D_{i,t}=0], \label{eq:ape_vote}
\end{align}
where \(W_{i,t}\) is a vector of all other control variables in equation \ref{eq:heterogeneity}.

Appendix table \ref{tab:heterogeneity} displays the estimates of equation \ref{eq:heterogeneity} for each outcome variable, and we report average marginal effects in the figures below.\footnote{Appendix table \ref{tab:heterogeneity_full} includes additional coefficient estimates for controls interacted with the treatment variable, and appendix table \ref{tab:heterogeneity_nocontrols} displays estimates from a robustness check omitting the controls interactecd with treatment.}  The standard errors are cluster-bootstrapped to incorporate the uncertainty due to sampling error from estimating the baseline thermostat setting.\footnote{The bootstrap procedure first draws observations with replacement from within household, day of week, and time period strata to estimate 100 different baseline temperatures for each household, day of week, and time of day combination.  It then samples from this empirical distribution and draws 100 bootstrap samples clustered at the city level.  Ultimately, these standard errors differ very little from clustered standard errors that ignore the uncertainty from the first-stage estimation.}  We discuss the vote share and baseline thermostat setting results separately in the following sections.

\subsubsection{Reference point effect}

\begin{figure}
\centering
\begin{subfigure}[t]{.49\textwidth}
    \centering
    \includegraphics[width=\linewidth]{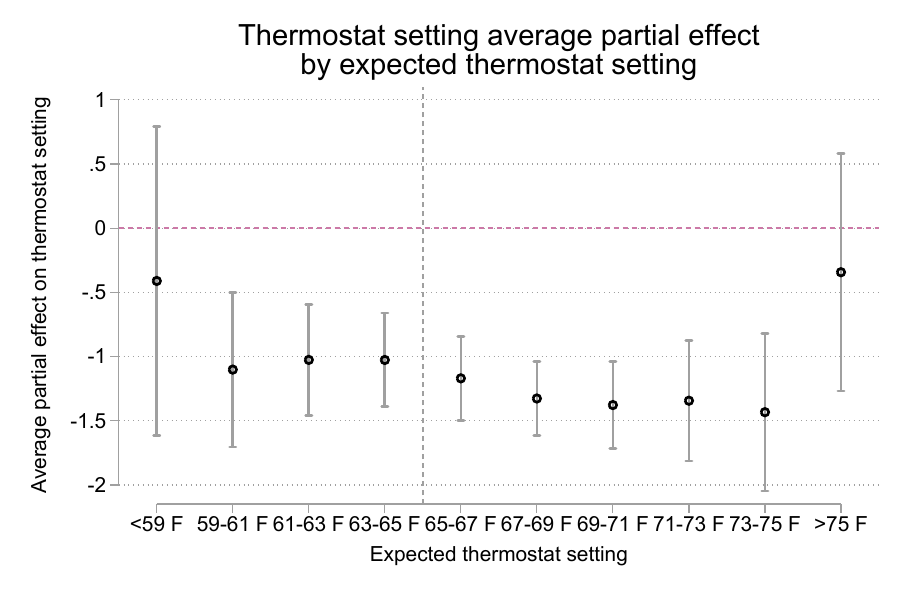}
            \caption{}\label{fig:thermref}
    \end{subfigure}
\begin{subfigure}[t]{.49\textwidth}
    \centering
    \includegraphics[width=\linewidth]{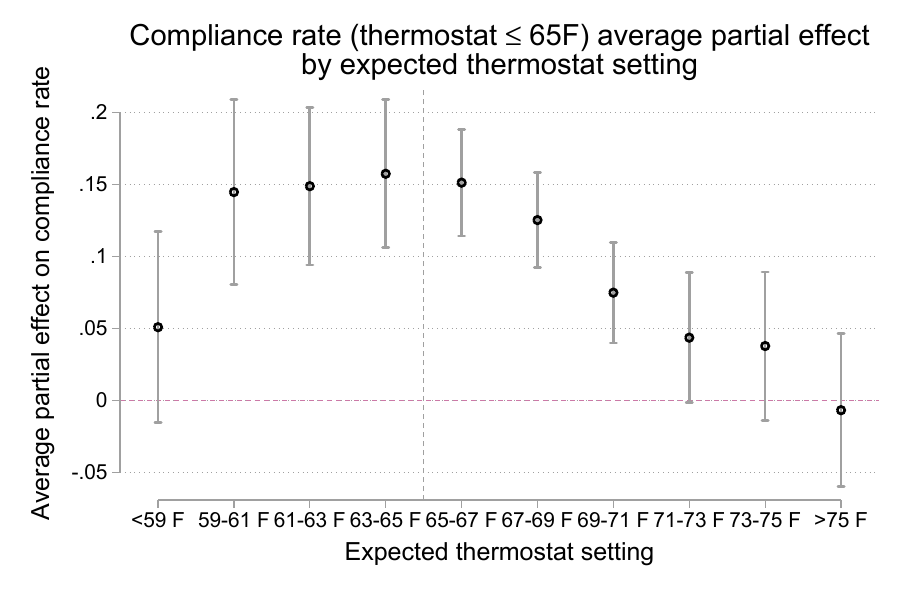}
        \caption{}\label{fig:complianceref}
\end{subfigure}
\medskip
\begin{subfigure}[t]{.49\textwidth}
    \centering
    \vspace{0pt}
    \includegraphics[width=\linewidth]{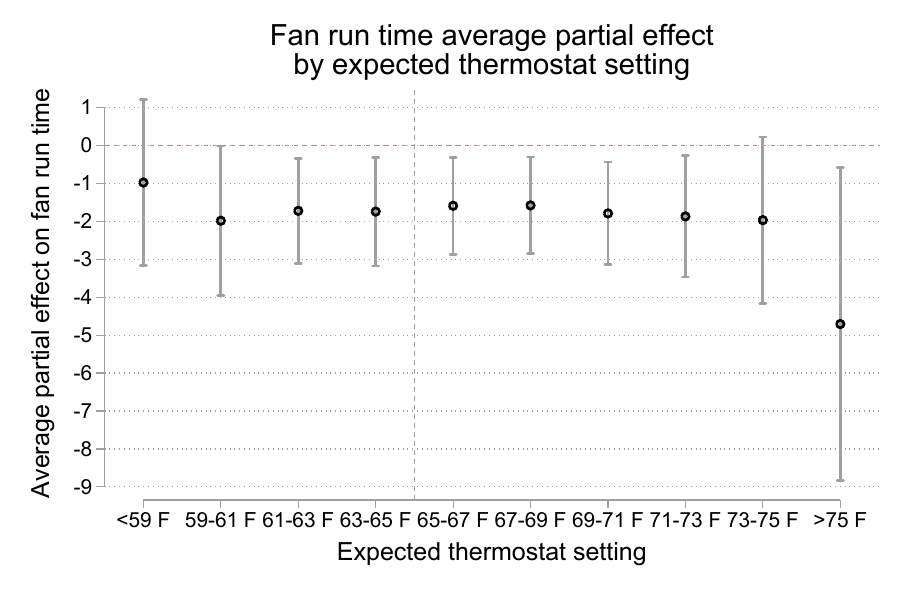}
        \caption{}\label{fig:fanref}
\end{subfigure}
\begin{minipage}[t]{.49\textwidth}
    \medskip \medskip
    \caption{Average partial effect (equation \ref{eq:ape_temp}) of the emergency request by expected thermostat setting estimated using equation \ref{eq:heterogeneity} with (\subref{fig:thermref}) thermostat setting, (\subref{fig:complianceref}) compliance, and (\subref{fig:fanref}) minutes of furnace fan run time as the dependent variables.  95 percent confidence intervals constructed from bootstrapped standard errors that account for first-stage estimation of expected thermostat setting and are cluster-robust to heteroskedasticity.} \label{fig:ref}
\end{minipage}
\end{figure}

Figure \ref{fig:ref} displays the average partial effects of treatment on the outcome variables by baseline thermostat setting.  When the expected thermostat setting is less than 65\(^\circ\)F, the appeal is slightly less effective and is not statistically significant in the coldest households, but we do not find support for a backfire effect as hypothesized in theory case 2.  Households just below the compliance target with baseline thermostat settings between 63-65\(^\circ\)F reduced thermostat settings by 1.02\(^\circ\)F on average. Households above the compliance target reduced thermostat settings more, with effects generally growing larger until households with thermostat settings above 75\(^\circ\)F whose treatment effect is statistically different than zero.  Baseline 65-67\(^\circ\)F households reduced thermostat settings by 1.16\(^\circ\)F, baseline 67-69\(^\circ\)F households reduced thermostat settings by 1.33\(^\circ\)F, baseline 69-71\(^\circ\)F reduced thermostat settings by 1.38\(^\circ\)F, baseline 71-73\(^\circ\)F households reduced thermostat settings by 1.34\(^\circ\)F, and baseline 73-75F households reduced thermostat settings by 1.43\(^\circ\)F.  The only statistically significant differences from the 63-65\(^\circ\)F treatment effect are the estimated effects for the baseline 67-69\(^\circ\)F and 69-71\(^\circ\)F households, which represent an approximately 30\% larger effect of treatment relative to the average treatment effect on the treated.  These treatment effect patterns are consistent with the appeal having different marginal impacts on either side of the 65\(^\circ\)F reference point as is consistent with the hypothesis in theory case 1.\footnote{At the recommendation of a thoughtful reviewer, we also sought to examine the hypothesis that the reference point effects differ by political affiliation.  We estimated a fully interacted model with expected thermostat setting and political affiliation.  Some coefficients in this model suffered from lack of precision or were not estimable due to not having sufficient observations in a particular vote share by thermostat setting bin.  Ultimately, we did not see clear evidence that the reference point effect varied systematically by vote share. The results of this model are available upon request.}  A possible alternative explanation for smaller effects for baseline thermostat settings below 65\(^\circ\)F might be that households face increasing costs of reducing the thermostat setting due to concerns about health or freezing pipes, which could generate similar results without reference dependence.

Households with baseline thermostat settings further from the compliance target were also less likely to fully comply with the reference level of 65\(^\circ\)F, which is in line with our model's prediction and may be caused by several mechanisms with different implications.  One potential mechanism is that the request may have appeared out of reach, which suggests that a reference level can induce larger contributions of effort for households near the reference level, but it discourages effort for households far away from that reference level.  For households with expected thermostat settings above 75\(^{\circ}\)F, the average partial effect on compliance rate is the lowest of those above the reference level; however, the average partial effect on fan run time was the largest, and in general higher baseline thermostat settings resulted in larger energy savings. This finding suggests that those households that did comply generated substantial energy savings, although the heterogeneity of this effect leads to a wide confidence interval. Another potential explanation may be that households that prefer warm temperatures have stronger preferences for deviating from their preferred thermostat setting.  We see this as unlikely, given the large average reductions in thermostat settings upon the request but cannot rule it out.  Without variation in the reference level, we are cautious not to draw more firm conclusions on the mechanisms.  The finding that high energy users generate the most savings or are most responsive is a common finding across conservation programs in several contexts \citep[see e.g.,][]{byrneetal2018,knittelstolper2021,brewer2023}.

The coefficients on the interaction between expected thermostat setting and treatment in appendix table \ref{tab:heterogeneity} are the difference in the average treatment effect by expected thermostat setting relative to households in time periods with an expected thermostat setting of 63-65\(^\circ\)F (the omitted category).  We find that households in time periods with baseline thermostat settings lower than the omitted category had statistically indistinguishable responses to the treatment relative to the omitted category.  Households with thermostat settings from 67-69\(^\circ\)F and 69-71\(^\circ\)F significantly reduced the thermostat by 0.3\(^\circ\)F and 0.35\(^\circ\)F more than the omitted category, roughly a 30\% increase relative to the average treatment effect. The estimated magnitude of the response increases and then decreases at higher temperatures but is not statistically significant.  Compliance with the emergency request falls with the distance from the compliance target.  The estimates for the fan run time variable show that for the coldest baseline thermostat setting, fan running times increased relative to the base category, although the difference is not statistically different from zero.  For baseline thermostat settings above 65\(^\circ\)F, fan run time decreased relative to the base category, with the largest effect (though not statistically significant) for the highest baseline thermostat settings.  Appendix table \ref{tab:heterogeneity_nocontrols} displays estimates of the coefficients when omitting the controls interacted with treatment.  We find that the reference point effect estimates are generally robust to omitting these controls.

One concern we had was whether these estimates were an artifact of statistical mean reversion rather than a meaningful pattern.\footnote{That is, do our estimates merely reflect that households with high or low temperatures in the past are mechanically more likely to have average temperatures when measured later?}  To test this alternative hypothesis, we estimate equation \ref{eq:heterogeneity} using the placebo cold wave event.  Appendix section \ref{sec:heterogeneityplacebo} displays the results of the placebo analysis.  The estimates from the placebo analysis are small, do not display similar patterns, and are statistically insignificant except for one coefficient, which leads us to reject mean reversion as an explanation of our estimated pattern in treatment effects on either side of the compliance target.  As an additional robustness check, we replicate the analysis on hourly data in appendix \ref{app:hourly} and find similar results to our primary analysis.  

\subsubsection{Vote share effect}

\begin{figure}
\centering
\begin{subfigure}[t]{.49\textwidth}
    \centering
    \includegraphics[width=\linewidth]{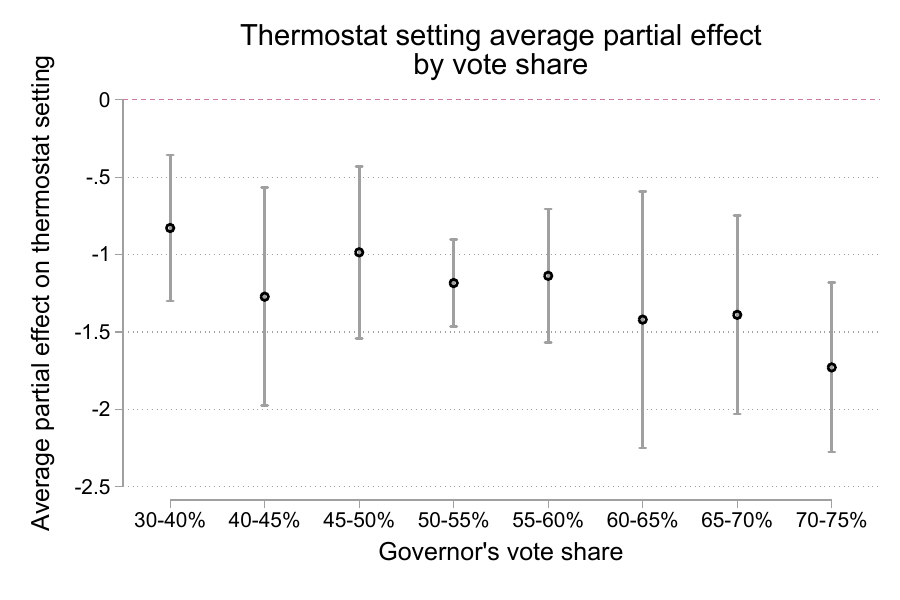}
            \caption{}\label{fig:defiancetemperature}
    \end{subfigure}
\begin{subfigure}[t]{.49\textwidth}
    \centering
    \includegraphics[width=\linewidth]{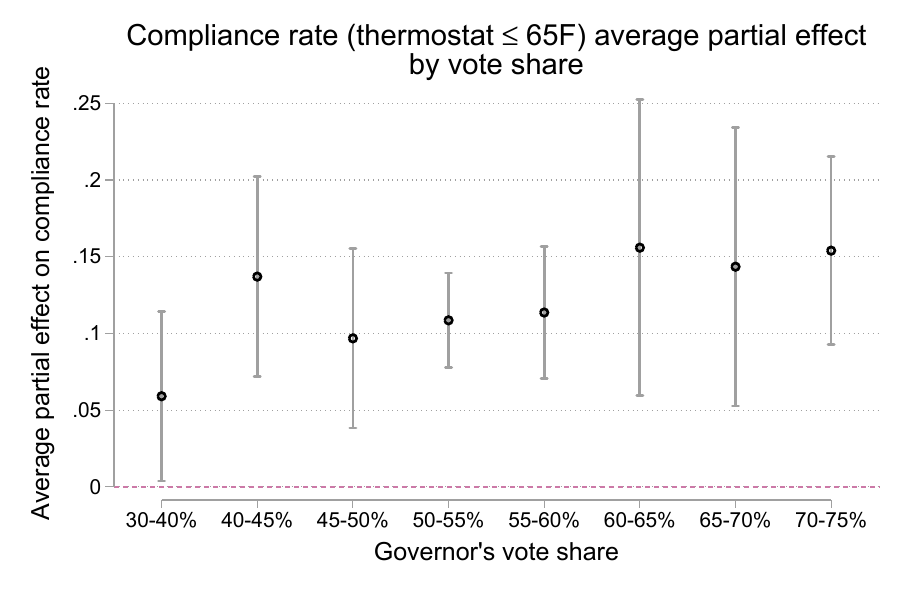}
        \caption{}\label{fig:defiancecompliance}
\end{subfigure}
\medskip
\begin{subfigure}[t]{.49\textwidth}
    \centering
    \vspace{0pt}
    \includegraphics[width=\linewidth]{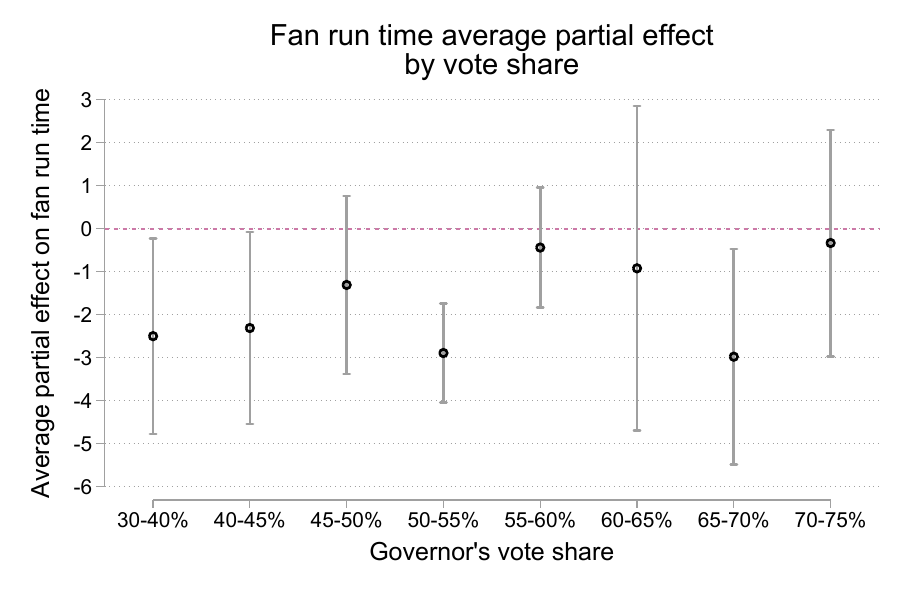}
        \caption{}\label{fig:defiancefan}
\end{subfigure}
\begin{minipage}[t]{.49\textwidth}
    \medskip \medskip \medskip
    \caption{Average partial effect (equation \ref{eq:ape_vote}) of the emergency request by Democratic Party Governor's vote share estimated using equation \ref{eq:heterogeneity} with (\subref{fig:defiancetemperature}) thermostat setting, (\subref{fig:defiancecompliance}) compliance, and (\subref{fig:defiancefan}) minutes of furnace fan run time  as the dependent variables.  95 percent confidence intervals constructed from standard errors cluster-robust to heteroskedasticity.} \label{fig:defiance}
\end{minipage}
\end{figure}

Figure \ref{fig:defiance} displays average partial effects of treatment on the outcome variables by the Democratic Party Governor's vote share. The average partial effect on thermostat setting and compliance is increasing in the Democratic Party Governor's vote share.  On average, the appeal induced about a 6 percent compliance rate in the counties most opposed to the Democratic Party Governor and a 15 percent compliance rate in the counties most in support of the Democratic Party Governor (difference = 9.5\%, p-value \(<\) 0.05).  In the counties most opposed to the Democratic Party Governor, the appeal induced a 0.8\(^\circ\)F decrease in thermostat settings while the most favorable counties saw a 1.7\(^\circ\)F decrease (difference = 0.9\(^\circ\)F, p-value \(<\) 0.05).

The coefficients on the interaction between county vote share and treatment in appendix table \ref{tab:heterogeneity} are the difference in the average treatment effect for households in counties with a given level of support for the Democratic Party Governor relative to households in counties where the Democratic Party Governor's vote share was 30-40\% (the omitted category).  The estimates show that the average thermostat reduction and compliance rate is higher in all counties relative to those with the lowest Democratic Party Governor's vote share. In addition, the treatment effects are generally increasing in magnitude as the Democratic Party Governor's vote share increases.  These differences relative to the baseline category are not statistically different from zero for the thermostat setting outcome except for the 70-75\% vote share category, while four out of seven coefficients from the compliance outcome are statistically significant.  The results indicate that households in the counties that supported the Democratic Party Governor the most reduced thermostats by up to 0.9\(^\circ\)F more on average and had a compliance rate 9.5 percentage points higher relative to the least supportive county.  Despite this, the estimates of the effect on fan run time are the opposite of the expected sign and are imprecise with all confidence intervals containing zero.  Given that fan run time is a function of underlying energy efficiency of the furnace and home, we believe that the fan estimates reflect unobserved differences in energy efficiency that are correlated with political affiliation.  

Our estimates in the appendix section \ref{sec:heterogeneityplacebo} show that the placebo analysis does not display similar patterns for thermostat setting and compliance, but that the fan run time estimates are similar to the main analysis, which supports our interpretation of the fan run time estimates as being driven by unobserved differences in energy efficiency causing differences in fan run time when temperatures drop.  Appendix table \ref{tab:heterogeneity_nocontrols} displays estimates of the coefficients when omitting the controls interacted with treatment.  We find that some of the estimates are sensitive to omitting these controls, which is because political affiliation is highly correlated with urban/rural status and a number of other demographics that likely affect responsiveness to emergency conservation requests.

Thus, we conclude that support for the Democratic Party Governor is weakly correlated with a stronger response to the public appeal, but these results are mixed and are sensitive to the inclusion of controls.  We caution against interpreting these estimates causally, but our findings are consistent with distrust arising out of affective political polarization or due to to differences in political ideology between parties.  In the increasingly polarized political environment of the United States \citep{iyengaretal2019}, a public appeal may be met with defiance rather than compliance. An alternative explanation is that political ideology may be correlated with willingness to contribute to public goods or thermostat setting behavior more broadly \citep{costakahn2013}.  Given our inability to distinguish the effect of polarization from political ideology, we cannot reject either explanation.

\section{Conclusion \label{sec:conclusion}}

This paper studies an acute natural gas shortage during the 2019 polar vortex in Michigan.  During near-record-low temperatures, a fire at a compressor plant resulted in natural gas demand that nearly outpaced supply.  In response, the utility requested that households voluntarily conserve natural gas, and the Michigan State Government subsequently issued an emergency text alert that requested households voluntarily reduce thermostat settings to 65\(^{\circ}\)F.

We use smart thermostat data to analyze consumer responses to the emergency request, finding robust evidence of voluntary compliance with the request. Using a difference-in-differences strategy with four control states, we obtain estimates of the average treatment effect on the treated.  On average, households lowered their thermostats by 1.1\(^{\circ}\)F, roughly a 25\% reduction of the typical variation in the average thermostat setting. 10 percent of households complied with the request fully by reducing their thermostats to 65\(^{\circ}\)F or lower, while 22 percent of household thermostat settings were already at or below the threshold.  Finally, we find evidence that the emergency request reduced furnace fan run times (our best proxy for natural gas consumption) by 1.5 minutes per hour; a 6\% decrease.  This reduction is smaller than the 10\% reduction in all consumption of natural gas we calculate using aggregate consumption data provided by the utility.  Our estimate is comparable with reductions in energy consumption observed in field experiments that use moral suasion to induce conservation (see e.g., \cite{brandonetal2018}), but falls short of field experiments that use price incentives to induce conservation (see e.g., \cite{itoidatanaka2018}).

Our analysis highlights the importance of wide-reaching emergency messaging for governments and utilities.  An event-study analysis reveals that the amplification of the utility's earlier request using the emergency alert system was critical for achieving compliance.  Prior to the emergency text alert, only 0.4\% of additional households reduced thermostat settings to 65\(^\circ\)F or less; after the phone alert, the additional compliance rate was as high as 20\%.  Using five-minute thermostat setting data (our highest frequency of obervation), we show that meaningful compliance with the emergency requests only began within five minutes of the cell phone alert.  The emergency directives communicated via social media and news media suffered from low visibility and were likely lost in the large amount of other content on these platforms.  As the emergency progressed, compliance with the request waned as households likely found it increasingly costly to maintain low thermostat settings.  In addition, we find evidence of persistence in low thermostat settings in the day after the emergency.  These habits appeared to be driven by thermostat setting changes left in place by households.

The particular phrasing of the emergency request around the reference point of 65\(^{\circ}\)F played a role in determining household behavior.  We identify two perverse effects of the reference point.  First, households that typically heat at 65\(^{\circ}\)F or lower responded less to the request than households that typically keep thermostats above the compliance target.  Households with baseline thermostat settings between 67-71\(^\circ\)F reduced their thermostat settings by approximately 30\% more than households with thermostat settings between 63-65\(^\circ\)F.  Second, those with the highest thermostat settings were less likely to comply with the request. For households with thermostat settings typically above 65\(^{\circ}\)F, the average treatment effect at first increases and then decreases with distance from the reference point.  These findings are consistent with our theoretical model, which suggests that a nudge with a compliance target will not achieve the least-cost reduction in energy consumption because of the difference in marginal incentives on either side of the reference point created by the target.

Setting a more aggressive target trades off a larger effect of compliance with the cost of compliance, which increases defiance.  This suggests that a particular reference point may induce a larger response.  While this natural experiment did not provide the necessary variation to determine the highest-impact reference points in this context, these could be determined via purposeful experimentation.   Furthermore, smart thermostat technology offers the possibility for targeted requests and automated compliance.  Alternatively, requests for a uniform reduction in thermostat setting may induce broader compliance than a uniform compliance goal because they avoid the problems driven by users in the tails of the energy consumption distribution.

Political affiliation also correlated with compliance.  Households in counties that voted least favorably for the 2018 Democratic Party Governor responded the least to the request, and households in counties that voted most favorably for the 2018 Democratic Party Governor had high levels of compliance.  We did not find evidence that this substantially impacted energy savings, so the results overall are mixed. We caution against a causal interpretation of these findings due to correlations between political affiliation and other unobserved characteristics of households, but our findings are consistent with prior findings on the influence of political ideology on the responsiveness of households to energy conservation requests \citep{costakahn2013}. 

Ultimately, the 2019 Michigan polar vortex crisis was resolved by a combination of residential and non-residential demand reductions and supply-side efforts.  After the crisis, the Governor of Michigan issued an executive order transferring energy emergency response management from the Michigan Energy Agency to the Michigan Public Service Commission (MPSC) and commissioned an assessment of Michigan energy resources from the MPSC \citep{mpsc2019c}. The report includes an overview of energy supply systems for natural gas, electricity, and propane, as well as a section on energy emergency management.  The section on emergency management states in general terms that utilities can pursue a variety of curtailment strategies to reduce demand, including voluntary requests and rate increases, but the guidelines are vague.

This paper shows that emergency demand response programs can help provide stability during times of crisis, but the efficacy of the program depends heavily on its design and implementation. While the emergency request in Michigan was largely successful and the worst-possible outcome was averted, the low overall level of compliance and perverse reference-point effects highlight the need for well-designed emergency measures to reduce energy demand.  To be successful, a voluntary emergency demand-response program needs a communication platform that enables it to reach households, can induce compliance from households that receive the request, and that has a demonstrated effectiveness so that utilities and balancing authorities know the size of the demand reductions to expect.  Rather than relying on voluntary requests with unknown efficacy, utilities should develop, test, and optimize programs that can be called upon when needed.

To avoid compliance issues, utilities can invest in voluntary programs that eliminate compliance barriers by purchasing centralized control of energy consumption long before emergency events occur.  Interruptible-load demand response programs are not new, but applications involving smart thermostats may provide a new opportunity to enhance emergency management.  For households who do not wish to surrender control or who have conventional thermostats, incentive-based emergency curtailment program design remains critical.

To assess the external validity of our estimates, we make use of the ``SANS'' conditions for generalizability suggested by \cite{list2020} in the appendix section \ref{sec:externalvalidity}.  In the SANS framework, the external validity of a study can be assessed by discussing selection, attrition, naturalness, and scaling.  In our context, there was very little attrition from the sample (we assess the sensitivity of the results to keeping and including those who enter or leave the sample early in appendix \ref{sec:robustness}), which leaves selection, naturalness, and scaling for discussion.  In our discussion, we find little evidence that selection plays a role in our results.  Given that the emergency request is a natural experiment and that similar emergency requests are made during other energy emergencies, we believe the intervention is natural and likely to be representative of interventions in other energy contexts.  Finally, we discuss the ability of this intervention to scale vertically to the grid (ISO) level, and horizontally across states and other energy emergencies.  We believe our results are likely to generalize to other energy emergencies based on the availability of a wireless emergency alert system, the political climate, and the frequency of repeated requests which may result in habituation to the alerts.

We see several results as particularly likely to generalize beyond energy consumption.  First, compliance with requests is likely to be higher when the message is conveyed through an official platform like the emergency text alert system.  This effect may be reduced by political polarization or distrust of institutions.  Second, a low-cost method of widespread emergency notification such as the cell phone alert system is key for communicating timely requests during a crisis.  Emergency communication via social media is likely to suffer from low reach and must compete with other content for visibility.  The incentives for compliance also matter.  Requests for uniform compliance goals with targets are likely to be too ambitious for some and too conservative for others.  Instead, a simple request for a marginal contribution to the public good or a menu of marginal contributions that can be dialed up or down avoids this problem without the need to explicitly tailor requests.  Finally, this event highlights the need for testing emergency planning and incorporating design elements that explicitly consider economic incentives and behavioral responses.

\bibliography{_ref}

\begin{thebibliography}{73}
\providecommand{\natexlab}[1]{#1}

\bibitem[{Abadie et~al.(2022)Abadie, Athey, Imbens, and
  Wooldridge}]{abadieatheyimbenswooldridge20}
Abadie, Alberto, Susan Athey, Guido~W Imbens, and Jeffrey~M Wooldridge. 2022.
\newblock When should you adjust standard errors for clustering?*.
\newblock \emph{The Quarterly Journal of Economics} 138~(1): 1--35.

\bibitem[{Akerlof and Kranton(2000)}]{akerlofkranton2000}
Akerlof, George~A., and Rachel~E. Kranton. 2000.
\newblock {Economics and Identity}.
\newblock \emph{The Quarterly Journal of Economics} 115~(3): 715--753.

\bibitem[{Allcott(2011)}]{allcott2011}
Allcott, Hunt. 2011.
\newblock Social norms and energy conservation.
\newblock \emph{Journal of Public Economics} 95~(9): 1082--1095.
\newblock Special Issue: The Role of Firms in Tax Systems.

\bibitem[{Allcott et~al.(2020)Allcott, Boxell, Conway, Gentzkow, Thaler, and
  Yang}]{allcottetal2020}
Allcott, Hunt, Levi Boxell, Jacob Conway, Matthew Gentzkow, Michael Thaler, and
  David Yang. 2020.
\newblock Polarization and public health: Partisan differences in social
  distancing during the coronavirus pandemic.
\newblock NBER working paper 26946, National Bureau of Economic Research.

\bibitem[{Allcott and Rogers(2014)}]{allcottrogers2014}
Allcott, Hunt, and Todd Rogers. 2014.
\newblock The short-run and long-run effects of behavioral interventions:
  Experimental evidence from energy conservation.
\newblock \emph{American Economic Review} 104~(10): 3003--37.

\bibitem[{Allcott and Taubinsky(2015)}]{allcotttaubinsky2015}
Allcott, Hunt, and Dmitry Taubinsky. 2015.
\newblock Evaluating behaviorally motivated policy: Experimental evidence from
  the lightbulb market.
\newblock \emph{American Economic Review} 105~(8): 2501--38.

\bibitem[{Anderson and Sallee(2011)}]{andersonsallee2011}
Anderson, Soren~T., and James~M. Sallee. 2011.
\newblock Using loopholes to reveal the marginal cost of regulation: The case
  of fuel-economy standards.
\newblock \emph{American Economic Review} 101~(4): 1375--1409.

\bibitem[{Barrios and Hochberg(2020)}]{barrioshochberg2020}
Barrios, John, and Yael Hochberg. 2020.
\newblock Risk perception through the lens of politics in the time of the
  covid-19 pandemic.
\newblock NBER working paper 27008, National Bureau of Economic Research.

\bibitem[{Beatty et~al.(2019)Beatty, Shimshack, and
  Volpe}]{beattyshimshackvolpe2019}
Beatty, Timothy K.~M., Jay~P. Shimshack, and Richard~J. Volpe. 2019.
\newblock Disaster preparedness and disaster response: Evidence from sales of
  emergency supplies before and after hurricanes.
\newblock \emph{Journal of the Association of Environmental and Resource
  Economists} 6~(4): 633--668.

\bibitem[{B\'enabou and Tirole(2006)}]{benaboutirole2006}
B\'enabou, Roland, and Jean Tirole. 2006.
\newblock Incentives and prosocial behavior.
\newblock \emph{American Economic Review} 96~(5): 1652--1677.

\bibitem[{Blackport and Screen(2020)}]{blackportscreen}
Blackport, R., and {J.A.} Screen. 2020.
\newblock Weakened evidence for mid-latitude impacts of {Arctic} warming.
\newblock \emph{Nature Climate Change} 10: 1065--1066.

\bibitem[{Blonz et~al.(2025)Blonz, Palmer, Wichman, and
  Wietelman}]{blonzetal2021}
Blonz, Joshua, Karen Palmer, Casey~J. Wichman, and Derek~C. Wietelman. 2025.
\newblock Smart thermostats, automation, and time-varying prices.
\newblock \emph{American Economic Journal: Applied Economics} 17~(1): 90–125.

\bibitem[{Brandon et~al.(2021)Brandon, Clapp, List, Metcalfe, and
  Price}]{brandonetal2021}
Brandon, Alec, Christopher~M. Clapp, John~A. List, Robert Metcalfe, and Michael
  Price. 2021.
\newblock Smart tech, dumb humans: The perils of scaling household
  technologies.
\newblock Working paper.

\bibitem[{Brandon et~al.(2018)Brandon, List, Metcalfe, and
  Rundhammer}]{brandonetal2018}
Brandon, Alec, John List, Robert Metcalfe, and Florian Rundhammer. 2018.
\newblock Testing for crowd out in social nudges: Evidence from a natural field
  experiment in the market for electricity.
\newblock \emph{Proceedings of the National Academy of Sciences} 116~(12):
  5293--5298.

\bibitem[{Brent et~al.(2015)Brent, Cook, and Olsen}]{brentetal2015}
Brent, Daniel~A., Joseph~H. Cook, and Skylar Olsen. 2015.
\newblock Social comparisons, household water use, and participation in utility
  conservation programs: Evidence from three randomized trials.
\newblock \emph{Journal of the Association of Environmental and Resource
  Economists} 2~(4): 597--627.

\bibitem[{Brewer(2022)}]{brewer2022}
Brewer, Dylan. 2022.
\newblock Equilibrium sorting and moral hazard in residential energy contracts.
\newblock \emph{Journal of Urban Economics} 129: 103424.

\bibitem[{Brewer(2023)}]{brewer2023}
---{}---{}---. 2023.
\newblock Household responses to winter heating costs: Implications for energy
  pricing policies and demand-side alternatives.
\newblock \emph{Energy Policy} 177: 113550.

\bibitem[{Breza et~al.(2021)Breza, Stanford, Alsan, Alsan, Banerjee,
  Chandrasekhar, Eichmeyer et~al.}]{brezaetal2021}
Breza, Emily, Fatima~Cody Stanford, Marcella Alsan, Burak Alsan, Abhijit
  Banerjee, Arun~G Chandrasekhar, Sarah Eichmeyer, et~al. 2021.
\newblock Effects of a large-scale social media advertising campaign on holiday
  travel and covid-19 infections: a cluster randomized controlled trial.
\newblock \emph{Nature Medicine} 27: 1622 -- 1628.

\bibitem[{Brown et~al.(2013)Brown, Johnstone, Haščič, Vong, and
  Barascud}]{brownetal2013}
Brown, Zachary, Nick Johnstone, Ivan Haščič, Laura Vong, and Francis
  Barascud. 2013.
\newblock Testing the effect of defaults on the thermostat settings of {OECD}
  employees.
\newblock \emph{Energy Economics} 39: 128--134.

\bibitem[{Burkhardt et~al.(2023)Burkhardt, Gillingham, and
  Kopalle}]{burkhardtetal2019}
Burkhardt, Jesse, Kenneth~T. Gillingham, and Praveen~K. Kopalle. 2023.
\newblock Field experimental evidence on the effect of pricing on residential
  electricity conservation.
\newblock \emph{Management Science} 69~(12): 7784--7798.

\bibitem[{Byrne et~al.(2018)Byrne, Nauze, and Martin}]{byrneetal2018}
Byrne, David~P., Andrea~La Nauze, and Leslie~A. Martin. 2018.
\newblock {Tell Me Something I Don’t Already Know: Informedness and the
  Impact of Information Programs}.
\newblock \emph{The Review of Economics and Statistics} 100~(3): 510--527.

\bibitem[{Cameron et~al.(2008)Cameron, Gelbach, and
  Miller}]{camerongelbachmiller2008}
Cameron, A.~Colin, Jonah~B. Gelbach, and Douglas~L. Miller. 2008.
\newblock {Bootstrap-Based Improvements for Inference with Clustered Errors}.
\newblock \emph{The Review of Economics and Statistics} 90~(3): 414--427.

\bibitem[{Carlsson et~al.(2021)Carlsson, Gravert, Johansson-Stenman, and
  Kurz}]{carlssonetal2021}
Carlsson, Fredrik, Christina Gravert, Olof Johansson-Stenman, and Verena Kurz.
  2021.
\newblock The use of green nudges as an environmental policy instrument.
\newblock \emph{Review of Environmental Economics and Policy} 15~(2): 216--237.

\bibitem[{Cialdini(2006)}]{cialdini2006}
Cialdini, Robert. 2006.
\newblock \emph{Influence: The Psychology of Persuasion, Revised Edition}.
\newblock Harper Business.

\bibitem[{{Consumers Energy Company}(2019)}]{mpsc2019b}
{Consumers Energy Company}. 2019.
\newblock Ray compressor station fire, {Jan.} 30, 2019.
\newblock Report submitted to Michigan Public Service Commission Case No.
  U-20463.

\bibitem[{Costa and Kahn(2013)}]{costakahn2013}
Costa, Dora~L., and Matthew~E. Kahn. 2013.
\newblock {Energy Conservation ``Nudges'' and Environmentalist Ideology:
  Evidence from a Randomized Residential Electricity Field Experiment}.
\newblock \emph{Journal of the European Economic Association} 11~(3): 680--702.

\bibitem[{Costa and Gerard(2021)}]{costagerard2021}
Costa, Francisco, and Francois Gerard. 2021.
\newblock Hysteresis and the welfare effect of corrective policies: Theory and
  evidence from an energy-saving program.
\newblock \emph{Journal of Political Economy} 129~(6): 1705--1743.

\bibitem[{{CQ Press}(2014-2019)}]{cqdatacounty}
{CQ Press}. 2014-2019.
\newblock Voting and elections collection.
\newblock Governor election returns, county detail by year,
  \url{http://library.cqpress.com/elections/download-data.php}, accessed
  05-2020.

\bibitem[{DellaVigna(2009)}]{dellavigna2009}
DellaVigna, Stefano. 2009.
\newblock Psychology and economics: Evidence from the field.
\newblock \emph{Journal of Economic Literature} 49~(2): 315--372.

\bibitem[{{DellaVigna} et~al.(2012){DellaVigna}, List, and
  Malmendier}]{dellavignaetal2012}
{DellaVigna}, Stefano, John List, and Ulrike Malmendier. 2012.
\newblock Testing for altruism and social pressure in charitable giving.
\newblock \emph{Quarterly Journal of Economics} 127~(1): 1--56.

\bibitem[{{DesOrmeau}(2019)}]{desormeau2019}
{DesOrmeau}, Taylor. 2019.
\newblock Michiganders answered call, cut gas usage 10 percent after emergency
  plea.

\bibitem[{Donald and Lang(2007)}]{donaldlang2007}
Donald, Stephen~G, and Kevin Lang. 2007.
\newblock {Inference with Difference-in-Differences and Other Panel Data}.
\newblock \emph{The Review of Economics and Statistics} 89~(2): 221--233.

\bibitem[{Edwards and List(2014)}]{EDWARDS2014}
Edwards, James~T., and John~A. List. 2014.
\newblock Toward an understanding of why suggestions work in charitable
  fundraising: Theory and evidence from a natural field experiment.
\newblock \emph{Journal of Public Economics} 114: 1--13.

\bibitem[{{EIA}(2019{\natexlab{a}})}]{eia2019b}
{EIA}. 2019{\natexlab{a}}.
\newblock Extreme cold in the {Midwest} led to high power demand and record
  natural gas demand.
\newblock United States Energy Information Administration, Today in Energy.

\bibitem[{{EIA}(2019{\natexlab{b}})}]{eia2019}
---{}---{}---. 2019{\natexlab{b}}.
\newblock Record cold temperatures in the upper midwest increase {U.S.} natural
  gas heating demand.
\newblock United States Energy Information Administration, Natural Gas Storage
  Dashboard Commentary.

\bibitem[{Engström et~al.(2015)Engström, Nordblom, Ohlsson, and
  Persson}]{engstrom2015}
Engström, Per, Katarina Nordblom, Henry Ohlsson, and Annika Persson. 2015.
\newblock Tax compliance and loss aversion.
\newblock \emph{American Economic Journal: Economic Policy} 7~(4): 132--64.

\bibitem[{Ferraro et~al.(2011)Ferraro, Miranda, and
  Price}]{ferraromirandaprice2011}
Ferraro, Paul~J., Juan~Jose Miranda, and Michael~K. Price. 2011.
\newblock The persistence of treatment effects with norm-based policy
  instruments: Evidence from a randomized environmental policy experiment.
\newblock \emph{American Economic Review} 101~(3): 318--22.

\bibitem[{Ferraro and Price(2013)}]{ferraroprice2013}
Ferraro, Paul~J., and Michael~K. Price. 2013.
\newblock {Using Nonpecuniary Strategies to Influence Behavior: Evidence from a
  Large-Scale Field Experiment}.
\newblock \emph{The Review of Economics and Statistics} 95~(1): 64--73.

\bibitem[{Ferraro and Tracy(2022)}]{ferrarotracy2022}
Ferraro, Paul~J., and {J. Dustin} Tracy. 2022.
\newblock A reassessment of the potential for loss‑framed incentive contracts
  to increase productivity: {A} meta‑analysis and a real‑effort experiment.
\newblock \emph{Experimental Economics} 25: 1441--1466.

\bibitem[{Foster et~al.(2009)Foster, Gutierrez, and Kumar}]{fosteretal2009}
Foster, Andrew, Emilio Gutierrez, and Naresh Kumar. 2009.
\newblock Voluntary compliance, pollution levels, and infant mortality in
  {Mexico}.
\newblock \emph{American Economic Review} 99~(2): 191--97.

\bibitem[{Foster and Gutierrez(2013)}]{fostergutierrez2013}
Foster, Andrew~D., and Emilio Gutierrez. 2013.
\newblock The informational role of voluntary certification: Evidence from the
  mexican clean industry program.
\newblock \emph{American Economic Review} 103~(3): 303--08.

\bibitem[{Ge and Ho(2019)}]{geho2018}
Ge, Qi, and Benjamin Ho. 2019.
\newblock Energy use and temperature habituation: Evidence from high frequency
  thermostat usage data.
\newblock \emph{Economic Inquiry} 57~(2): 1196--1214.

\bibitem[{Gray(2019)}]{freep2019}
Gray, Kathleen. 2019.
\newblock How the {Consumers Energy} polar vortex emergency unfolded.
\newblock The Detroit Free Press,
  \href{https://www.freep.com/story/news/politics/2019/02/20/consumers-energy-alert-fire-emergency/2929762002/}{https://www.freep.com/story/news/politics/2019/02/20/consumers-energy-alert-fire-emergency/2929762002/}
  accessed 01-2025.

\bibitem[{Hallsworth et~al.(2017)Hallsworth, List, Metcalfe, and
  Vlaev}]{hallsworthetal2017}
Hallsworth, Michael, John~A. List, Robert~D. Metcalfe, and Ivo Vlaev. 2017.
\newblock The behavioralist as tax collector: Using natural field experiments
  to enhance tax compliance.
\newblock \emph{Journal of Public Economics} 148: 14--31.

\bibitem[{Harding and Hsiaw(2014)}]{harding2014}
Harding, Matthew, and Alice Hsiaw. 2014.
\newblock Goal setting and energy conservation.
\newblock \emph{Journal of Economic Behavior \& Organization} 107: 209--227.

\bibitem[{Holladay et~al.(2015)Holladay, Price, and
  Wanamaker}]{holladayetal2015}
Holladay, J.~Scott, Michael~K. Price, and Marianne Wanamaker. 2015.
\newblock The perverse impact of calling for energy conservation.
\newblock \emph{Journal of Economic Behavior \& Organization} 110: 1--18.

\bibitem[{Ito et~al.(2018)Ito, Ida, and Tanaka}]{itoidatanaka2018}
Ito, Koichiro, Takanori Ida, and Makoto Tanaka. 2018.
\newblock Moral suasion and economic incentives: Field experimental evidence
  from energy demand.
\newblock \emph{American Economic Journal: Economic Policy} 10~(1): 240--67.

\bibitem[{Iyengar et~al.(2019)Iyengar, Lelkes, Levendusky, Malhotra, and
  Westwood}]{iyengaretal2019}
Iyengar, Shanto, Yphtach Lelkes, Matthew Levendusky, Neil Malhotra, and Sean~J.
  Westwood. 2019.
\newblock The origins and consequences of affective polarization in the {United
  States}.
\newblock \emph{Annual Review of Political Science} 22~(1): 129--146.

\bibitem[{Kim and Oh(2015)}]{KimOh2015}
Kim, Jungbu, and Seong~Soo Oh. 2015.
\newblock Confidence, knowledge, and compliance with emergency evacuation.
\newblock \emph{Journal of Risk Research} 18~(1): 111--126.

\bibitem[{Knittel and Stolper(2021)}]{knittelstolper2021}
Knittel, Christopher~R., and Samuel Stolper. 2021.
\newblock Machine learning about treatment effect heterogeneity: The case of
  household energy use.
\newblock \emph{AEA Papers and Proceedings} 111: 440--444.

\bibitem[{Levitt and List(2007)}]{levittlist2007}
Levitt, Steven~D., and John~A. List. 2007.
\newblock What do laboratory experiments measuring social preferences reveal
  about the real world?
\newblock \emph{Journal of Economic Perspectives} 21~(2): 153--174.

\bibitem[{List(2020)}]{list2020}
List, John~A. 2020.
\newblock Non est disputandum de generalizability? {A} glimpse into the
  external validity trial.
\newblock Working Paper 27535, National Bureau of Economic Research.

\bibitem[{Luyben(1982)}]{luyben1982}
Luyben, Paul~D. 1982.
\newblock Prompting thermostat setting behavior: Public response to a
  presidential appeal for conservation.
\newblock \emph{Environment and Behavior} 14~(1): 113--128.

\bibitem[{MacKinnon and Webb(2017)}]{mackinnonwebb}
MacKinnon, James~G., and Matthew~D. Webb. 2017.
\newblock Wild bootstrap inference for wildly different cluster sizes.
\newblock \emph{Journal of Applied Econometrics} 32~(2): 233--254.

\bibitem[{Meier et~al.(2019)Meier, Tsuyoshi, Pritoni, Rainer, Daken, and
  Baldewicz}]{meieretal2019}
Meier, Alan, Ueno Tsuyoshi, Marco Pritoni, Leo Rainer, Abigail Daken, and Dan
  Baldewicz. 2019.
\newblock What can connected thermostats tell us about {American} heating and
  cooling habits?
\newblock In \emph{{ECEEE} Summer Study Proceedings}. European Council for an
  Energy Efficient Economy.

\bibitem[{Montero(1999)}]{montero1999}
Montero, Juan‐Pablo. 1999.
\newblock Voluntary compliance with market‐based environmental policy:
  Evidence from the {U.S.} acid rain program.
\newblock \emph{Journal of Political Economy} 107~(5): 998--1033.

\bibitem[{{MPSC}(2019{\natexlab{a}})}]{mpsc2019c}
{MPSC}. 2019{\natexlab{a}}.
\newblock Michigan statewide energy assessment.
\newblock Michigan Public Service Commission.

\bibitem[{{MPSC}(2019{\natexlab{b}})}]{mpsc2019a}
---{}---{}---. 2019{\natexlab{b}}.
\newblock Michigan's statewide energy assessment fact sheet.
\newblock Michigan Public Service Commission.

\bibitem[{{NOAA}(2019)}]{noaa2019}
{NOAA}. 2019.
\newblock The science behind the polar vortex: You might want to put on a
  sweater.
\newblock Technical report, United States Department of Commerce National
  Oceanic and Atmospheric Administration.

\bibitem[{{NOAA}(2021)}]{noaadisasters}
---{}---{}---. 2021.
\newblock {U.S.} billion-dollar weather and climate disasters.
\newblock \url{https://www.ncdc.noaa.gov/billions/} accessed 09-06-2021).

\bibitem[{{Perryman Group}(2021)}]{perryman2021}
{Perryman Group}. 2021.
\newblock Preliminary estimates of economic costs of the february 2021 {Texas}
  winter storm.
\newblock Technical report, Perryman Group.
\newblock
  \href{https://www.perrymangroup.com/media/uploads/brief/perryman-preliminary-estimates-of-economic-costs-of-the-february-2021-texas-winter-storm-02-25-21.pdf}{https://www.perrymangroup.com/media/uploads/brief/perryman-preliminary-estimates-of-economic-costs-of-the-february-2021-texas-winter-storm-02-25-21.pdf},
  accessed 05-24-2021.

\bibitem[{Reiss and White(2008)}]{reisswhite2008}
Reiss, Peter, and Matthew White. 2008.
\newblock What changes energy consumption? {Prices} and public pressures.
\newblock \emph{{RAND} Journal of Economics} 39~(3): 636--663.

\bibitem[{Seibold(2021)}]{seibold2021}
Seibold, Arthur. 2021.
\newblock Reference points for retirement behavior: Evidence from {German}
  pension discontinuities.
\newblock \emph{American Economic Review} 111~(4): 1126--65.

\bibitem[{Sigman and Chang(2011)}]{sigmanchang2011}
Sigman, Hilary, and Howard~F. Chang. 2011.
\newblock The effect of allowing pollution offsets with imperfect enforcement.
\newblock \emph{American Economic Review} 101~(3): 268--72.

\bibitem[{Thaler and Sunstein(2008)}]{thalersunstein2008}
Thaler, Richard, and Cass Sunstein. 2008.
\newblock \emph{Nudge: Improving Decisions about Health, Wealth, and
  Happiness}.
\newblock Yale University Press.

\bibitem[{Tversky and Kahneman(1981)}]{kahnemantversky1981}
Tversky, Amos, and Daniel Kahneman. 1981.
\newblock The framing of decisions and the psychology of choice.
\newblock \emph{Science} 211~(4481): 453 -- 458.

\bibitem[{{United States Census Bureau}(2023)}]{households}
{United States Census Bureau}. 2023.
\newblock Table {HH-4}. households by size: 1960 to present.
\newblock Historical Households Tables,
  \href{https://www.census.gov/data/tables/time-series/demo/families/households.html}{https://www.census.gov/data/tables/time-series/demo/families/households.html},
  version last revised 21 November 202, accessed September 2024.

\bibitem[{{{US} Census Bureau}(2019)}]{acs2019}
{{US} Census Bureau}. 2019.
\newblock American community survey 1-year sample.

\bibitem[{Whitehead et~al.(2000)Whitehead, Edwards, Willigen, Maiolo, Wilson,
  and Smith}]{whiteheadetal2000}
Whitehead, John~C., Bob Edwards, Marieke~Van Willigen, John~R. Maiolo, Kenneth
  Wilson, and Kevin~T. Smith. 2000.
\newblock Heading for higher ground: {Factors} affecting real and hypothetical
  hurricane evacuation behavior.
\newblock \emph{Global Environmental Change Part B: Environmental Hazards}
  2~(4): 133--142.

\bibitem[{Wichman et~al.(2016)Wichman, Taylor, and {von
  Haefen}}]{wichmantaylorvonhaefen2016}
Wichman, Casey~J., Laura~O. Taylor, and Roger~H. {von Haefen}. 2016.
\newblock Conservation policies: Who responds to price and who responds to
  prescription?
\newblock \emph{Journal of Environmental Economics and Management} 79:
  114--134.

\bibitem[{Wooldridge(2007)}]{wooldridge2007}
Wooldridge, Jeffrey~M. 2007.
\newblock Inverse probability weighted estimation for general missing data
  problems.
\newblock \emph{Journal of Econometrics} 141~(2): 1281--1301.

\bibitem[{Wooldridge(2010)}]{wooldridge2010}
---{}---{}---. 2010.
\newblock \emph{Econometric Analysis of Cross Section and Panel Data}.
\newblock 2 edition. The {MIT} Press.

\bibitem[{Zou(2021)}]{zou2021}
Zou, Eric~Yongchen. 2021.
\newblock Unwatched pollution: The effect of intermittent monitoring on air
  quality.
\newblock \emph{American Economic Review} 111~(7): 2101--26.

\end{thebibliography}
\bibliographystyle{jaere}

\begin{appendices}
\appendixpage

\section{Phone alert and social media posts}

\begin{figure}[ht]
    \includegraphics{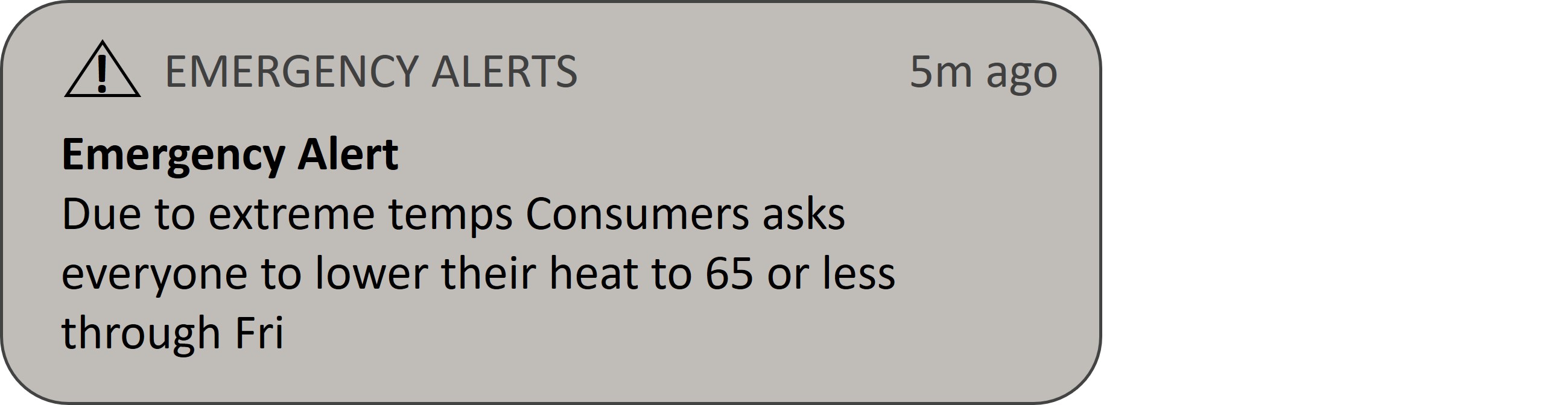}
    \footnotesize{\href{https://www.mlive.com/news/2019/01/heat-interruptions-possible-if-residents-dont-turn-down-thermostats-utility-says.html}{https://www.mlive.com/news/2019/01/heat-interruptions-possible-if-residents-dont-turn-down-thermostats-utility-says.html}}
    \caption{Recreation of cell phone alert } \label{fig:alert}
\end{figure}

\begin{figure}
\centering
\begin{subfigure}[t]{.49\textwidth}
    \centering
    \includegraphics[width=\linewidth]{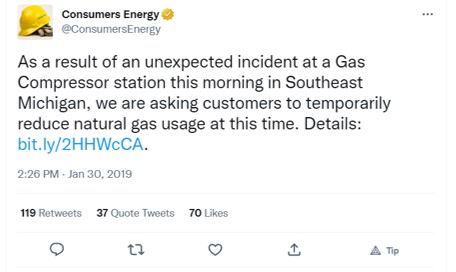}
            \caption{\href{https://twitter.com/ConsumersEnergy/status/1090692811081551885}{https://twitter.com/ConsumersEnergy/ \break status/1090692811081551885}}\label{fig:tweet1}
    \end{subfigure}
\begin{subfigure}[t]{.49\textwidth}
    \centering
    \includegraphics[width=\linewidth]{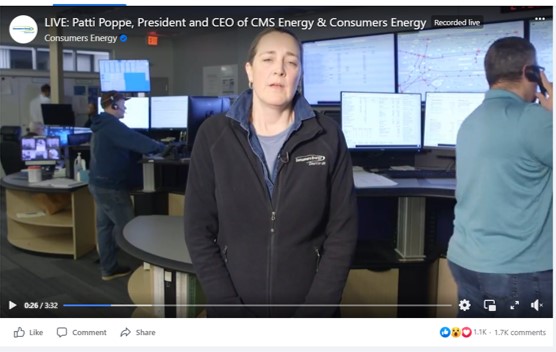}
        \caption{\href{https://www.facebook.com/85543026043/videos/357785638397997/}{https://www.facebook.com/85543026043/ \break videos/357785638397997/}}\label{fig:facebooklive}
\end{subfigure}
\medskip
\begin{subfigure}[t]{.49\textwidth}
    \centering
    \vspace{0pt}
    \includegraphics[width=\linewidth]{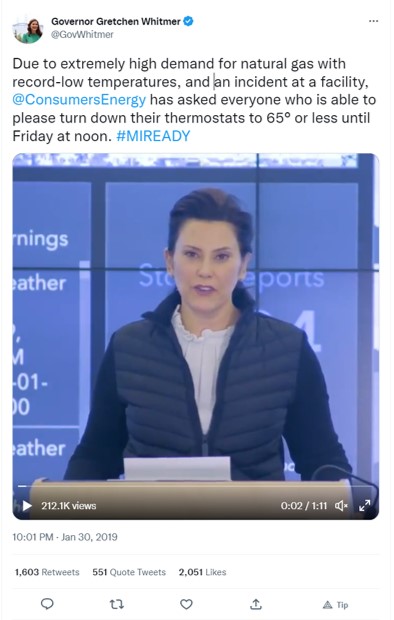}
        \caption{\href{https://twitter.com/GovWhitmer/status/1090807363811065857}{https://twitter.com/GovWhitmer/ \break status/1090807363811065857}}\label{fig:tweet2}
\end{subfigure}
\begin{subfigure}[t]{.49\textwidth}
    \centering
    \vspace{9pt}
    \includegraphics[width=\linewidth]{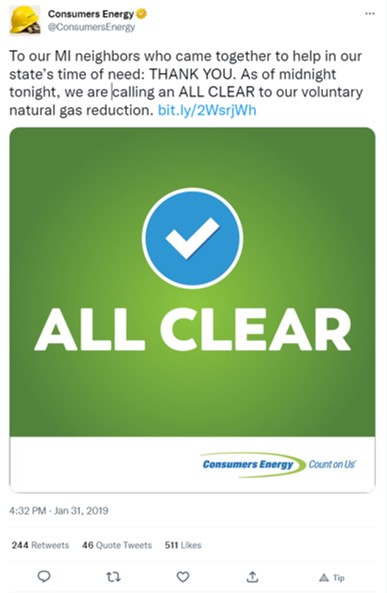}
       \\ \caption{\href{https://twitter.com/ConsumersEnergy/status/1091086892592959495}{https://twitter.com/ConsumersEnergy/ \break status/1091086892592959495}}\label{fig:tweet3}
\end{subfigure}
    \caption{Social media appeals.} \label{fig:socialmedia}
\end{figure}

\clearpage

\section{Tests of alternative behavioral responses} \label{app:alternates}

Here, we consider potential alternative behavioral responses to the emergency request outside of the thermostat setting.  Households may have altered the thermostat mode or turned the furnace off in order to comply with the emergency request.  In addition, the emergency request may have either induced households in Michigan to stay at home during the emergency or to seek shelter elsewhere, potentially differently than households in the control states.  The smart thermostat data contain variables on the thermostat's mode and whether the thermostat's motion sensor detects motion. 
 Our thermostat mode variables contain information on whether the furnace is turned off, set to ``hold'' (which keeps a constant thermostat setting), or set to a smart automation setting.  The Ecobee thermostat's smart automation feature ``smart recovery'' can pre-heat or cool a home based on a household's typical behavior (for example, if a person typically arrives home from work at 6:00 pm and increases the thermostat setting, the smart recovery feature may automatically begin heating the home at 5:45 pm in anticipation of arrival).

 We test whether the emergency event induced households to set the thermostat to ``hold,'' use the automation feature, turn the furnace off, and spend more or less time at home by using these variables as outcomes in our main specification (equation \ref{eq:twfe}).  Each of these variables is a binary indicator equal to one if the thermostat was in the indicated mode or detected motion during the period, making these linear probability models.  Table \ref{tab:alt_outcome} displays these estimates.  We estimate a 2 percentage point increase in thermostat settings on ``hold'' and a 3 percentage point increase in the use of automation, which we interpret as small changes.  We find no statistically significant change in whether the furnace was turned off or whether motion was detected by the thermostat.

 Although we do not see large responses via these channels, we test the robustness of our thermostat setting estimates to controlling for thermostat mode and the motion sensor indicator in the thermostat setting, compliance, and fan run time regressions.  Table \ref{tab:robust_mode} contains these estimates.  While the furnace fan estimate is slightly larger than the estimates in the main text (table \ref{tab:twfe}), the estimates are overall similar.

\begin{table}
    \centering
    \begin{threeparttable}
    \caption{Potential alternative behavioral responses \label{tab:alt_outcome}}
    \begin{tabular}{lcccc} \hline
 & (1) & (2) & (3) & (4) \\
VARIABLES & Thermostat hold & Thermostat automation & Furnace off & Motion detected \\ \hline
 &  &  &  &  \\
Michigan x Post & 0.02* & 0.03* & -0.00 & -0.00 \\
 & (0.01) & (0.01) & (0.00) & (0.00) \\
 &  &  &  &  \\
Observations & 2,144,796 & 2,144,796 & 2,144,796 & 2,144,796 \\
R-squared & 0.51 & 0.30 & 0.69 & 0.55 \\
Weather controls & YES & YES & YES & YES \\
Household FE & YES & YES & YES & YES \\
Time FE & YES & YES & YES & YES \\
DOW $\times$ HOD $\times$ state & YES & YES & YES & YES \\
 Pre-treatment mean & 0.271 & 0.199 & 0.00927 & 0.574 \\ \hline
\end{tabular}

    \begin{tablenotes}
        \footnotesize Estimates of regressions from equation \ref{eq:twfe} using alternate potential outcome variables.  The sample includes 4-hour-average household observations from January 2nd-January 31st. Pre-treatment means reported for Michigan households.  Standard errors clustered at the state level.  ** p$<$0.01, * p$<$0.05
    \end{tablenotes}
\end{threeparttable}
\end{table}

\begin{table}
    \centering
    \begin{threeparttable}
    \caption{Controlling for thermostat mode and motion sensor \label{tab:robust_mode}}
    \begin{tabular}{lccc} \hline
 & (1) & (2) & (3) \\
VARIABLES & Thermostat setting & Thermostat $\leq$ 65F & Fan run time \\ \hline
 &  &  &  \\
Michigan x Post & -1.07** & 0.10** & -1.63* \\
 & (0.12) & (0.01) & (0.50) \\
 &  &  &  \\
Observations & 2,126,336 & 2,126,336 & 2,135,114 \\
R-squared & 0.74 & 0.53 & 0.76 \\
Weather controls & YES & YES & YES \\
Household FE & YES & YES & YES \\
Time FE & YES & YES & YES \\
Day of week $\times$ hour of day $\times$ state & YES & YES & YES \\
Thermostat modes & YES & YES & YES \\
Motion detected & YES & YES & YES \\
 Pre-treatment mean & 66.87 & 0.270 & 18.09 \\ \hline
\end{tabular}

    \begin{tablenotes}
        \footnotesize Estimates of regressions from equation \ref{eq:twfe} controlling for thermostat mode and motion sensor.  The sample includes 4-hour-average household observations from January 2nd-January 31st. Pre-treatment means reported for Michigan households.  Standard errors clustered at the state level.  ** p$<$0.01, * p$<$0.05
    \end{tablenotes}
\end{threeparttable}
\end{table}

\section{Heterogeneity analysis additional figures and tables}

\begin{figure}[h]
    \centering
    \begin{subfigure}[t]{.49\textwidth}
        \centering
        \includegraphics[width=\linewidth]{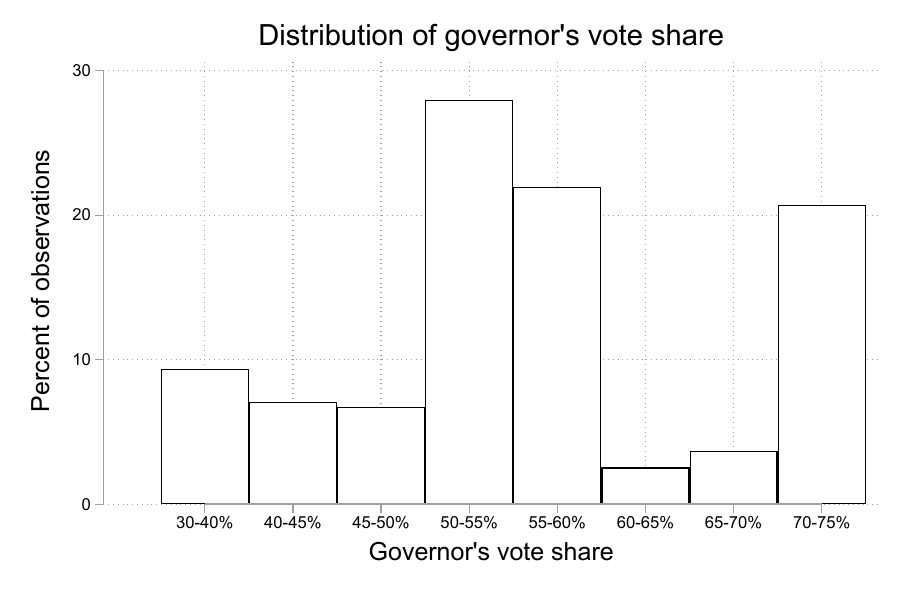}
                \caption{}\label{fig:voteshare_distribution}
        \end{subfigure}
    \begin{subfigure}[t]{.49\textwidth}
        \centering
        \includegraphics[width=\linewidth]{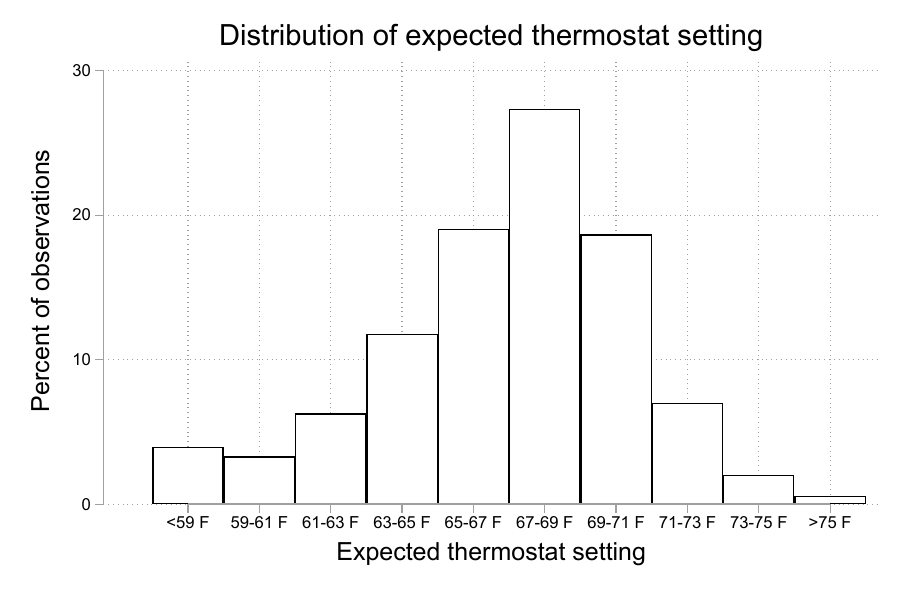}
            \caption{}\label{fig:expected_distribution}
    \end{subfigure}

        \caption{Distribution of variables analyzed in the heterogeneity analysis.  Panel (\subref{fig:voteshare_distribution}) plots displays the distribution of the Democratic Party Governor's vote shares in Michigan, panel (\subref{fig:expected_distribution}) plots the distribution of expected thermostat settings in pre-treatment Michigan.} \label{fig:heterogeneity_distribution}
\end{figure}

\begin{table}
    \centering
    \small
    \begin{threeparttable}
    \caption{Heterogeneity analysis}
    {\def\sym#1{\ifmmode^{#1}\else\(^{#1}\)\fi} \begin{tabular}{l*{3}{c}} \hline
                &\multicolumn{1}{c}{(1)}&\multicolumn{1}{c}{(2)}&\multicolumn{1}{c}{(3)}\\
                &\multicolumn{1}{c}{Thermostat setting}&\multicolumn{1}{c}{Thermostat $\leq$ 65F}&\multicolumn{1}{c}{Fan run time}\\
\hline
59 F or lower expected X Treatment&     0.62        &    -0.11\sym{**}&     0.77        \\
                &   (0.58)        &  (0.032)        &   (1.04)        \\
59-61 F expected X Treatment&   -0.075        &   -0.013        &    -0.24        \\
                &   (0.26)        &  (0.026)        &   (0.82)        \\
61-63 F expected X Treatment&  0.00062        &  -0.0086        &    0.020        \\
                &   (0.16)        &  (0.023)        &   (0.55)        \\
65-67 F expected X Treatment&    -0.14        &  -0.0062        &     0.15        \\
                &   (0.10)        &  (0.019)        &   (0.43)        \\
67-69 F expected X Treatment&    -0.30\sym{*} &   -0.032        &     0.16        \\
                &   (0.13)        &  (0.023)        &   (0.41)        \\
69-71 F expected X Treatment&    -0.35\sym{**}&   -0.083\sym{**}&   -0.047        \\
                &   (0.13)        &  (0.022)        &   (0.43)        \\
71-73 F expected X Treatment&    -0.32        &    -0.11\sym{**}&    -0.13        \\
                &   (0.21)        &  (0.024)        &   (0.64)        \\
73-75 F expected X Treatment&    -0.41        &    -0.12\sym{**}&    -0.23        \\
                &   (0.28)        &  (0.026)        &   (0.98)        \\
Higher than 75 F expected X Treatment&     0.68        &    -0.16\sym{**}&    -2.97        \\
                &   (0.45)        &  (0.027)        &   (2.01)        \\
40-45\% Democrat X Treatment&    -0.44        &    0.078\sym{**}&     0.19        \\
                &   (0.29)        &  (0.026)        &   (0.80)        \\
45-50\% Democrat X Treatment&    -0.16        &    0.038        &     1.19        \\
                &   (0.25)        &  (0.024)        &   (0.91)        \\
50-55\% Democrat X Treatment&    -0.36        &    0.049\sym{*} &    -0.39        \\
                &   (0.21)        &  (0.023)        &   (1.07)        \\
55-60\% Democrat X Treatment&    -0.31        &    0.055        &     2.06        \\
                &   (0.36)        &  (0.038)        &   (1.59)        \\
60-65\% Democrat X Treatment&    -0.59        &    0.097\sym{*} &     1.58        \\
                &   (0.40)        &  (0.045)        &   (2.11)        \\
65-70\% Democrat X Treatment&    -0.56        &    0.084        &    -0.48        \\
                &   (0.39)        &  (0.051)        &   (1.80)        \\
70-75\% Democrat X Treatment&    -0.90\sym{*} &    0.095\sym{*} &     2.17        \\
                &   (0.41)        &  (0.047)        &   (2.07)        \\
                &                 &                 &                 \\
Observations    &1,959,762        &1,959,762        &1,960,072        \\
R-squared       &     0.82        &     0.70        &     0.76        \\
Household FE    &      YES        &      YES        &      YES        \\
Time FE         &      YES        &      YES        &      YES        \\
Controls $\times$ treatment&      YES        &      YES        &      YES        \\
Day of week $\times$ hour of day $\times$ state&      YES        &      YES        &      YES        \\
Expected thermostat $\times$ time FE&      YES        &      YES        &      YES        \\
Pre-treatment mean&     66.9        &     0.27        &     18.1        \\
\hline \end{tabular} }

    \label{tab:heterogeneity}
    \begin{tablenotes}
        \footnotesize Results from the estimation of equation \ref{eq:heterogeneity}. The sample includes 4-hour-average observations from January 2nd-January 31st. Pre-treatment means reported for Michigan households. Standard errors cluster-bootstrapped to incorporate the sampling error from estimation of the baseline thermostat setting.  ** p$<$0.01, * p$<$0.05
    \end{tablenotes}
\end{threeparttable}
\end{table}

\begin{table}
    \centering
    \small
    \begin{threeparttable}
    \caption{Robustness check: Heterogeneity analysis without controls \(\times\) treatment}
    {\def\sym#1{\ifmmode^{#1}\else\(^{#1}\)\fi} \begin{tabular}{l*{3}{c}} \hline
                &\multicolumn{1}{c}{(1)}&\multicolumn{1}{c}{(2)}&\multicolumn{1}{c}{(3)}\\
                &\multicolumn{1}{c}{Thermostat setting}&\multicolumn{1}{c}{Thermostat $\leq$ 65F}&\multicolumn{1}{c}{Fan run time}\\
\hline
59 F or lower expected X Treatment&     0.48        &   -0.098\sym{**}&     0.77        \\
                &   (0.58)        &  (0.032)        &   (0.97)        \\
59-61 F expected X Treatment&   -0.072        &   -0.013        &   -0.049        \\
                &   (0.25)        &  (0.026)        &   (0.81)        \\
61-63 F expected X Treatment&   -0.083        &   0.0039        &     0.14        \\
                &   (0.16)        &  (0.022)        &   (0.50)        \\
65-67 F expected X Treatment&    -0.14        &  -0.0037        &     0.27        \\
                &  (0.094)        &  (0.017)        &   (0.41)        \\
67-69 F expected X Treatment&    -0.33\sym{**}&   -0.027        &     0.13        \\
                &   (0.12)        &  (0.022)        &   (0.40)        \\
69-71 F expected X Treatment&    -0.40\sym{**}&   -0.077\sym{**}&    -0.13        \\
                &   (0.12)        &  (0.021)        &   (0.42)        \\
71-73 F expected X Treatment&    -0.32        &    -0.12\sym{**}&     0.32        \\
                &   (0.20)        &  (0.023)        &   (0.65)        \\
73-75 F expected X Treatment&    -0.54\sym{*} &    -0.12\sym{**}&    -0.55        \\
                &   (0.27)        &  (0.024)        &   (0.92)        \\
Higher than 75 F expected X Treatment&     0.40        &    -0.16\sym{**}&    -3.71        \\
                &   (0.44)        &  (0.025)        &   (2.00)        \\
40-45\% Democrat X Treatment&    -0.45\sym{*} &    0.063\sym{**}&    0.061        \\
                &   (0.20)        &  (0.020)        &   (0.67)        \\
45-50\% Democrat X Treatment&    -0.11        &    0.026        &     0.36        \\
                &   (0.19)        &  (0.022)        &   (0.90)        \\
50-55\% Democrat X Treatment&    -0.21        &    0.044\sym{**}&    -0.81        \\
                &   (0.14)        &  (0.016)        &   (0.57)        \\
55-60\% Democrat X Treatment&   -0.033        &    0.028        &     1.36\sym{*} \\
                &   (0.13)        &  (0.017)        &   (0.65)        \\
60-65\% Democrat X Treatment&    0.073        &    0.030        &     0.80        \\
                &   (0.26)        &  (0.037)        &   (1.34)        \\
65-70\% Democrat X Treatment&    -0.38        &    0.090\sym{*} &    -2.18\sym{*} \\
                &   (0.20)        &  (0.036)        &   (0.89)        \\
70-75\% Democrat X Treatment&   -0.011        &    0.034        &     1.87\sym{**}\\
                &   (0.16)        &  (0.020)        &   (0.63)        \\
                &                 &                 &                 \\
Observations    &2,125,591        &2,125,591        &2,125,905        \\
R-squared       &     0.82        &     0.70        &     0.76        \\
Household FE    &      YES        &      YES        &      YES        \\
Time FE         &      YES        &      YES        &      YES        \\
Day of week $\times$ hour of day $\times$ state&      YES        &      YES        &      YES        \\
Expected thermostat $\times$ time FE&      YES        &      YES        &      YES        \\
Pre-treatment mean&     66.9        &     0.27        &     18.1        \\
\hline \end{tabular} }

    \label{tab:heterogeneity_nocontrols}
    \begin{tablenotes}
        \footnotesize Results from the estimation of equation \ref{eq:heterogeneity} without controls interacted with treatment. The sample includes 4-hour-average observations from January 2nd-January 31st. Pre-treatment means reported for Michigan households. Standard errors cluster-bootstrapped to incorporate the sampling error from estimation of the baseline thermostat setting.  ** p$<$0.01, * p$<$0.05
    \end{tablenotes}
\end{threeparttable}
\end{table}

\begin{table}
    \centering
    \small
    \begin{threeparttable}
    \caption{Additional coefficients from heterogeneity analysis}
    {\def\sym#1{\ifmmode^{#1}\else\(^{#1}\)\fi} \begin{tabular}{l*{3}{c}} \hline
                &\multicolumn{1}{c}{(1)}&\multicolumn{1}{c}{(2)}&\multicolumn{1}{c}{(3)}\\
                &\multicolumn{1}{c}{Thermostat setting}&\multicolumn{1}{c}{Thermostat $\leq$ 65F}&\multicolumn{1}{c}{Fan run time}\\
\hline
Treatment X Detached home&   -0.055        &   0.0075        &    -0.29        \\
                &  (0.074)        &  (0.010)        &   (0.32)        \\
Treatment X Floor area (sq. ft.)& 0.000054        &-0.0000045        & -0.00017        \\
                &(0.000052)        &(0.0000062)        &(0.00015)        \\
Treatment X Age of home (years)&  0.00058        & 0.000087        &   0.0094        \\
                & (0.0016)        &(0.00015)        & (0.0053)        \\
Treatment X Number of occupants&   -0.016        & -0.00035        &    -0.29\sym{**}\\
                &  (0.019)        & (0.0029)        &  (0.083)        \\
Treatment X Median income& 0.000019        &-0.0000018        &  0.00016\sym{**}\\
                &(0.000010)        &(0.0000012)        &(0.000043)        \\
Treatment X Median age&  0.00017        &  -0.0034        &    0.037        \\
                &  (0.031)        & (0.0026)        &  (0.082)        \\
Treatment X Total population&-0.00000033        &0.000000026        &-0.0000024        \\
                &(0.00000027)        &(0.000000034)        &(0.0000014)        \\
Treatment X Fraction male&    -0.99        &    0.032        &     7.27        \\
                &   (11.1)        &   (1.03)        &   (39.6)        \\
Treatment X Fraction white&    -5.31\sym{**}&     0.47\sym{*} &    -15.9        \\
                &   (1.76)        &   (0.22)        &   (9.24)        \\
Treatment X Fraction with high school education&     2.44        &    -0.30        &     37.9\sym{**}\\
                &   (3.56)        &   (0.34)        &   (11.4)        \\
Treatment X Fraction non-US citizens&    -6.59        &     0.49        &     26.7        \\
                &   (8.06)        &   (0.72)        &   (24.1)        \\
                &                 &                 &                 \\
Observations    &1,959,762        &1,959,762        &1,960,072        \\
R-squared       &     0.82        &     0.70        &     0.76        \\
Household FE    &      YES        &      YES        &      YES        \\
Time FE         &      YES        &      YES        &      YES        \\
Controls $\times$ treatment&      YES        &      YES        &      YES        \\
Day of week $\times$ hour of day $\times$ state&      YES        &      YES        &      YES        \\
Expected thermostat $\times$ time FE&      YES        &      YES        &      YES        \\
Pre-treatment mean&     66.9        &     0.27        &     18.1        \\
\hline \end{tabular} }

    \label{tab:heterogeneity_full}
    \begin{tablenotes}
        \footnotesize Additional coefficient estimates from the estimation of equation \ref{eq:heterogeneity} (see table \ref{tab:heterogeneity} for main coefficient estimates). The sample includes 4-hour-average observations from January 2nd-January 31st. Pre-treatment means reported for Michigan households. Standard errors cluster-bootstrapped to incorporate the sampling error from estimation of the baseline thermostat setting.  ** p$<$0.01, * p$<$0.05
    \end{tablenotes}
\end{threeparttable}
\end{table}

\clearpage

\section{Hourly analysis} \label{app:hourly}

\begin{figure}
\centering
\begin{subfigure}[t]{.49\textwidth}
    \centering
    \includegraphics[width=\linewidth]{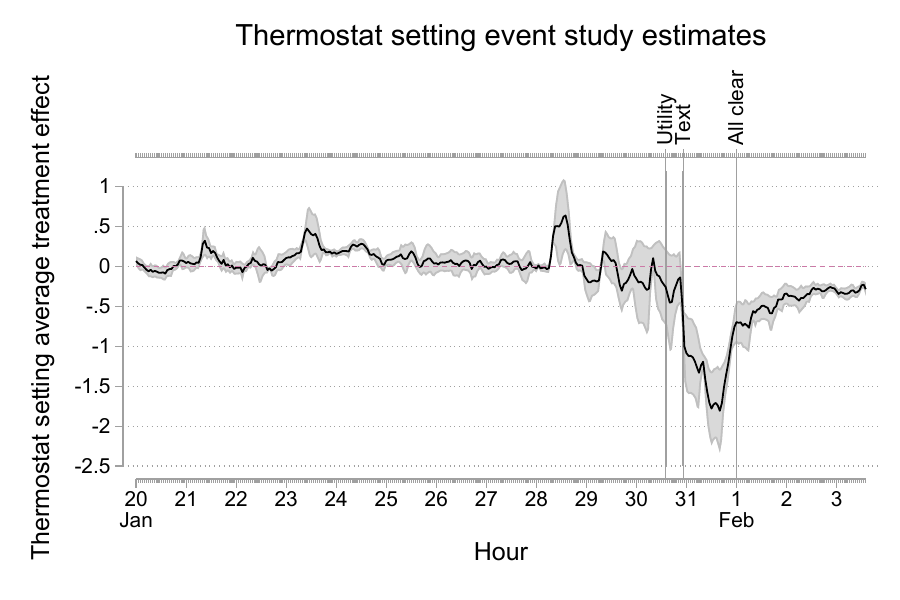}
            \caption{}\label{fig:eventtemperature_hour}
    \end{subfigure}
\begin{subfigure}[t]{.49\textwidth}
    \centering
    \includegraphics[width=\linewidth]{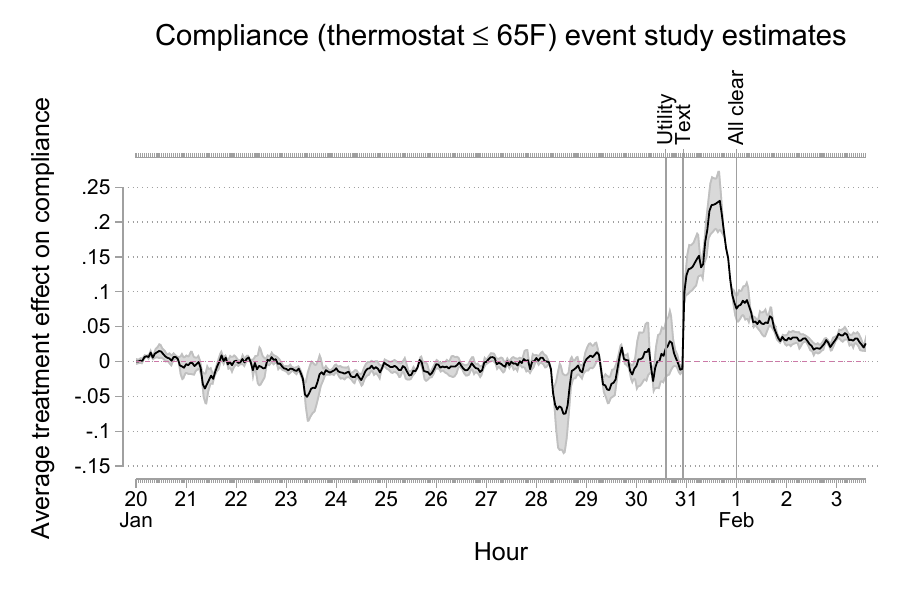}
        \caption{}\label{fig:eventcompliance_hour}
\end{subfigure}
\medskip
\begin{subfigure}[t]{.49\textwidth}
    \centering
    \vspace{0pt}
    \includegraphics[width=\linewidth]{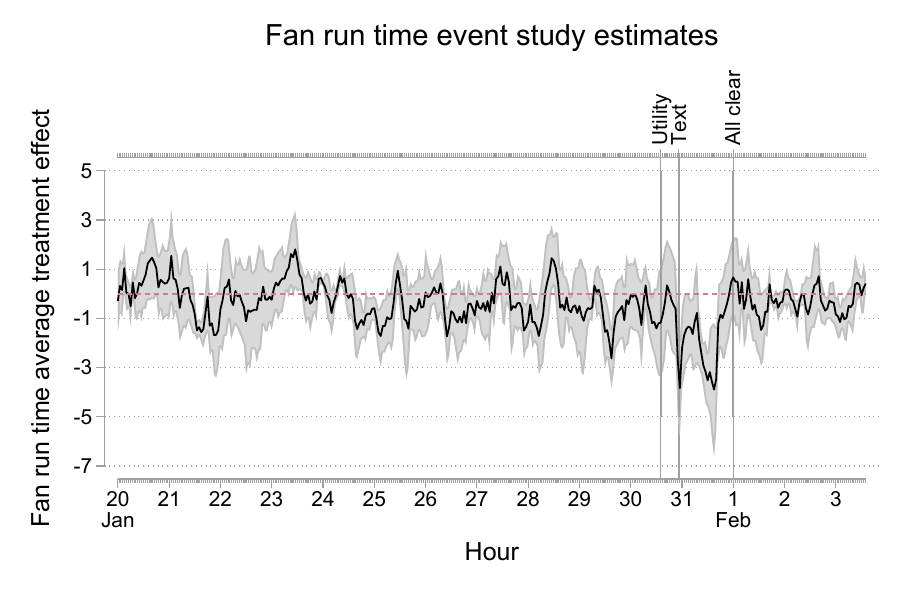}
        \caption{}\label{fig:eventfan_hour}
\end{subfigure}
\begin{minipage}[t]{.49\textwidth}
    \medskip \medskip \medskip
    \caption{Event-study coefficients estimated on hourly data using regression equation \ref{eq:eventstudy} with (\subref{fig:eventtemperature}) thermostat setting, (\subref{fig:eventcompliance}) compliance, and (\subref{fig:eventfan}) minutes of furnace fan run time as the dependent variables.  95 percent confidence intervals constructed from standard errors cluster-robust to heteroskedasticity.} \label{fig:eventstudy_hour}
\end{minipage}
\end{figure}

In this section, we replicate the analyses from the main text using hourly data rather than data aggregated to four-hour intervals.  Table \ref{tab:twfe_hour} contains average treatment effect estimates, figure \ref{fig:eventstudy_hour} displays the event-study estimates, and table \ref{tab:heterogeneity_hour} displays the heterogeneity estimates. We find that the choice of time frequency makes very little difference for the main estimates, but the four-hour intervals reduce noise in the pre-period of the event-study analysis, resulting in cleaner pre-trends.  Given the reduction in noise, we favor the four-hour analysis displayed in the main text.

\begin{table}
    \centering
    \begin{threeparttable}
    \caption{Treatment effect estimates using hourly data \label{tab:twfe_hour}}
    \begin{tabular}{lccc} \hline
 & (1) & (2) & (3) \\
VARIABLES & Thermostat setting & Thermostat $\leq$ 65F & Fan run time \\ \hline
 &  &  &  \\
Michigan x Post & -1.116** & 0.126** & -1.420 \\
 & (0.134) & (0.008) & (0.521) \\
 &  &  &  \\
Observations & 8,474,273 & 8,474,273 & 8,509,460 \\
R-squared & 0.666 & 0.476 & 0.631 \\
Weather controls & YES & YES & YES \\
Household FE & YES & YES & YES \\
Time FE & YES & YES & YES \\
Day of week $\times$ hour of day $\times$ state & YES & YES & YES \\
 Pre-treatment mean & 66.87 & 0.291 & 18.13 \\ \hline
\end{tabular}

    \begin{tablenotes}
        \footnotesize  Estimates of regressions from equation \ref{eq:twfe}. The sample includes hourly household observations from January 2nd-January 31st. Pre-treatment means reported for Michigan households.  Standard errors clustered at the state level.  ** p$<$0.01, * p$<$0.05
    \end{tablenotes}
\end{threeparttable}
\end{table}

\begin{table}
    \centering
    \small
    \begin{threeparttable}
    \caption{Hourly heterogeneity analysis}
    {\def\sym#1{\ifmmode^{#1}\else\(^{#1}\)\fi} \begin{tabular}{l*{3}{c}} \hline
                &\multicolumn{1}{c}{(1)}&\multicolumn{1}{c}{(2)}&\multicolumn{1}{c}{(3)}\\
                &\multicolumn{1}{c}{Thermostat setting}&\multicolumn{1}{c}{Thermostat $\leq$ 65F}&\multicolumn{1}{c}{Fan run time}\\
\hline
59 F or lower expected X Treatment&     0.50        &    -0.12\sym{**}&     0.42        \\
                &   (0.47)        &  (0.026)        &   (0.94)        \\
59-61 F expected X Treatment&   0.0096        &   -0.043        &     0.38        \\
                &   (0.24)        &  (0.024)        &   (0.76)        \\
61-63 F expected X Treatment&    0.049        &   -0.025        &   0.0047        \\
                &   (0.16)        &  (0.020)        &   (0.54)        \\
65-67 F expected X Treatment&    -0.11        &   0.0043        &     0.37        \\
                &  (0.080)        &  (0.016)        &   (0.40)        \\
67-69 F expected X Treatment&    -0.28\sym{**}&   -0.012        &     0.24        \\
                &   (0.10)        &  (0.019)        &   (0.45)        \\
69-71 F expected X Treatment&    -0.33\sym{**}&   -0.072\sym{**}&    -0.12        \\
                &   (0.11)        &  (0.017)        &   (0.46)        \\
71-73 F expected X Treatment&    -0.29        &    -0.12\sym{**}&   0.0046        \\
                &   (0.17)        &  (0.019)        &   (0.62)        \\
73-75 F expected X Treatment&    -0.39        &    -0.11\sym{**}&    -0.24        \\
                &   (0.27)        &  (0.021)        &   (0.98)        \\
Higher than 75 F expected X Treatment&     0.82        &    -0.17\sym{**}&    -3.01        \\
                &   (0.49)        &  (0.025)        &   (2.25)        \\
40-45\% Democrat X Treatment&    -0.44        &    0.068\sym{*} &     0.28        \\
                &   (0.27)        &  (0.029)        &   (0.71)        \\
45-50\% Democrat X Treatment&    -0.13        &    0.036        &     1.27        \\
                &   (0.27)        &  (0.029)        &   (1.00)        \\
50-55\% Democrat X Treatment&    -0.35        &    0.056\sym{*} &    -0.23        \\
                &   (0.23)        &  (0.027)        &   (0.97)        \\
55-60\% Democrat X Treatment&    -0.28        &    0.053        &     2.39        \\
                &   (0.30)        &  (0.038)        &   (1.56)        \\
60-65\% Democrat X Treatment&    -0.53        &    0.086        &     2.05        \\
                &   (0.40)        &  (0.055)        &   (1.86)        \\
65-70\% Democrat X Treatment&    -0.58        &    0.094        &    -0.28        \\
                &   (0.43)        &  (0.059)        &   (1.76)        \\
70-75\% Democrat X Treatment&    -0.86\sym{*} &    0.093        &     2.63        \\
                &   (0.42)        &  (0.051)        &   (1.83)        \\
                &                 &                 &                 \\
Observations    &7,811,227        &7,811,227        &7,812,552        \\
R-squared       &     0.81        &     0.70        &     0.65        \\
Household FE    &      YES        &      YES        &      YES        \\
Time FE         &      YES        &      YES        &      YES        \\
Controls $\times$ treatment&      YES        &      YES        &      YES        \\
Day of week $\times$ hour of day $\times$ state&      YES        &      YES        &      YES        \\
Expected thermostat $\times$ time FE&      YES        &      YES        &      YES        \\
Pre-treatment mean&     66.9        &     0.29        &     18.1        \\
\hline \end{tabular} }

    \label{tab:heterogeneity_hour}
    \begin{tablenotes}
        \footnotesize Results from the estimation of equation \ref{eq:heterogeneity} using hourly data. The sample includes hourly observations from January 2nd-January 31st. Pre-treatment means reported for Michigan households. Standard errors cluster-bootstrapped to incorporate the sampling error from estimation of the baseline thermostat setting.  ** p$<$0.01, * p$<$0.05
    \end{tablenotes}
\end{threeparttable}
\end{table}

\section{Robustness checks \label{sec:robustness}}

\begin{table}
    \centering \small
    \begin{threeparttable}
    \caption{Alternate difference-in-differences specifications with temperature as the outcome variable \label{tab:didapp}}
    \begin{tabular}{lcccccc} \hline
 & (1) & (2) & (3) & (4) & (5) & (6) \\
VARIABLES & Model 1 & Model 2 & Model 3 & Model 4 & Model 5 & Model 6 \\ \hline
 &  &  &  &  &  &  \\
Michigan & -0.588 & -0.591 &  &  &  &  \\
 & (0.356) & (0.349) &  &  &  &  \\
Post & 1.203** & 1.054* & 0.858** & 0.675** &  &  \\
 & (0.180) & (0.287) & (0.149) & (0.110) &  &  \\
Michigan x Post & -0.992** & -1.171** & -1.196** & -1.100** & -1.052** & -1.063** \\
 & (0.180) & (0.223) & (0.161) & (0.135) & (0.129) & (0.131) \\
 &  &  &  &  &  &  \\
Observations & 2,126,410 & 2,126,338 & 2,126,336 & 2,126,336 & 2,126,336 & 2,121,536 \\
R-squared & 0.008 & 0.012 & 0.687 & 0.705 & 0.708 & 0.841 \\
Pre-treatment mean & 66.87 & 66.87 & 66.87 & 66.87 & 66.87 & 66.87 \\
Weather controls &  & YES & YES & YES & YES & YES \\
Household FE &  &  & YES & YES & YES & YES \\
Time FE &  &  &  &  & YES & YES \\
Day of week $\times$ hour of day $\times$ FE &  &  &  &  &  & YES \\
Day of week &  &  &  & YES &  &  \\
 Hour of day &  &  &  & YES &  &  \\ \hline
\end{tabular}

    \begin{tablenotes}
        \footnotesize The sample includes 4-hour-average household observations from January 2nd-January 31st. Pre-treatment means reported for Michigan households.  Standard errors clustered at the state level.  ** p$<$0.01, * p$<$0.05
    \end{tablenotes}
    \end{threeparttable}
\end{table}

\begin{table}
    \centering \small
    \begin{threeparttable}
    \caption{Alternate difference-in-differences specifications with compliance as the outcome variable \label{tab:lpmapp}}
    \begin{tabular}{lcccccc} \hline
 & (1) & (2) & (3) & (4) & (5) & (6) \\
VARIABLES & Model 1 & Model 2 & Model 3 & Model 4 & Model 5 & Model 6 \\ \hline
 &  &  &  &  &  &  \\
Michigan & 0.049 & 0.049 &  &  &  &  \\
 & (0.025) & (0.025) &  &  &  &  \\
Post & -0.081** & -0.074* & -0.061** & -0.043** &  &  \\
 & (0.010) & (0.023) & (0.009) & (0.007) &  &  \\
Michigan x Post & 0.093** & 0.108** & 0.111** & 0.102** & 0.098** & 0.100** \\
 & (0.010) & (0.014) & (0.007) & (0.008) & (0.007) & (0.007) \\
 &  &  &  &  &  &  \\
Observations & 2,126,410 & 2,126,338 & 2,126,336 & 2,126,336 & 2,126,336 & 2,121,536 \\
R-squared & 0.004 & 0.007 & 0.471 & 0.497 & 0.502 & 0.777 \\
Pre-treatment mean & 0.270 & 0.270 & 0.270 & 0.270 & 0.270 & 0.270 \\
Weather controls &  & YES & YES & YES & YES & YES \\
Household FE &  &  & YES & YES & YES & YES \\
Time FE &  &  &  &  & YES & YES \\
Day of week $\times$ hour of day $\times$ FE &  &  &  &  &  & YES \\
Day of week &  &  &  & YES &  &  \\
 Hour of day &  &  &  & YES &  &  \\ \hline
\end{tabular}

    \begin{tablenotes}
        \footnotesize The sample includes 4-hour-average household observations from January 2nd-January 31st. Pre-treatment means reported for Michigan households.  Standard errors clustered at the state level.  ** p$<$0.01, * p$<$0.05
    \end{tablenotes}
\end{threeparttable}
\end{table}

\begin{table}
    \centering \small
    \begin{threeparttable}
    \caption{Alternate difference-in-differences specifications with fan run time as the outcome variable \label{tab:fanapp}}
    \begin{tabular}{lcccccc} \hline
 & (1) & (2) & (3) & (4) & (5) & (6) \\
VARIABLES & Model 1 & Model 2 & Model 3 & Model 4 & Model 5 & Model 6 \\ \hline
 &  &  &  &  &  &  \\
Michigan & -2.503* & -2.997* &  &  &  &  \\
 & (0.559) & (0.894) &  &  &  &  \\
Post & 10.254** & 1.409* & 0.661** & 0.898** &  &  \\
 & (0.735) & (0.419) & (0.128) & (0.102) &  &  \\
Michigan x Post & -2.206* & -0.933 & -1.338* & -1.450* & -1.483* & -1.471* \\
 & (0.735) & (0.591) & (0.451) & (0.498) & (0.483) & (0.511) \\
 &  &  &  &  &  &  \\
Observations & 2,135,188 & 2,135,116 & 2,135,114 & 2,135,114 & 2,135,114 & 2,130,280 \\
R-squared & 0.020 & 0.071 & 0.737 & 0.749 & 0.751 & 0.851 \\
Pre-treatment mean & 18.09 & 18.09 & 18.09 & 18.09 & 18.09 & 18.09 \\
Weather controls &  & YES & YES & YES & YES & YES \\
Household FE &  &  & YES & YES & YES & YES \\
Time FE &  &  &  &  & YES & YES \\
Day of week $\times$ hour of day $\times$ FE &  &  &  &  &  & YES \\
Day of week &  &  &  & YES &  &  \\
 Hour of day &  &  &  & YES &  &  \\ \hline
\end{tabular}

    \begin{tablenotes}
        \footnotesize The sample includes 4-hour-average household observations from January 2nd-January 31st. Pre-treatment means reported for Michigan households.  Standard errors clustered at the state level.  ** p$<$0.01, * p$<$0.05
    \end{tablenotes}
\end{threeparttable}
\end{table}

In this section, we test the sensitivity of the average treatment effect estimates to specification, sample selection, and potential spillovers.  We find that the estimates are not affected by these potential confounders.

First, we examine the effect of specification choice on the difference-in-difference estimates.  In the main text, we display results using a two-way fixed effects approach with weather controls day of week times hour of sample times state indicators.  Tables \ref{tab:didapp} - \ref{tab:fanapp} display alternative specification coefficient estimates for each outcome variable.  Column one displays estimates using a standard difference in differences specification with indicator variables for being in Michigan, being in the post-treatment period, and the interaction of these two.  Column two adds time-varying controls for temperature and humidity.  Column three replaces the treatment group indicator with household fixed effects. Column four replaces the post-treatment indicator with day-of-week and time of day indicator variables.  Column five is a two-way fixed effects specification.  Column six uses day-of-week by hour-of-day by household indicators to control for household-specific temporal patterns.  The results do not differ substantially across all specifications for each outcome variable.

\begin{table}[h]
    \centering
    \begin{threeparttable}
    \caption{Balanced panel robustness check \label{tab:balance}}
    \begin{tabular}{lccc} \hline
 & (1) & (2) & (3) \\
VARIABLES & Thermostat setting & Thermostat $\leq$ 65F & Fan run time \\ \hline
 &  &  &  \\
Michigan x Post & -1.09** & 0.10** & -1.53* \\
 & (0.13) & (0.01) & (0.52) \\
 &  &  &  \\
Observations & 1,850,760 & 1,850,760 & 1,850,760 \\
R-squared & 0.71 & 0.50 & 0.75 \\
Weather controls & YES & YES & YES \\
Household FE & YES & YES & YES \\
Time FE & YES & YES & YES \\
Day of week $\times$ hour of day $\times$ state & YES & YES & YES \\
 Pre-treatment mean & 66.88 & 0.269 & 18.33 \\ \hline
\end{tabular}

    \begin{tablenotes}
        \footnotesize Estimates of regressions from equation \ref{eq:twfe} using a balanced panel of households.  The sample includes 4-hour-average household observations from January 2nd-January 31st. Pre-treatment means reported for Michigan households.  Standard errors clustered at the state level.  ** p$<$0.01, * p$<$0.05
    \end{tablenotes}
\end{threeparttable}
\end{table}

Next, we consider the possibility of sample selection.  In the sample, 6.1\% of households enter late or leave early.  This is best thought of as a sample selection problem as we do not observe the households before or after these points.  In addition, 5.2\% of observations are missing data on thermostat setting or weather data.  Because entry, exit, and missingness are unrelated to the treatment, we consider the missing observations to be ``missing completely at random'' and are therefore unrelated to the error term \citep{wooldridge2007}.  Nonetheless, we interpolate missing values and drop households that enter the sample late or leave early and estimate the average treatment effect using a balanced panel of households using the two-way fixed effects specification of equation \ref{eq:twfe}.  Table \ref{tab:balance} displays the results of this estimation.  The estimates are slightly smaller than those in the main text but are not substantially different.

\begin{table}
    \centering
    \begin{threeparttable}
    \caption{Potential spillovers to bordering counties \label{tab:spillover}}
    \begin{tabular}{lccc} \hline
 & (1) & (2) & (3) \\
VARIABLES & Thermostat setting & Thermostat $\leq$ 65F & Fan run time \\ \hline
 &  &  &  \\
Michigan x Post & -1.08** & 0.10** & -1.47* \\
 & (0.13) & (0.01) & (0.52) \\
Border county x Post & -0.12 & 0.00 & -0.07 \\
 & (0.22) & (0.01) & (0.60) \\
 &  &  &  \\
Observations & 2,126,336 & 2,126,336 & 2,135,114 \\
R-squared & 0.71 & 0.50 & 0.75 \\
Weather controls & YES & YES & YES \\
Household FE & YES & YES & YES \\
Time FE & YES & YES & YES \\
Day of week $\times$ hour of day $\times$ state & YES & YES & YES \\
 Pre-treatment mean & 66.87 & 0.270 & 18.09 \\ \hline
\end{tabular}

    \begin{tablenotes}
        \footnotesize Estimates of regressions from equation \ref{eq:spillover} allowing for spillovers to bordering counties.  The sample includes 4-hour-average household observations from January 2nd-January 31st. Pre-treatment means reported for Michigan households.  Standard errors clustered at the state level.  ** p$<$0.01, * p$<$0.05
    \end{tablenotes}
\end{threeparttable}
\end{table}

Our next robustness check allows for the possibility of spillovers to counties bordering Michigan.  Because the text alerts go to cellphones based upon the closest cell phone tower, it is possible that households living near the border in Indiana and Ohio were also treated.  Illinois and Wisconsin do not border Michigan's lower peninsula.  3.4 percent of Ohio and 11.6 percent of Indiana sample households live in counties bordering Michigan.  We estimate the following regression, which allows for a spillover treatment effect for households living in border counties:
\begin{align} \label{eq:spillover}
    Y_{i,t} = \alpha_i + \lambda_t + \beta D_{i,t} + \sigma S_{i,t} + \gamma X_{i,t} + \delta_{s,h,d} + \varepsilon_{i,t},
\end{align}
where \(S_{i,t}\) is a treatment variable equal to one for households in counties that border Michigan's lower peninsula during the post-treatment period.  The estimated coefficient \(\sigma\) on \(S_{i,t}\) should be equal to zero if there are no spillovers into the bordering counties.  Table \ref{tab:spillover} displays the estimated coefficients using all three outcome variables.  In each regression, the estimated spillover coefficient is small and the confidence interval contains zero.  The coefficient on the main treatment variable is not substantially different from the estimates in the main text.  We have experimented with modeling potential spillovers as far as two counties away from the border and we do not see much difference in the estimated coefficients on the main treatment variable, so we do not display the results here.  In addition to these empirical results, searches of the archives of Toledo, Ohio's main newspaper, \textit{The Blade}, using the keywords ``polar vortex'' and ``natural gas'' does not turn up any news coverage of the Michigan event.  These results suggest that any potential spillover effects are not affecting the estimates.

\section{Placebo analyses \label{sec:placebo}}

Ten days before the polar vortex, Michigan experienced a similar cold wave that did not coincide with a supply-side emergency causing an emergency request for voluntary thermostat reductions.  Figure \ref{fig:temps_a} displays mean daily temperatures for sample households in Michigan in January 2019.  Temperatures on January 20-21 dropped from 25\(^\circ\)F to below 10\(^\circ\)F, making these days a good placebo event for the January 30-31 emergency.  Because there was no emergency request to reduce thermostat settings on the 20th and 21st, we would expect to see no difference in heating behavior for Michigan and control states.

\subsection{Average treatment effects}

\begin{figure}
\centering
\begin{subfigure}[t]{.49\textwidth}
    \centering
    \includegraphics[width=\linewidth]{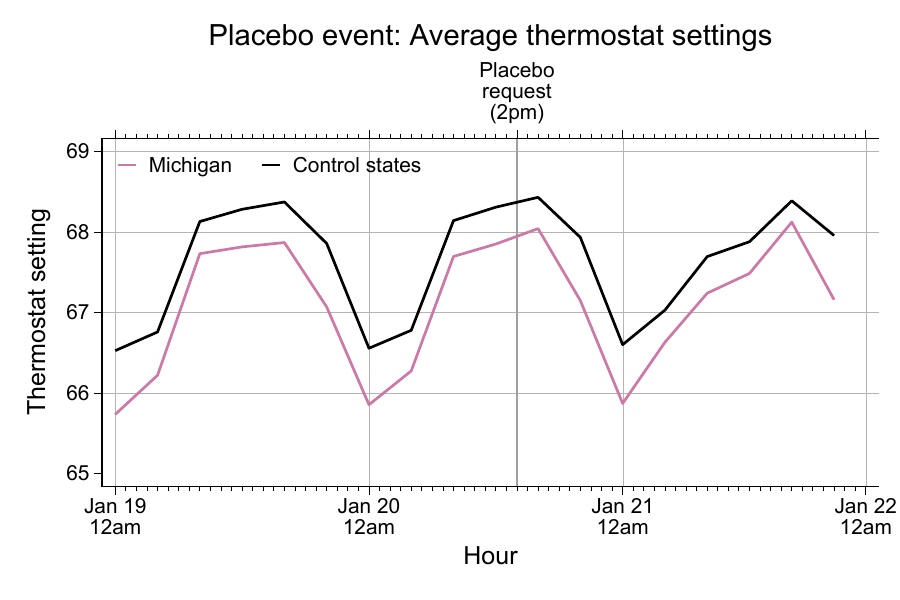}
            \caption{}\label{fig:thermostatparallel_placebo}
    \end{subfigure}
\begin{subfigure}[t]{.49\textwidth}
    \centering
    \includegraphics[width=\linewidth]{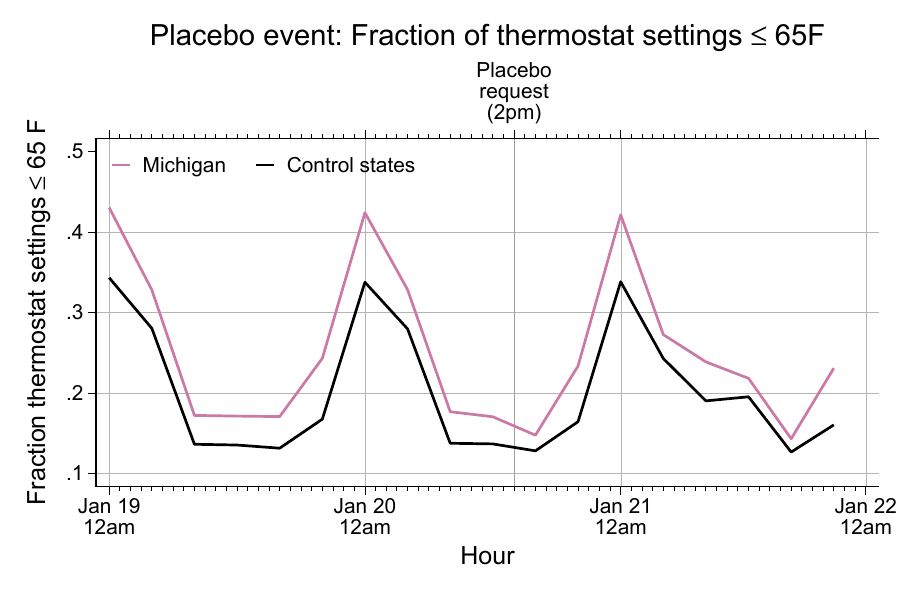}
        \caption{}\label{fig:complianceparallel_placebo}
\end{subfigure}
\medskip
\begin{subfigure}[t]{.49\textwidth}
    \centering
    \vspace{0pt}
    \includegraphics[width=\linewidth]{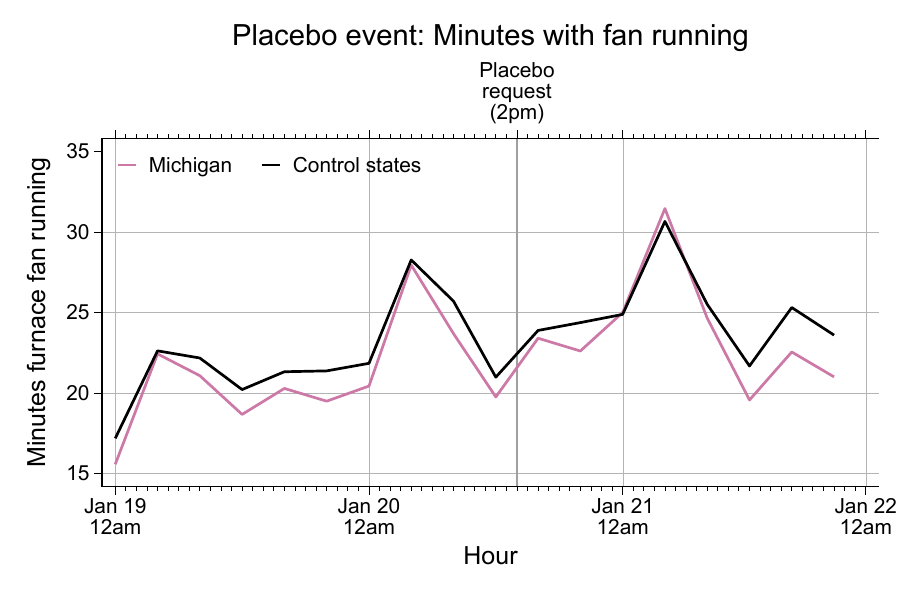}
        \caption{}\label{fig:fanparallel_placebo}
\end{subfigure}
\begin{minipage}[t]{.49\textwidth}
    \medskip \medskip
    \caption{Sample average values of the outcome variables for treatment and control households, January 19 - 22.  Panel (\subref{fig:thermostatparallel_placebo}) plots average thermostat settings, panel (\subref{fig:complianceparallel_placebo}) plots the fraction of households with thermostat settings at or below 65\(^\circ\)F, and panel (\subref{fig:fanparallel_placebo}) plots the average furnace fan run time in each time period. The vertical dashed line indicates 2:00 pm, the beginning of the placebo event.} \label{fig:paralleltrends_placebo}
\end{minipage}
\end{figure}

Figure \ref{fig:paralleltrends_placebo} displays average thermostat settings, fraction of households with thermostat settings at or below 65\(^{\circ}\)F, and fan running times for Michigan and control states from January 19-22.  Unlike in the main text, we do not see a change in heating behavior between Michigan and the controls states after the placebo treatment time.

We then estimate the two-way fixed-effects specification of regression equation \ref{eq:twfe} for the placebo event.  To do so, we include all observations observed between January 1st and January 21st, 2019 and treat 2:00 pm on January 20th as the placebo treatment time.  Table \ref{tab:placebo} displays the results of this estimation using thermostat setting, an indicator variable for having the thermostat at or below 65\(^\circ\)F, and fan running time as outcome variables.  As expected, the estimates are close to zero.  The only statistically significant estimate suggests that Michigan households increased thermostat settings by 0.06\(^\circ\)F, which we interpret as a precisely estimated zero.  This placebo procedure demonstrates that households in Michigan respond similarly to cold spells as households in the control states absent a request to reduce energy consumption.  

\begin{table}
    \centering
    \begin{threeparttable}
    \caption{Placebo analysis of January 20th cold wave \label{tab:placebo}}
    \begin{tabular}{lccc} \hline
 & (1) & (2) & (3) \\
VARIABLES & Thermostat setting & Thermostat $\leq$ 65F & Fan run time \\ \hline
 &  &  &  \\
Michigan x Post & 0.06* & -0.00 & -0.08 \\
 & (0.02) & (0.00) & (0.39) \\
 &  &  &  \\
Observations & 1,413,411 & 1,413,411 & 1,419,364 \\
R-squared & 0.72 & 0.52 & 0.77 \\
Weather controls & YES & YES & YES \\
Household FE & YES & YES & YES \\
Time FE & YES & YES & YES \\
Day of week $\times$ hour of day $\times$ state & YES & YES & YES \\
 Pre-treatment mean & 66.73 & 0.281 & 16.45 \\ \hline
\end{tabular}

    \begin{tablenotes}
        \footnotesize The sample includes 4-hour-average household observations from January 1st-January 21st. Pre-treatment means reported for Michigan households. We expect the estimates from this placebo estimation to be close to zero.  Standard errors clustered at the state level.  ** p$<$0.01, * p$<$0.05
    \end{tablenotes}
\end{threeparttable}
\end{table}

\subsection{Five-minute thermostat settings} \label{sec:fiveminplacebo}

\begin{figure}
    \begin{subfigure}[t]{.49\textwidth}
        \centering
        \includegraphics[width=\linewidth]{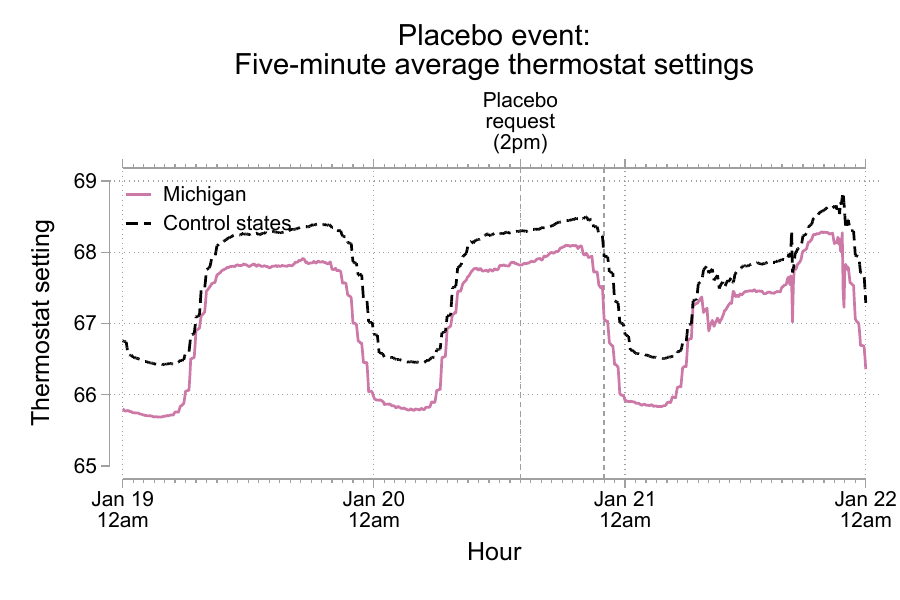}
        \caption{}\label{fig:fiveminp_a}
    \end{subfigure}
    \begin{subfigure}[t]{.49\textwidth}
        \centering
        \includegraphics[width=\linewidth]{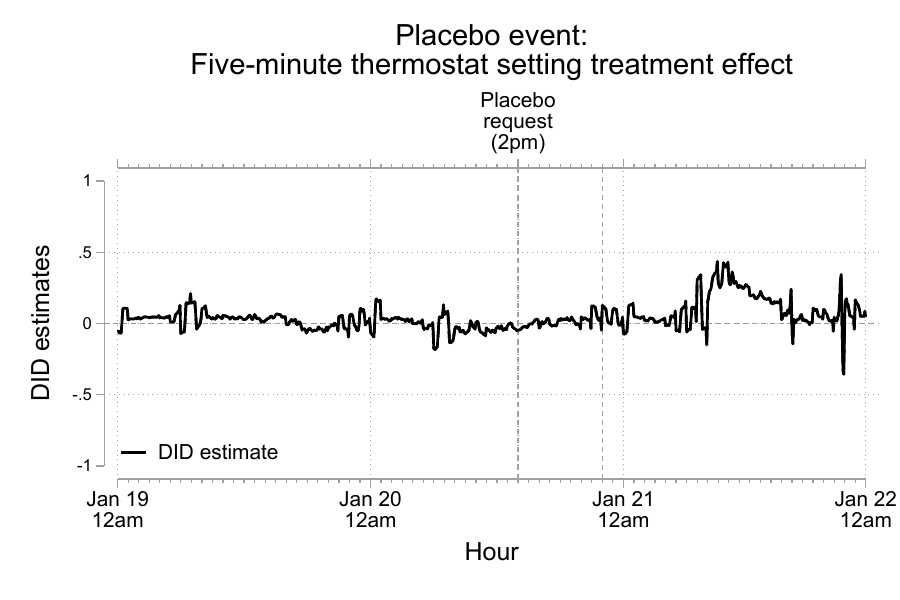}
        \caption{}\label{fig:fiveminp_b}
    \end{subfigure}
    \caption{(a): Five-minute sample mean thermostat settings for Michigan and control households from 12:00 pm January 20th - 11:59 pm January 21st.  (b): Five-minute difference-in-differences estimate.}
    \label{fig:fiveminp}
\end{figure}

In figure \ref{fig:fiveminp_a}, we plot five-minute thermostat setting data for Michigan and the control states between 12:00 pm on January 20th and 11:59 pm on January 21st.  In figure \ref{fig:fiveminp_b}, we plot a difference-in-differences estimate of the placebo treatment effect, which we construct as the difference between five-minute thermostat setting for Michigan and control states during the event minus the average difference in same time-of-day and day-of-week five-minute thermostat settings before the event.  In these figures, we see a discrete increase in thermostat setting in the five-minute periods beginning at 7:00 pm and 12:00 am for both treatment and control households.  These discrete changes correspond to commonly programmed times for the thermostat to automatically change.  Other than in a few five-minute periods on January 21st, the measured treatment effect is zero in the placebo period.  These treatment effects are driven by seemingly spurious five-minute jumps in the average thermostat setting for Michigan households.  Overall, these placebo plots demonstrate the validity of the difference-in-differences approach and verify that discrete increases at 7:00 pm and 12:00 am are common and not artifacts of the emergency event.

\subsection{Heterogeneity analysis} \label{sec:heterogeneityplacebo}

\begin{table}
    \centering
    \begin{threeparttable}
    \caption{Placebo January 20th cold wave heterogeneity analysis \label{tab:reference_placebo}}
    {\def\sym#1{\ifmmode^{#1}\else\(^{#1}\)\fi} \begin{tabular}{l*{3}{c}} \hline
                &\multicolumn{1}{c}{(1)}&\multicolumn{1}{c}{(2)}&\multicolumn{1}{c}{(3)}\\
                &\multicolumn{1}{c}{Thermostat setting}&\multicolumn{1}{c}{Thermostat $\leq$ 65F}&\multicolumn{1}{c}{Fan run time}\\
\hline
59 F or lower expected X Treatment&     0.19        &   -0.012        &    -0.53        \\
                &   (0.43)        &  (0.028)        &   (0.85)        \\
59-61 F expected X Treatment&    -0.37        &    0.012        &    -0.78        \\
                &   (0.29)        &  (0.027)        &   (0.73)        \\
61-63 F expected X Treatment&  -0.0053        &   -0.014        &    -0.80        \\
                &   (0.13)        &  (0.020)        &   (0.55)        \\
65-67 F expected X Treatment&    0.010        &   -0.017        &   -0.035        \\
                &  (0.080)        &  (0.018)        &   (0.42)        \\
67-69 F expected X Treatment&    0.061        &   -0.028        &     0.26        \\
                &  (0.083)        &  (0.017)        &   (0.45)        \\
69-71 F expected X Treatment&    0.082        &   -0.027        &    -0.18        \\
                &  (0.099)        &  (0.017)        &   (0.42)        \\
71-73 F expected X Treatment&     0.13        &   -0.028        &     0.11        \\
                &   (0.13)        &  (0.017)        &   (0.63)        \\
73-75 F expected X Treatment&    -0.26        &   -0.023        &    -0.28        \\
                &   (0.20)        &  (0.016)        &   (1.02)        \\
Higher than 75 F expected X Treatment&     0.62        &   -0.047\sym{*} &    -0.12        \\
                &   (0.49)        &  (0.021)        &   (1.95)        \\
40-45\% Democrat X Treatment&     0.12        &  0.00069        &     0.76        \\
                &   (0.21)        &  (0.020)        &   (0.58)        \\
45-50\% Democrat X Treatment&    -0.17        &   -0.018        &     0.70        \\
                &   (0.19)        &  (0.019)        &   (0.70)        \\
50-55\% Democrat X Treatment&    -0.20        &    0.010        &     0.57        \\
                &   (0.18)        &  (0.018)        &   (0.67)        \\
55-60\% Democrat X Treatment&   -0.085        &   0.0086        &     1.45        \\
                &   (0.26)        &  (0.026)        &   (1.15)        \\
60-65\% Democrat X Treatment&    0.020        &   0.0034        &     2.42        \\
                &   (0.36)        &  (0.030)        &   (1.65)        \\
65-70\% Democrat X Treatment&    -0.38        &  -0.0094        &     1.33        \\
                &   (0.38)        &  (0.033)        &   (1.36)        \\
70-75\% Democrat X Treatment&    -0.41        &    0.029        &     1.60        \\
                &   (0.41)        &  (0.038)        &   (1.50)        \\
                &                 &                 &                 \\
Observations    &1,302,219        &1,302,219        &1,302,511        \\
R-squared       &     0.85        &     0.75        &     0.78        \\
FE              &      YES        &      YES        &      YES        \\
Hour            &      YES        &      YES        &      YES        \\
Controls        &      YES        &      YES        &      YES        \\
Expected thermostat level&      YES        &      YES        &      YES        \\
Pre-treatment mean&     66.7        &     0.28        &     16.4        \\
\hline \end{tabular} }

    \begin{tablenotes}
        \footnotesize Estimates of regressions from equation \ref{eq:heterogeneity} using placebo data from January 1st-January 21st.  Standard errors cluster-bootstrapped to incorporate the sampling error from estimation of the baseline thermostat setting.  ** p$<$0.01, * p$<$0.05
    \end{tablenotes}
\end{threeparttable}
\end{table}

One concern that we had was that the estimates from the heterogeneity analysis in the main text reflect statistical reversion to the mean rather than a meaningful pattern of results for the sub-groups (particularly for the baseline thermostat-setting results).  To test this, we estimate the heterogeneity regression analysis from equation \ref{eq:heterogeneity} during the placebo period.  Table \ref{tab:reference_placebo} displays the estimated coefficients and bootstrapped standard errors that account for the first-stage estimation of the baseline thermostat setting. The magnitude of the estimated coefficients in the placebo analysis are small and do not display systematic trends as you move away from the 65\(^\circ\)F compliance target.  Thus, we conclude that mean reversion is not driving our heterogeneity estimates in the main paper.

\section{Donald and Lang inference} \label{app:donaldlang}

Here, we use an aggregation approach inspired by \cite{donaldlang2007} to estimate the average treatment effects and provide valid inference for five clusters.  In our approach, we average our outcome variables to the state level \(s\) and estimate the following ordinary-least-squares regression:
\begin{align} \label{eq:DL}
    Y_{s,t} =  \alpha + \lambda_t + \beta D_{s,t} + \gamma X_{s,t} + \delta_{s,h,d} + \varepsilon_{s,t},
\end{align}
where \(Y_{s,t} = \Bar{Y}_{i,t}\) and \(X_{s,t} =  \Bar{X}_{i,t}\) are the state sample averages.  Under standard exogeneity assumptions and large state-cluster group sizes (implying normality of \(\varepsilon_{s,t}\) via the central limit theorem) estimation of \(\beta\) in equation \ref{eq:DL} is consistent and standard inference is valid \citep{wooldridge2010}.  When there are equal cluster sizes and a balanced panel, the aggregated Donald and Lang estimates are exactly equal to the estimates from the individual-level regression, but in our case the unbalanced panel and unequal cluster sizes will result in a slight difference in estimates.

Table \ref{tab:donaldlang} displays the Donald and Lang estimates of the average treatment effect for each outcome variable.  The estimated average treatment effects are slightly smaller than those in the main text, but are not substantially different.  Importantly, each estimated effect is statistically significant even under the conservative Donald and Lang inference test, which alleviates concerns that clustering at the state level may result in over-rejection of the null hypothesis.

\begin{table}
    \centering
    \begin{threeparttable}
    \caption{Donald and Lang estimation of average treatment effects \label{tab:donaldlang}}
    \begin{tabular}{lccc} \hline
 & (1) & (2) & (3) \\
VARIABLES & Thermostat setting & Thermostat $\leq$ 65F & Fan run time \\ \hline
 &  &  &  \\
Michigan x Post & -1.008** & 0.104** & -0.981** \\
 & (0.052) & (0.005) & (0.344) \\
 &  &  &  \\
Observations & 900 & 900 & 900 \\
R-squared & 0.990 & 0.989 & 0.985 \\
Weather controls & YES & YES & YES \\
Household FE & YES & YES & YES \\
Time FE & YES & YES & YES \\
Day of week $\times$ hour of day $\times$ state & YES & YES & YES \\
 Pre-treatment mean & 66.87 & 0.270 & 18.08 \\ \hline
\end{tabular}

    \begin{tablenotes}
        \footnotesize Estimates of regressions from equation \ref{eq:DL}.  The sample includes 4-hour state average observations from January 2nd-January 31st. Pre-treatment means reported for Michigan households.  Donald and Lang standard errors valid for finite number of clusters.  ** p$<$0.01, * p$<$0.05
    \end{tablenotes}
\end{threeparttable}
\end{table}

\clearpage

\section{External validity} \label{sec:externalvalidity}

Here, we discuss how likely our results are to generalize to non-smart thermostat households and to future energy crises.  Our research design takes advantage of the exceptional nature of the emergency event and availability of the smart thermostat data during this period to construct well-identified estimates of the effect of the emergency request on behavior, but it may be unclear to what extent these unique data and event are representative of other energy users in other events.  To orient our discussion, we make use of the ``SANS'' conditions for generalizability suggested by \cite{list2020}.  In the SANS framework, the external validity of a study can be assessed by discussing selection, attrition, naturalness, and scaling.  In our context, there was very little attrition from the sample (we assess the sensitivity of the results to keeping and including those who enter or leave the sample early in appendix \ref{sec:robustness}), which leaves selection, naturalness, and scaling for discussion.  In our discussion, we find little evidence that selection plays a role in our results.  Given that the emergency request is a natural experiment and that similar emergency requests are made during other energy emergencies, we believe the intervention is natural and likely to be representative of interventions in other energy contexts.  Finally, we discuss the ability of this intervention to scale vertically to the grid (ISO) level, and horizontally across states and other energy emergencies.  We believe our results are likely to generalize to other energy emergencies based on the availability of a wireless emergency alert system, the political climate, and the frequency of repeated requests which may result in habituation to the alerts.

\subsection{Selection}

\begin{table}
    \centering
    \small
    \begin{threeparttable}
    \caption{Estimates by adoption date \label{tab:tempadopters} }
    \begin{tabular}{lcccc} \hline
 & (1) & (2) & (3) & (4) \\
VARIABLES & Pre-2016 adopters & 2016 adopters & 2017 adopters & 2018 adopters \\ \hline
 &  &  &  &  \\
Michigan x Post & -1.152** & -1.223** & -1.027** & -1.081** \\
 & (0.111) & (0.107) & (0.146) & (0.151) \\
 &  &  &  &  \\
Observations & 163,033 & 372,872 & 714,402 & 846,655 \\
R-squared & 0.696 & 0.701 & 0.717 & 0.710 \\
Weather controls & YES & YES & YES & YES \\
Household FE & YES & YES & YES & YES \\
Time FE & YES & YES & YES & YES \\
Day of week $\times$ hour of day $\times$ state & YES & YES & YES & YES \\
 Pre-treatment mean & 66.20 & 66.83 & 66.72 & 67.10 \\ \hline
\end{tabular}

    \begin{tablenotes}
        \footnotesize Each column is an estimate of equation \ref{eq:twfe} by smart thermostat adoption date. The sample includes 4-hour-average household observations from January 2nd-January 31st. Pre-treatment means reported for Michigan households.  Standard errors are clustered at the state level. ** p$<$0.01, * p$<$0.05.
    \end{tablenotes}
\end{threeparttable}
    
\end{table}

Our primary concern about generalizability is the selection into owning an Ecobee thermostat and sharing data used for this study.  As other studies have noted \citep{burkhardtetal2019,blonzetal2021}, most papers in the residential energy literature rely on data from self-selected households---this paper is no different.  Given that our treatment is unrelated to selection, selection into the sample does not affect the internal validity of the research design, but may be relevant to the external validity of the estimates. Previous work has found that, on observable characteristics, the Ecobee Donate-Your-Data households are comparable to the average household in the nationally representative Residential Energy Consumption Survey sample, though the Ecobee households have slightly more members \citep{meieretal2019}.  Another study argues that the Ecobee smart thermostat features do not differ substantially from other smart thermostats \citep{blonzetal2021}. One notable difference in our case is that in Michigan, 99.89\% of Ecobee househoulds heat with natural gas compared with 75\% of households overall.  Despite being comparable on most observable characteristics, there remains a concern that households that choose to share data may be more willing to contribute to other public goods such as compliance during an emergency.  Furthermore, the Ecobee is marketed as being eco-friendly, which may also be correlated with pro-social attitudes.  Finally, smart thermostat features differ from traditional thermostats.  The largest and most relevant difference is that smart thermostat users can adjust the thermostat remotely via app, which reduces the cost of compliance.  \cite{brandonetal2021} find little evidence that smart thermostats on their own cause households to consume energy differently relative to households with conventional thermostats, but the concern remains that these households may find it easier to comply with the request. If these hypotheses about selection are correct, our estimates from smart-thermostat household behavior may overstate compliance relative to non-smart-thermostat households.

We assess the degree to which selection might affect our estimates in two ways.  First, we compare our average treatment effect to estimates of the reduction in residential natural gas consumption calculated using aggregate data provided by Consumers Energy, and second we consider differential responses to the emergency request by early adopters of the Ecobee thermostat.  During the event, forecasted natural gas demand using realized weather conditions was 3.3 billion cubic feet on January 30th and 2.9 billion cubic feet on January 31st.  After all reductions in consumption were accounted for, actual consumption was 3.0 billion cubic feet on January 30th and 2.6 billion cubic feet on January 31st, implying a 10.7\% and 10.5\% reduction in daily consumption from all sources (residential and non-residential). In a mostly-residential area of Detroit, Consumers Energy claimed to see a 10\% reduction in energy consumption.  Using furnace fan run time as a proxy variable for energy consumption, we estimate a 6\% reduction.  This estimate is similar but smaller in magnitude, which is the opposite of what we would expect if Ecobee users were complying more strongly with the request than the average household, although we caveat that the utility's estimate does include some non-residential consumption that may have been subject to curtailment.

The second test of selection examines whether early adopters of the smart thermostats responded differently to the request than late adopters.  In the smart thermostat data, we observe the date the smart thermostat account became active.  We hypothesize that early adopters of smart thermostats are more highly selected relative to the general population, whereas late adopters are more representative.  If early adopters have differential responses to the emergency request, this evidence would suggest that smart thermostat users are not representative of the general population.  

Table \ref{tab:tempadopters} displays the results of estimating the main specification from equation \ref{eq:twfe} on samples segmented by smart thermostat adoption year using thermostat setting as the outcome variable.\footnote{We pool households who adopted in 2016 or earlier, and we do not include households that adopted the smart thermostat in January 2019, as any differences may be attributed to lack of experience with the thermostat.}  The estimates are similar across adoption years, and none of the differences are statistically different from zero.  Thus, to the extent that early adopters are more highly selected, the treatment effect does not vary by that selection.

Overall, our diagnostics suggest a limited role for selection.  Hypothetically, all of the selection effects point to a potentially larger responsiveness of households in our data to the requests, but we do not find any evidence that suggests our estimates will not generalize out of sample.

\subsection{Naturalness}

Given that this is a natural experiment, the treatment subjects are subjected to a natural source of variation in the emergency request.  Moreover, emergency appeals with requested compliance levels are a standard response in the toolkit of public utilities responding to emergency shortages.  For example, California's Flex Alert system regularly includes a thermostat setting target of 78\(^\circ\)F.\footnote{See \href{https://flexalert.org/}{https://flexalert.org/}.}  Since this polar vortex in 2019, thermostat-targeted emergency appeals have been used in Texas and California in response to extreme  heat events in 2021 and 2022.\footnote{Texas ERCOT request to increase thermostats to 78\(^\circ\)F: \href{https://www.ercot.com/news/release?id=8b772e9e-51d0-4c3c-e653-1e5079f28e89}{https://www.ercot.com/news/release?id=8b772e9e-51d0-4c3c-e653-1e5079f28e89}.  California Governor request to increase thermostats to 78\(^\circ\)F: \href{https://twitter.com/CAgovernor/status/1567316274849660928}{https://twitter.com/CAgovernor/status/1567316274849660928}.}  Emergency appeals of this form are used widely, though future research should analyze the effect of variation in the reference level and requests for fixed reductions from the baseline (e.g., ``please reduce thermostat settings by 5\(^\circ\)F'').

\subsection{Scaling}

The Michigan emergency appeal was moderately successful at the state level, suggesting that similar appeals can be successful at a large scale.  Here, we discuss which policy design elements are necessary for similar emergency appeals to scale vertically to the grid (ISO) level, horizontally from state to state, and to future emergencies.  Our analysis focuses on the availability and use of the wireless emergency alert system, the political climate, and the frequency of requests.

Our event-study analyses in section \ref{sec:eventstudy} suggest that the use of the wireless emergency alert system were critical to achieving compliance.  Previous studies of requests for voluntary reductions have demonstrated low effectiveness of emergency requests when the request is made only by the utility or by local news media \citep{holladayetal2015}.  Thus, a wide-reaching message from an authority figure is crucial for scaling this effect.  In other contexts, these appeals will be limited by the reach of the emergency communications systems.  Areas with low media and technology penetration or with limited cell phone service may see limited success of emergency appeals.  Similarly, the size of the treatment effect peaked mid-day, after people had time to respond, suggesting that the timing of the messaging is pivotal in future emergencies.  ISOs spanning multiple states should have emergency communications relationships with state emergency agencies to gain access to key communications infrastructure.

Finally, we caution that similar emergency appeals may not scale when made repeatedly.  \cite{itoidatanaka2018} find that repeated conservation nudges result in habituation or desensitization, reducing their effectiveness over time.  The Michigan polar vortex appeal was a unique emergency and to our knowledge was the only such appeal in recent years.  Thus, the stimulus was novel, likely increasing the salience of the request.  In states such as Texas and California where requests are relatively commonplace, habituation may decrease the request's effectiveness.

\end{appendices}
\end{document}